\documentclass[11pt]{article}

\usepackage[T1]{fontenc}
\usepackage[utf8]{inputenc}
\usepackage{lmodern}
\usepackage{microtype}
\usepackage{amsmath,amssymb,amsthm}
\usepackage{booktabs}
\usepackage{graphicx}
\usepackage{xcolor}
\usepackage{longtable}
\usepackage{float}
\usepackage{placeins}
\usepackage{enumitem}
\usepackage[round]{natbib}
\usepackage[hidelinks]{hyperref}
\usepackage{geometry}
\newtheorem{theorem}{Theorem}

\newtheorem{corollary}{Corollary}
\newtheorem{proposition}{Proposition}
\theoremstyle{remark}
\newtheorem{remark}{Remark}

\DeclareMathOperator{\E}{E}
\DeclareMathOperator{\Var}{Var}
\DeclareMathOperator{\Cov}{Cov}
\DeclareMathOperator{\argmin}{arg\,min}
\DeclareMathOperator{\argmax}{arg\,max}
\newcommand{\R}{\mathbb{R}}
\renewcommand{\P}{\mathbb{P}}
\newcommand{\ind}{\mathbf{1}}
\newcommand{\revisedfigure}[2]{\begingroup\setlength{\fboxsep}{2pt}\setlength{\fboxrule}{0.6pt}\fcolorbox{white}{white}{\includegraphics[width=\dimexpr#1-2\fboxsep-2\fboxrule\relax]{#2}}\endgroup}

\title{\textbf{Moment-Calibrated Conformal Prediction with Side Information}}
\author{
\begin{minipage}{0.96\textwidth}
  \centering
  \linespread{1}\selectfont
  Guanghui Wang\textsuperscript{1}, Liuhua Peng\textsuperscript{2}, Liu Liu\textsuperscript{3}, and Changliang Zou\textsuperscript{1}\\[0.6em]
  {\small\textsuperscript{1}School of Statistics and Data Science, Nankai University,}\\
  {\small Tianjin, China}\\[0.25em]
  {\small\textsuperscript{2}School of Mathematics and Statistics, University of Melbourne,}\\
  {\small Victoria, Australia}\\[0.25em]
  {\small\textsuperscript{3}School of Mathematical Science, Chengdu University of Technology,}\\
  {\small Chengdu, China}
\end{minipage}
}
\hypersetup{pdfauthor={Guanghui Wang, Liuhua Peng, Liu Liu, and Changliang Zou}}
\date{}

\begin{document}

\maketitle

\begin{abstract}
  For a fixed conformal score function, standard split conformal prediction calibrates a prediction set using the empirical distribution of the conformal scores, without using available side information about the data. We propose a moment-calibrated approach that uses such information to reduce variability across calibration data sets. Side information involving the covariates, the response, or both is expressed as moment restrictions, which are enforced through empirical likelihood weights on the calibration data points. The resulting weighted empirical distribution of the scores determines the weighted empirical quantile without changing the score function. We establish finite-sample bounds on training-conditional and marginal coverage of the moment-calibrated prediction set. We derive asymptotic distributions for training-conditional coverage and average prediction-set size, show that their asymptotic variances are no larger than the corresponding variances under standard split conformal prediction, and characterize when the reductions are strict. In the semi-supervised setting, we identify an optimal moment function of a given covariate representation, develop implementations using prespecified moment functions or a function learned by cross-fitting, and quantify the effect of the unlabeled sample size. We also extend the method to coverage conditions indexed by a class of functions of the covariates through weighted quantile regression. Numerical studies demonstrate the effectiveness of the proposed methods.
\end{abstract}

\noindent\textbf{Keywords:} training-conditional coverage; moment restrictions; prediction-set size; semi-supervised learning; weighted quantiles.

\section{Introduction}
\label{sec:introduction}

Conformal prediction constructs prediction sets with distribution-free coverage guarantees \citep{Vovk2005}. Let \(Z=(X,Y)\), where \(X\in\mathcal X\) denotes the covariates and \(Y\in\mathcal Y\) the response. Let \(s:\mathcal X\times\mathcal Y\to\mathbb R\) be a conformal score function constructed using independent training data, and write \(S=s(X,Y)\). Larger scores indicate less conformity. For a nominal miscoverage level \(\alpha\in(0,1)\), standard split conformal prediction uses an adjusted empirical \((1-\alpha)\)-quantile \(\widehat q_{\alpha}^{\,\mathrm{sc}}\) of the conformal scores computed from the calibration data and returns the prediction set
\[
  C_{\widehat q_{\alpha}^{\,\mathrm{sc}}}(x)
  =
  \{y:s(x,y)\le\widehat q_{\alpha}^{\,\mathrm{sc}}\}.
\]
Exchangeability of the calibration and test scores gives the finite-sample marginal coverage guarantee of standard split conformal prediction \citep{LeiGsellRinaldoTibshiraniWasserman2018,AngelopoulosBarberBates2026}. Even with \(s\) fixed, the threshold varies with the calibration data, leading to variation in the coverage probability conditional on those data, as well as in the average prediction-set size. Marginal coverage describes coverage averaged over calibration data sets, rather than its variability across them.

We study whether additional information about the data distribution can be used in the calibration step to reduce this variability. In many applications, summaries of the data distribution are available in addition to the training and calibration data. We refer to these summaries as \emph{side information} and express them as moment restrictions \(\E\{h(Z)\}=0\), where \(h(Z)\in\R^r\). For example, in credit-default prediction, let \(Y=1\) denote default and \(Y=0\) otherwise. Let \(G\subseteq\mathcal X\) denote the set of covariate values for borrowers whose credit scores fall below a prespecified cutoff. If \(\pi_G=\P\{X\in G\}\) is known, then \(h(Z)=\ind(X\in G)-\pi_G\). For a fitted default-probability model \(\eta\), the score \(S=\{Y-\eta(X)\}^2\) also gives a moment function \(h(Z)=S-\mu_S\) when its mean \(\mu_S=\E\{S\}\) is known. Such summaries may be supplied by historical records or external validation studies.

For calibration data points \(Z_1,\ldots,Z_n\), we choose positive weights \(w_1,\ldots,w_n\) by empirical likelihood subject to
\[
  \sum_{i=1}^n w_i=1,
  \qquad
  \sum_{i=1}^n w_i h(Z_i)=0.
\]
The weighted empirical distribution of the conformal scores then determines the weighted quantile, changing the calibration step while leaving the score function and the conformal scores unchanged. If the constraints are infeasible, we use uniform weights. We analyze the resulting prediction sets in terms of finite-sample coverage and the variability of training-conditional coverage and average prediction-set size across calibration data sets.

Our main contributions are as follows.
\begin{itemize}[leftmargin=*,itemsep=0.35em,topsep=0.35em]
  \item \textbf{Coverage and variance comparisons.} For known moments, we establish finite-sample bounds for training-conditional and marginal coverage. Under regularity conditions, the limiting variances of training-conditional coverage and average prediction-set size are no larger than those under standard split conformal prediction, and we characterize when each reduction is strict.
  \item \textbf{Optimal moment functions and unlabeled data.} In the semi-supervised setting, we identify an optimal function of a given covariate representation for reducing limiting variance. We develop implementations using prespecified moment functions or a function learned by cross-fitting, and quantify how the unlabeled sample size affects the reduction.
  \item \textbf{Extension to conditional coverage.} We construct covariate-dependent thresholds for the conformal score through weighted quantile regression to address coverage conditions indexed by a class of functions of the covariates   \citep{GibbsCherianCandes2025}. For fixed finite-dimensional linear classes, we establish finite-sample coverage bounds and show that the limiting covariance matrices of the fitted coefficients and coverage moments are no larger than those under ordinary quantile regression over the same class.
\end{itemize}

\subsection{Related work}

Weighted conformal prediction addresses covariate shift by accounting for differences between the calibration and test distributions \citep{TibshiraniBarberCandesRamdas2019}. Local and conditional coverage have also been studied through localization and covariate-dependent calibration \citep{Guan2023LCP,HoreBarber2025LocalWeights,GibbsCherianCandes2025}. In this paper, the calibration data and test point follow the same distribution, and the weights assigned to the calibration data points enforce known moment restrictions to reduce variability in training-conditional coverage.

Calibration constraints are used in survey sampling to incorporate known finite-population totals \citep{DevilleSarndal1992}. Empirical likelihood provides a general framework for inference under moment restrictions \citep{Owen1988,Qin1994EL,Owen2001EL}. Our construction uses empirical likelihood to assign weights to the calibration data points and estimate the score distribution. The analysis concerns the coverage and average size of prediction sets formed from its quantiles.

Our semi-supervised setting is related to inference using unlabeled covariates \citep{ZhangBrownCai2019,ChakraborttyCai2018,SongLinZhou2024} and prediction-powered inference based on predictions from pretrained or cross-fitted models \citep{AngelopoulosEtAl2023PPI,ZrnicCandes2024}. Efficiency improvements in prediction-powered inference use tuning and learned adjustments \citep{AngelopoulosDuchiZrnic2023,GanLiangZou2024,MiaoEtAl2025,JiLeiZrnic2025}. The closest connection is to \citet{WenJiaRenWangZou2025}, who develop a cross-fitted semi-supervised distribution estimator based on conditional distribution estimation. Such a distribution estimator could also be used to estimate a quantile of the score distribution. Our method constructs a weighted empirical distribution by matching moments and studies the resulting coverage and prediction-set size. It can use prespecified moment functions or learn a function tailored to the \((1-\alpha)\)-quantile of the score distribution.

Section~\ref{sec:marginal} presents the method for known moments, with coverage guarantees and asymptotic variance comparisons. Section~\ref{sec:semi-supervised} considers the semi-supervised setting, identifying optimal moment functions and developing practical implementations. Section~\ref{sec:conditional} extends the method to function-class conditional coverage. Sections~\ref{sec:numerical} and~\ref{sec:applications} present simulation studies and real-data analyses, respectively. Proofs and additional results are provided in the Supplementary Material.

\section{Moment Restrictions for Conformal Prediction}
\label{sec:marginal}

\subsection{Weighted empirical distributions and prediction sets}
\label{sec:marginal-method}

Throughout, probabilities and expectations are conditional on the training data used to construct \(s\), which is therefore treated as fixed. Let \(P\) denote the distribution of \(Z\). Let \(Z_i=(X_i,Y_i)\), \(i=1,\ldots,n\), be i.i.d. calibration data points from \(P\), and let \(Z_{n+1}=(X_{n+1},Y_{n+1})\) be an independent test point from \(P\), with \(Y_{n+1}\) unobserved. Write \(\mathcal D_n=(Z_1,\ldots,Z_n)\) and \(S_i=s(X_i,Y_i)\), \(i=1,\ldots,n+1\). Let \(F(q)=\P\{S\le q\}\) be the cumulative distribution function (CDF) of the score. We assume that the calibration and test scores have no ties almost surely.

For \(\alpha\in(0,1)\), the standard split conformal quantile is
\[
  \widehat q_{\alpha}^{\,\mathrm{sc}}
  =\inf\left\{q:\frac1n\sum_{i=1}^n\ind(S_i\le q)
  \ge\frac{(n+1)(1-\alpha)}n\right\},
\]
with \(\inf\varnothing=+\infty\) \citep{Vovk2005,AngelopoulosBarberBates2026}. For any threshold \(q\), define the prediction set \(C_q(x)=\{y:s(x,y)\le q\}\). Standard split conformal prediction uses \(C_{\widehat q_{\alpha}^{\,\mathrm{sc}}}(x)\).

Let \(h:\mathcal X\times\mathcal Y\to\R^r\) be a moment function, fixed conditional on the training data, for which the restriction \(\E\{h(Z)\}=0\) is known. We choose the weights \(\widehat w=(\widehat w_1,\ldots,\widehat w_n)\) assigned to the calibration data points by empirical likelihood \citep{Owen1988,Qin1994EL}:
\begin{equation}
  \label{eq:weight-optimization}
  \widehat w
  \in
  \argmax_{w_1,\ldots,w_n>0}
  \sum_{i=1}^n\log w_i
  \quad
  \text{subject to}
  \quad
  \sum_{i=1}^n w_i=1,
  \quad
  \sum_{i=1}^n w_i h(Z_i)=0.
\end{equation}
If the constraints are infeasible, set \(\widehat w_i=1/n\). Define the weighted empirical CDF and weighted quantile by
\[
  \widehat F_w(q)=\sum_{i=1}^n\widehat w_i\ind(S_i\le q),
  \qquad
  \widehat q_\alpha=\inf\left\{q:\widehat F_w(q)\ge\frac{(n+1)(1-\alpha)}n\right\}.
\]
The moment-calibrated prediction set is
\[
  \widehat C(x)=C_{\widehat q_\alpha}(x)=\{y:s(x,y)\le\widehat q_\alpha\}.
\]
The adjusted quantile level yields the standard split conformal prediction set under uniform weights, which arise when \(h\equiv0\).

\subsection{Finite-sample coverage guarantees}
\label{sec:marginal-finite}

The training-conditional coverage of \(\widehat C\) is
\[
  \P\{Y_{n+1}\in\widehat C(X_{n+1})\mid\mathcal D_n\}=F(\widehat q_\alpha)
\]
\citep[e.g.,][]{AngelopoulosBarberBates2026}. It varies with \(\mathcal D_n\), with the score function fixed by conditioning on the training data. Let \(\Sigma_h=\E\{h(Z)h(Z)^\top\}\). For \(B>0\) and \(\delta\in(0,1)\), write
\[
  \varepsilon_{\mu,n}(\delta)=B\left\{\frac{2r\log(6r/\delta)}n\right\}^{1/2},
  \qquad
  \varepsilon_{\Sigma,n}(\delta)=B^2r\left\{\frac{2\log(6r^2/\delta)}n\right\}^{1/2}.
\]
These quantities bound the deviations of the sample mean and sample second-moment matrix of \(h(Z)\) under the assumptions below.

\begin{theorem}[Training-conditional coverage]
  \label{thm:finite-perturbation}
  Assume \(\E\{h(Z)\}=0\), \(\|h(Z)\|_2\le B\) almost surely, and \(\lambda_{\min}(\Sigma_h)\ge\kappa>0\). For \(\delta\in(0,1)\), suppose \(\varepsilon_{\mu,n}(\delta)\le\kappa/(16B)\) and \(\varepsilon_{\Sigma,n}(\delta)\le\kappa/2\). Then, with probability at least \(1-\delta\), the optimization in Eq.~\eqref{eq:weight-optimization} has a solution, and
  \begin{equation*}
    \scalebox{0.96}{$\displaystyle
    -\left\{\sqrt{\frac{\log(6/\delta)}{2n}}+\frac{16B}{\kappa}\varepsilon_{\mu,n}(\delta)\right\}
    \le \P\{Y_{n+1}\in\widehat C(X_{n+1})\mid\mathcal D_n\}-(1-\alpha)
    \le \sqrt{\frac{\log(6/\delta)}{2n}}+\frac{16B}{\kappa}\varepsilon_{\mu,n}(\delta)+\frac3n.
    $}
  \end{equation*}
\end{theorem}

For fixed \(r\), \(B\), \(\kappa\), and \(\delta\), the magnitudes of both bounds in Theorem~\ref{thm:finite-perturbation} are of order \(n^{-1/2}\). For comparison, with probability at least \(1-\delta\), standard split conformal prediction satisfies
\[
  \P\!\left\{Y_{n+1}\in
  C_{\widehat q_{\alpha}^{\,\mathrm{sc}}}(X_{n+1})\mid\mathcal D_n\right\}
  -(1-\alpha)
  \ge
  -\sqrt{\frac{\log(1/\delta)}{2n}};
\]
see Theorem~4.1 of \citet{AngelopoulosBarberBates2026}. The additional term in Theorem~\ref{thm:finite-perturbation} controls the deviation of the weights from uniform weights and, for fixed \(r\), \(B\), \(\kappa\), and \(\delta\), is also of order \(n^{-1/2}\).

Marginal coverage is the expectation of training-conditional coverage over the calibration data. Equivalently,
\[
  \E\!\left[\P\{Y_{n+1}\in\widehat C(X_{n+1})\mid\mathcal D_n\}-(1-\alpha)\right]=\P\{Y_{n+1}\in\widehat C(X_{n+1})\}-(1-\alpha).
\]
\begin{corollary}[Marginal coverage]
  \label{cor:finite-marginal}
  Assume \(\E\{h(Z)\}=0\), \(\|h(Z)\|_2\le B\) almost surely, and \(\lambda_{\min}(\Sigma_h)\ge\kappa>0\). Then
  \begin{equation*}
    -\left(
    \sqrt{\frac{\pi}{2n}}
    +
    \frac{16B^2}{\kappa\sqrt n}
    +
    2p_{\mu,\Sigma,n}
    \right)
    \le
    \P\{Y_{n+1}\in \widehat C(X_{n+1})\}-(1-\alpha)
    \le
    \sqrt{\frac{\pi}{2n}}
    +
    \frac{16B^2}{\kappa\sqrt n}
    +
    \frac3n
    +
    3p_{\mu,\Sigma,n},
  \end{equation*}
  where \(p_{\mu,\Sigma,n}=2r\exp(-n\kappa^2/(512B^4r))+2r^2\exp(-n\kappa^2/(8B^4r^2))\).
\end{corollary}

For fixed \(r\), \(B\), and \(\kappa\), the magnitudes of both bounds in Corollary~\ref{cor:finite-marginal} are of order \(n^{-1/2}\). For comparison, standard split conformal prediction satisfies
\[
  0
  \le
  \P\!\left\{Y_{n+1}\in
  C_{\widehat q_{\alpha}^{\,\mathrm{sc}}}(X_{n+1})\right\}-(1-\alpha)
  \le
  \frac1{n+1};
\]
see \citet{LeiGsellRinaldoTibshiraniWasserman2018} and \citet{AngelopoulosBarberBates2026}. For standard split conformal prediction, the score of the test point has a uniform rank among the calibration and test scores, and the coverage event is determined by this rank and a fixed cutoff. In contrast, the weighted quantile \(\widehat q_\alpha\) depends on the data-dependent weights assigned to the calibration data points, so the same rank calculation does not directly give its coverage probability.

\begin{remark}[Components of the bounds]
  The bounds combine the Dvoretzky--Kiefer--Wolfowitz inequality \citep{Massart1990DKW} with control of the weights assigned to the calibration data points. In the proof of Theorem~\ref{thm:finite-perturbation}, the sample mean and sample second-moment matrix satisfy concentration bounds determined by \(\varepsilon_{\mu,n}(\delta)\) and \(\varepsilon_{\Sigma,n}(\delta)\), respectively. On this event, the conditions of the theorem ensure feasibility of the moment constraints and yield \(\sum_{i=1}^n\left|\widehat w_i-1/n\right|\le(16B/\kappa)\varepsilon_{\mu,n}(\delta)\). The \(3/n\) term accounts for the adjusted quantile level and the largest jump of \(\widehat F_w\). The terms involving \(p_{\mu,\Sigma,n}\) account for failure of the concentration event for the sample mean and sample second-moment matrix used in the proof of Corollary~\ref{cor:finite-marginal}. See Section~S1 of the Supplementary Material for the proofs.
\end{remark}

\subsection{Asymptotic distributions and variance comparisons}
\label{sec:marginal-stability}

We derive asymptotic distributions for training-conditional coverage and average prediction-set size, and compare their limiting variances with those under standard split conformal prediction. The dimension \(r\) is fixed in the asymptotic analysis. Let \(q_\alpha=\inf\{q:F(q)\ge1-\alpha\}\), and define
\[
  \psi_\alpha(Z)=\ind(S\le q_\alpha)-(1-\alpha),
  \qquad
  c_\alpha=\E\{h(Z)\psi_\alpha(Z)\}.
\]
Under the continuity assumption below, \(\psi_\alpha\) is centred and \(c_\alpha\) is its cross-covariance with \(h(Z)\).

\begin{theorem}[Asymptotic distribution of training-conditional coverage]
  \label{thm:marginal-stability}
  Assume that \(\Sigma_h\) is positive definite, \(\E\{\|h(Z)\|_2^2\}<\infty\), and \(F\) is continuously differentiable in a neighborhood of \(q_\alpha\), with \(f(q_\alpha)=F'(q_\alpha)>0\). Then
  \begin{equation*}
    \sqrt n[
    \P\{Y_{n+1}\in\widehat C(X_{n+1})\mid\mathcal D_n\}
    -(1-\alpha)]
    =
    -n^{-1/2}
    \sum_{i=1}^n
    \{\psi_\alpha(Z_i)
    -c_\alpha^\top\Sigma_h^{-1}h(Z_i)\}
    +o_p(1).
  \end{equation*}
  Consequently,
  \begin{equation*}
    \sqrt n[
    \P\{Y_{n+1}\in\widehat C(X_{n+1})\mid\mathcal D_n\}-(1-\alpha)]
    \xrightarrow{d}
    N(0,\alpha(1-\alpha)-c_\alpha^\top\Sigma_h^{-1}c_\alpha).
  \end{equation*}
\end{theorem}

For standard split conformal prediction,
\[
  \sqrt n[\P\{Y_{n+1}\in C_{\widehat q_\alpha^{\,\mathrm{sc}}}(X_{n+1})\mid\mathcal D_n\}
  -(1-\alpha)]=-n^{-1/2}\sum_{i=1}^n\psi_\alpha(Z_i)+o_p(1),
\]
with limiting variance \(\alpha(1-\alpha)\). Thus, moment calibration reduces the limiting variance by \(c_\alpha^\top\Sigma_h^{-1}c_\alpha\), which is strictly positive exactly when \(c_\alpha\ne0\).

The condition \(c_\alpha\ne0\) means that \(\ind(S\le q_\alpha)\) has nonzero covariance with at least one coordinate of \(h(Z)\). The two examples in Section~\ref{sec:introduction} make this condition explicit. In the group-proportion example, \(c_\alpha=\pi_G(1-\pi_G)[\P\{S\le q_\alpha\mid X\in G\}-\P\{S\le q_\alpha\mid X\notin G\}]\). Thus, for \(0<\pi_G<1\), \(c_\alpha\ne0\) exactly when these two conditional probabilities differ. In the example based on the known mean of \(S\), \(c_\alpha=\alpha(1-\alpha)[\E\{S\mid S\le q_\alpha\}-\E\{S\mid S>q_\alpha\}]<0\) under the conditions of Theorem~\ref{thm:marginal-stability}.

\begin{remark}[Projection interpretation]
  Write \(\|g\|_{P,2}=[\E\{g(Z)^2\}]^{1/2}\), and let \(\Pi_h\) be the orthogonal projection in \(L_2(P)\) onto the span of the coordinates of \(h(Z)\). Then \(\Pi_h\psi_\alpha=c_\alpha^\top\Sigma_h^{-1}h(Z)\). The summand in Theorem~\ref{thm:marginal-stability} is the orthogonal residual \(\psi_\alpha-\Pi_h\psi_\alpha\), whose squared norm is
  \[
    \|\psi_\alpha-\Pi_h\psi_\alpha\|_{P,2}^2
    =\alpha(1-\alpha)-c_\alpha^\top\Sigma_h^{-1}c_\alpha.
  \]
  The reduction therefore depends on how well the coordinates of \(h(Z)\) linearly explain the centred indicator \(\psi_\alpha(Z)\).
\end{remark}

For a fixed threshold \(q\), define the average prediction-set size by \(\mathcal L(q)=\E_X\{\lambda_{\mathcal Y}(C_q(X))\}\), where \(\lambda_{\mathcal Y}\) is a fixed measure on \(\mathcal Y\). For the estimated threshold, \(\mathcal L(\widehat q_\alpha)\) averages prediction-set size over test covariates conditional on \(\mathcal D_n\) and remains random across calibration data sets.

\begin{corollary}[Asymptotic distribution of average prediction-set size]
  \label{cor:length-stability}
  Under the conditions of Theorem~\ref{thm:marginal-stability}, if \(\mathcal L\) is differentiable at \(q_\alpha\), then
  \begin{equation*}
    \sqrt n\{\mathcal L(\widehat q_\alpha)-\mathcal L(q_\alpha)\}\xrightarrow{d}N\left(0,\left\{\frac{\mathcal L'(q_\alpha)}{f(q_\alpha)}\right\}^2\left\{\alpha(1-\alpha)-c_\alpha^\top\Sigma_h^{-1}c_\alpha\right\}\right).
  \end{equation*}
\end{corollary}

For standard split conformal prediction, the limiting variance is \(\{\mathcal L'(q_\alpha)/f(q_\alpha)\}^2\alpha(1-\alpha)\). The reduction in Corollary~\ref{cor:length-stability} is strict exactly when \(c_\alpha\ne0\) and \(\mathcal L'(q_\alpha)\ne0\).

For a real-valued response, let \(\lambda_{\mathcal Y}\) be Lebesgue measure and consider the score function \(s(x,y)=|y-\mu(x)|/\sigma(x)\), where \(\mu\) and \(\sigma>0\) are fixed location and scale functions with \(\E_X\{\sigma(X)\}<\infty\). For \(q\ge0\), \(\mathcal L(q)=2q\,\E_X\{\sigma(X)\}\), and hence \(\mathcal L'(q)=2\E_X\{\sigma(X)\}\) for \(q>0\) \citep{LeiGsellRinaldoTibshiraniWasserman2018}. For a response with \(K\) classes, let \(\lambda_{\mathcal Y}\) be counting measure and consider the score function \(s(x,y)=1-\eta_y(x)\), where \(\eta_y(x)\) is the fitted probability of class \(y\), treated as fixed. If \(\eta_y(X)\) has a density \(g_y\) that is continuous in a neighborhood of \(1-q_\alpha\) for each \(y\), then, for \(q\) near \(q_\alpha\), \(\mathcal L(q)=\sum_{y=1}^K\int_{1-q}^1g_y(u)\,du\) and \(\mathcal L'(q_\alpha)=\sum_{y=1}^K g_y(1-q_\alpha)\). These calculations verify the differentiability condition in Corollary~\ref{cor:length-stability}.

\section{Optimal Moment Functions and Unlabeled Data}
\label{sec:semi-supervised}
\label{sec:marginal-design}

We now consider functions of the covariates whose means are unknown. In addition to \(\mathcal D_n\), assume that we observe only the covariates \(X_1^{\mathrm u},\ldots,X_m^{\mathrm u}\), sampled i.i.d. from the marginal distribution of \(X\) and independently of the calibration data and test point.

Let \(A\) be a covariate representation obtained as a fixed function of \(X\), with \(A=X\) allowed. The representation may be specified in advance or learned from the training data, on which we continue to condition. Examples include \(\{\mu(X),\sigma(X)\}^{\top}\), fitted class-probability vectors, and learned low-dimensional representations. Write \(A_i\) and \(A_j^{\mathrm u}\) for the representations of \(X_i\) and \(X_j^{\mathrm u}\). Section~\ref{sec:ss-prespecified} uses prespecified moment functions of \(A\), while Section~\ref{sec:optimal-side-information-features} learns a moment function from the calibration data.

\subsection{Prespecified moment functions and optimality}
\label{sec:ss-prespecified}

Let \(h(A)\in\R^r\) be a prespecified moment function. Unlike \(h(Z)\) in Section~\ref{sec:marginal}, \(h(A)\) need not be centred, and its mean will be estimated. For scalar \(A\), examples include \(h(A)=A\), \(h(A)=(A,A^2)^\top\), and a vector of selected spline basis functions. For multivariate \(A\), \(h(A)\) may collect selected coordinates together with their transformations and interactions. With \(N=n+m\), we estimate \(\E\{h(A)\}\) by the pooled mean
\[
  \bar h_N=\frac1N\left\{\sum_{i=1}^n h(A_i)+\sum_{j=1}^m h(A^{\mathrm u}_j)\right\}.
\]
We apply Eq.~\eqref{eq:weight-optimization} with \(h(Z_i)\) replaced by \(h(A_i)-\bar h_N\), and write \(\widehat w^{\,\mathrm{ss}}=(\widehat w_1^{\,\mathrm{ss}},\ldots,\widehat w_n^{\,\mathrm{ss}})\) for the resulting weights assigned to the calibration data points. When the optimization has a solution, these weights satisfy \(\sum_{i=1}^n\widehat w_i^{\,\mathrm{ss}}h(A_i)=\bar h_N\). If it has no solution, we set \(\widehat w_i^{\,\mathrm{ss}}=1/n\) for \(i=1,\ldots,n\). These weights define the weighted empirical CDF \(\widehat F_{\mathrm{ss}}(q)=\sum_{i=1}^n\widehat w_i^{\,\mathrm{ss}}\ind(S_i\le q)\). The semi-supervised weighted quantile is \(\widehat q_{\alpha}^{\,\mathrm{ss}}=\inf\{q:\widehat F_{\mathrm{ss}}(q)\ge(n+1)(1-\alpha)/n\}\), and the corresponding prediction set is \(\widehat C_{\mathrm{ss}}(x)=C_{\widehat q_{\alpha}^{\,\mathrm{ss}}}(x)=\{y:s(x,y)\le\widehat q_{\alpha}^{\,\mathrm{ss}}\}\).

For the theoretical analysis, write \(H=h(A)-\E\{h(A)\}\), \(\Sigma_A=\E\{HH^\top\}\), and \(c_{\alpha,A}=\E\{\psi_\alpha(Z)H\}\). Also write \(\bar h_n=n^{-1}\sum_{i=1}^n h(A_i)\), and let \(\mathcal D_n^{\mathrm{ss}}=(\mathcal D_n,\{A_j^{\mathrm u}\}_{j=1}^m)\) denote the data used for calibration.

\begin{theorem}[Semi-supervised asymptotic distribution of training-conditional coverage]
  \label{thm:ss-marginal-stability}
  Let \(n\to\infty\) with \(n/N\to\gamma\in[0,1]\). Assume that \(\Sigma_A\) is positive definite, \(\E\{\|H\|_2^2\}<\infty\), and \(F\) is continuously differentiable in a neighborhood of \(q_\alpha\), with \(f(q_\alpha)=F'(q_\alpha)>0\). Then
  \begin{equation*}
    \sqrt n\left[
    \P\{Y_{n+1}\in\widehat C_{\mathrm{ss}}(X_{n+1})\mid\mathcal D_n^{\mathrm{ss}}\}
    -(1-\alpha)\right]
    =
    -\frac1{\sqrt n}\sum_{i=1}^n\psi_\alpha(Z_i)
    +c_{\alpha,A}^\top\Sigma_A^{-1}\sqrt n(\bar h_n-\bar h_N)
    +o_p(1).
  \end{equation*}
  Consequently,
  \begin{equation*}
    \sqrt n\left[
    \P\{Y_{n+1}\in\widehat C_{\mathrm{ss}}(X_{n+1})\mid\mathcal D_n^{\mathrm{ss}}\}
    -(1-\alpha)\right]
    \xrightarrow{d}
    N\left(0,\alpha(1-\alpha)
    -(1-\gamma)c_{\alpha,A}^\top\Sigma_A^{-1}c_{\alpha,A}\right).
  \end{equation*}
\end{theorem}

Estimating the means from pooled covariates multiplies the variance reduction obtained with known means by \(1-\gamma=\lim_{n\to\infty}m/N\). The reduction is strict exactly when \(\gamma<1\) and \(c_{\alpha,A}\ne0\). When \(m/n\to\infty\), \(\gamma=0\) and the known-mean reduction is recovered, whereas when \(m/n\to0\), \(\gamma=1\) and the limiting variance equals that of standard split conformal prediction. The corresponding result for average prediction-set size is as follows.

\begin{corollary}[Semi-supervised asymptotic distribution of average prediction-set size]
  \label{cor:ss-length-stability}
  Under the conditions of Theorem~\ref{thm:ss-marginal-stability}, if \(\mathcal L\) is differentiable at \(q_\alpha\), then
  \begin{equation*}
    \sqrt n\{\mathcal L(\widehat q_{\alpha}^{\,\mathrm{ss}})-\mathcal L(q_\alpha)\}\xrightarrow{d}N\left(0,\left\{\frac{\mathcal L'(q_\alpha)}{f(q_\alpha)}\right\}^2\left\{\alpha(1-\alpha)-(1-\gamma)c_{\alpha,A}^\top\Sigma_A^{-1}c_{\alpha,A}\right\}\right).
  \end{equation*}
\end{corollary}

For a given representation \(A\), we compare the variance reductions obtained from different moment functions. For \(j=1,2\), let \(h_j(A)\in\R^{r_j}\) be a square-integrable prespecified moment function. Define \(\Sigma_{A,j}=\Var\{h_j(A)\}\) and \(c_{\alpha,A,j}=\Cov\{h_j(A),\psi_\alpha(Z)\}\), and let \(\mathcal V_j\) be the linear span in \(L_2(P)\) of the coordinates of \(h_j(A)-\E\{h_j(A)\}\). To characterize the largest reduction in limiting variance available from \(A\), define
\begin{equation*}
  h^*_\alpha(A)=\E\{\psi_\alpha(Z)\mid A\}=\P\{S\le q_\alpha\mid A\}-(1-\alpha).
\end{equation*}

\begin{proposition}[Nested linear spans and optimality]
  \label{prop:ss-basis-optimality}
  Suppose that \(\Sigma_{A,j}\) is positive definite for \(j=1,2\), and that \(\mathcal V_1\subseteq\mathcal V_2\). Then
  \[
    c_{\alpha,A,1}^\top\Sigma_{A,1}^{-1}c_{\alpha,A,1}
    \le c_{\alpha,A,2}^\top\Sigma_{A,2}^{-1}c_{\alpha,A,2}
    \le \Var\{h^*_\alpha(A)\}.
  \]
  For \(j=1,2\), \(c_{\alpha,A,j}^\top\Sigma_{A,j}^{-1}c_{\alpha,A,j}=\Var\{h^*_\alpha(A)\}\) holds exactly when \(\mathcal V_j\) contains \(h^*_\alpha(A)\).
\end{proposition}

Among square-integrable moment functions of the given representation \(A\) satisfying the assumptions of Theorem~\ref{thm:ss-marginal-stability}, the largest reduction is \((1-\gamma)\Var\{h^*_\alpha(A)\}\). If \(\Var\{h^*_\alpha(A)\}>0\), the scalar choice \(h=h^*_\alpha\) attains it. A prespecified moment function attains the same bound when the linear span of its centred coordinates contains \(h^*_\alpha\).

\subsection{Learning a moment function by probabilistic classification}
\label{sec:optimal-side-information-features}

We can learn \(h^*_\alpha\) by treating \(\ind(S\le q_\alpha)\) as a binary class label and fitting a probabilistic classifier with \(A\) as the predictor. Because \(q_\alpha\) is unknown, we replace it with a standard split conformal quantile. For a fixed \(K\ge2\), we use \(K\)-fold cross-fitting \citep[e.g.,][]{ChernozhukovEtAl2018DML} to separate classifier fitting from each fold's contribution to calibration of the final prediction set.
\begin{itemize}[leftmargin=*,itemsep=2pt]
  \item Independently of the observed data, partition the calibration indices into \(K\) nearly equal folds \(\mathcal I_1,\ldots,\mathcal I_K\), and partition the unlabeled indices into corresponding nearly equal folds \(\mathcal J_1,\ldots,\mathcal J_K\).
  \item For each \(k\), compute the standard split conformal quantile \(\widehat q_\alpha^{(-k)}\) from \(\{S_i:i\notin\mathcal I_k\}\) at the adjusted level \((1-\alpha)\{1+1/(n-|\mathcal I_k|)\}\). For the same indices \(i\notin\mathcal I_k\), use \(\ind\{S_i\le\widehat q_\alpha^{(-k)}\}\) as the class labels and \(A_i\) as the predictors to fit a probabilistic classifier. Let \(\widehat r^{(-k)}(a)\) denote the estimated probability that the label equals one at \(A=a\), and define \(\widehat h^{(-k)}(a)=\widehat r^{(-k)}(a)-(1-\alpha)\).
  \item Evaluate \(\widehat h^{(-k)}\) on the held-out calibration fold \(\mathcal I_k\) and the corresponding unlabeled fold \(\mathcal J_k\). For \(i\in\mathcal I_k\), centre \(\widehat h^{(-k)}(A_i)\) by the pooled mean of these evaluations:
  \[
    \widehat H_i
    =
    \widehat h^{(-k)}(A_i)
    -
    \frac1{|\mathcal I_k|+|\mathcal J_k|}
    \left\{
    \sum_{\ell\in\mathcal I_k}\widehat h^{(-k)}(A_\ell)
    +
    \sum_{\ell\in\mathcal J_k}\widehat h^{(-k)}(A^{\mathrm u}_\ell)
    \right\}.
  \]
  \item After all \(\widehat H_i\) have been defined, apply Eq.~\eqref{eq:weight-optimization} once with \(h(Z_i)\) replaced by \(\widehat H_i\), and write \(\widehat w^{\,\mathrm{cf}}=(\widehat w_1^{\,\mathrm{cf}},\ldots,\widehat w_n^{\,\mathrm{cf}})\) for the resulting weights assigned to the calibration data points. If the optimization has no solution, set \(\widehat w_i^{\,\mathrm{cf}}=1/n\) for \(i=1,\ldots,n\). These weights define the weighted empirical CDF \(\widehat F_{\mathrm{cf}}(q)=\sum_{i=1}^n\widehat w_i^{\,\mathrm{cf}}\ind(S_i\le q)\). The cross-fitted weighted quantile is \(\widehat q_\alpha^{\,\mathrm{cf}}=\inf\{q:\widehat F_{\mathrm{cf}}(q)\ge(n+1)(1-\alpha)/n\}\), and the corresponding prediction set is \(\widehat C_{\mathrm{cf}}(x)=C_{\widehat q_\alpha^{\,\mathrm{cf}}}(x)=\{y:s(x,y)\le\widehat q_\alpha^{\,\mathrm{cf}}\}\).
\end{itemize}

For the analysis below, condition on the fold partitions and any auxiliary randomization used to fit the classifiers. These are generated independently of the data, so the conditioning on \(\mathcal D_n^{\mathrm{ss}}\) is understood to include them.

\begin{theorem}[Asymptotic distributions with a learned moment function of \(A\)]
  \label{thm:ss-adaptive}
  Suppose there exists a nonrandom square-integrable scalar function \(h^\dagger\) such that, for an independent draw \(A\) from its marginal distribution,
  \[
    \max_{1\le k\le K}
    \E\{|\widehat h^{(-k)}(A)-h^\dagger(A)|^2\mid \widehat h^{(-k)}\}
    \xrightarrow{p}0.
  \]
  If the conditions of Theorem~\ref{thm:ss-marginal-stability} hold with \(h^\dagger(A)\) in place of \(h(A)\), then the expansion and limiting distribution in that theorem hold with \(\widehat C_{\mathrm{cf}}\) in place of \(\widehat C_{\mathrm{ss}}\). If, in addition, \(\mathcal L\) is differentiable at \(q_\alpha\), the limiting distribution in Corollary~\ref{cor:ss-length-stability} holds with \(\widehat q_\alpha^{\,\mathrm{cf}}\) in place of \(\widehat q_\alpha^{\,\mathrm{ss}}\). In both results, \(H\), \(\Sigma_A\), \(c_{\alpha,A}\), \(\bar h_n\), and \(\bar h_N\) are defined using \(h^\dagger\). If \(h^\dagger=h^*_\alpha\), the limiting variance in Theorem~\ref{thm:ss-marginal-stability} is \(\alpha(1-\alpha)-(1-\gamma)\operatorname{Var}\{h^*_\alpha(A)\}\).
\end{theorem}

Under the conditions of Theorem~\ref{thm:ss-adaptive}, convergence to \(h^\dagger\) gives the same limiting variance as using that function in the procedure using prespecified moment functions. In particular, \(h^\dagger=h^*_\alpha\) attains the optimal reduction for the given \(A\). Consistency for \(h^*_\alpha\) is not required for the limiting variance to be no larger than that under standard split conformal prediction.

\section{Side Information for Conditional Coverage}
\label{sec:conditional}

\subsection{Weighted quantile regression for conditional coverage}

The construction in Section~\ref{sec:marginal} uses a common threshold for the conformal score. We now allow the threshold to depend on \(X\) and study coverage conditions indexed by functions of the covariates.

Exact finite-sample distribution-free coverage conditional on every value of \(X\) cannot be attained nontrivially over unrestricted distributions \citep{Vovk2012Conditional,BarberCandesRamdasTibshirani2021Limits}. Following \citet{GibbsCherianCandes2025}, let \(\mathcal F\) be a chosen class of functions of \(X\), and seek a threshold \(g\in\mathcal F\) satisfying the coverage conditions indexed by \(f\in\mathcal F\):
\begin{equation}
  \label{eq:conditional-class}
  \E\!\left[
  f(X)\{\ind(S\le g(X))-(1-\alpha)\}
  \right]=0,\qquad f\in\mathcal F.
\end{equation}
If \(1\in\mathcal F\), the choice \(f\equiv1\) gives marginal coverage. If \(\ind(\cdot\in G)\in\mathcal F\) for a prespecified subgroup \(G\) with \(\P\{X\in G\}>0\), the corresponding condition gives \(\P\{S\le g(X)\mid X\in G\}=1-\alpha\). If \(\mathcal F\) contains all bounded measurable functions of \(X\), Eq.~\eqref{eq:conditional-class} is equivalent to \(\P\{S\le g(X)\mid X\}=1-\alpha\) almost surely.

Throughout this section, \(h\) again denotes a fixed function with \(\E\{h(Z)\}=0\), and \(\widehat w_1,\ldots,\widehat w_n\) are the weights assigned to the calibration data points by Eq.~\eqref{eq:weight-optimization}. Uniform weights are used when the optimization has no solution. For a linear function space \(\mathcal F\), define \(\widehat g\) by weighted quantile regression:
\[
  \widehat g
  \in\argmin_{g\in\mathcal F}
  \sum_{i=1}^n\widehat w_i\rho_{1-\alpha}(S_i-g(X_i)),
\]
where \(\rho_{1-\alpha}(u)=u\{1-\alpha-\ind(u<0)\}\) is the check loss and the minimum is assumed to be attained. The fitted threshold \(\widehat g\) defines the prediction set \(\widehat C_{\mathrm{cond}}(x)=\{y:s(x,y)\le\widehat g(x)\}\). The weights are computed and the threshold is fitted once using the calibration data.

Under uniform weights, the optimization reduces to ordinary quantile regression over \(\mathcal F\) at level \(1-\alpha\). The construction of \citet{GibbsCherianCandes2025} instead augments the calibration data with a candidate score for the test point in each quantile-regression fit, so the fitted threshold depends on the candidate score.

\subsection{Finite-sample coverage guarantees}

For theoretical analysis, let \(\Phi:\mathcal X\to\R^d\) be a vector of fixed functions, and take \(\mathcal F=\{x\mapsto a^\top\Phi(x):a\in\R^d\}\). Each \(a^\top\Phi\) indexes a coverage condition, while \(\Phi(x)^\top\beta\) parameterizes the threshold for the conformal score. Coordinates of \(\Phi\) may be prespecified subgroup indicators, selected covariates, spline basis functions, or products of these functions. The fitted threshold defined above can be written as \(\widehat g(x)=\Phi(x)^\top\widehat\beta\), where
\begin{equation}
  \label{eq:conditional-weighted-estimator}
  \widehat\beta
  \in
  \argmin_{\beta\in\R^d}
  \sum_{i=1}^n\widehat w_i
  \rho_{1-\alpha}\bigl(S_i-\Phi(X_i)^\top\beta\bigr).
\end{equation}
The corresponding prediction set is \(\widehat C_{\mathrm{cond}}(x)=\{y:s(x,y)\le\Phi(x)^\top\widehat\beta\}\).

We now study the coverage conditions in Eq.~\eqref{eq:conditional-class} for the fitted threshold \(\widehat g\). For \(K\ge1\), consider the class
\[
  \mathcal F_K^+
  =
  \{f\in\mathcal F:0\le f(X)\le K\ \text{a.s.},\ \E\{f(X)\}=1\}.
\]
The bound \(f(X)\le K\) controls how narrowly coverage can be localized. For a subgroup \(G\) with positive probability and indicator function in \(\mathcal F\), the normalized indicator \(f(x)=\ind(x\in G)/\P\{X\in G\}\) belongs to \(\mathcal F_K^+\) exactly when \(\P\{X\in G\}\ge1/K\). Thus larger \(K\) allows coverage conditions for smaller subgroups represented in \(\mathcal F\). For a universal constant \(C\), let
\[
  \varepsilon^{\mathrm{cond}}_{n,K}(\delta)
  =
  CK\sqrt{\frac{d+\log(3/\delta)}{n}}
  +\frac{16KB}{\kappa}\varepsilon_{\mu,n}(\delta) \text{~~and~~}
  \bar\varepsilon^{\mathrm{cond}}_{n,K}
  =
  CK\sqrt{\frac dn}
  +K\left(
  \frac{16B^2}{\kappa\sqrt n}
  +2p_{\mu,\Sigma,n}
  \right).
\]

\begin{theorem}[Finite-sample control of indexed coverage conditions]
  \label{thm:conditional-finite}
  Assume \(\mathcal F_K^+\ne\varnothing\). For \(\delta\in(0,1)\), suppose the conditions of Theorem~\ref{thm:finite-perturbation} hold. Then, with probability at least \(1-\delta\),
  \[
    \inf_{f\in\mathcal F_K^+}
    \E\!\left[
    f(X_{n+1})
    \{\ind(Y_{n+1}\in\widehat C_{\mathrm{cond}}(X_{n+1}))-(1-\alpha)\}
    \,\middle|\, \mathcal D_n
    \right]
    \ge
    -\varepsilon^{\mathrm{cond}}_{n,K}(\delta).
  \]
  If, in addition, the conditional CDF of \(S\) given \(X\) is continuous almost surely, then with probability at least \(1-\delta\),
  \[
    \sup_{f\in\mathcal F_K^+}
    \left|
    \E\!\left[
    f(X_{n+1})
    \{\ind(Y_{n+1}\in\widehat C_{\mathrm{cond}}(X_{n+1}))-(1-\alpha)\}
    \,\middle|\, \mathcal D_n
    \right]
    \right|
    \le
    \varepsilon^{\mathrm{cond}}_{n,K}(\delta).
  \]
  Averaging over the calibration data gives
  \[
    \inf_{f\in\mathcal F_K^+}
    \E\!\left[
    f(X_{n+1})
    \{\ind(Y_{n+1}\in\widehat C_{\mathrm{cond}}(X_{n+1}))-(1-\alpha)\}
    \right]
    \ge
    -\bar\varepsilon^{\mathrm{cond}}_{n,K}.
  \]
  Under the same continuity condition,
  \[
    \sup_{f\in\mathcal F_K^+}
    \left|
    \E\!\left[
    f(X_{n+1})
    \{\ind(Y_{n+1}\in\widehat C_{\mathrm{cond}}(X_{n+1}))-(1-\alpha)\}
    \right]
    \right|
    \le
    \bar\varepsilon^{\mathrm{cond}}_{n,K}.
  \]
\end{theorem}

Under the continuity and permutation-invariance assumptions of their Theorem~2, \citet{GibbsCherianCandes2025} bound the absolute coverage moments for \(f\in\mathcal F_K^+\) by \(Kd/(n+1)\). For fixed dimensions and constants, this is an \(O(n^{-1})\) bound, whereas the marginal bounds in Theorem~\ref{thm:conditional-finite} are \(O(n^{-1/2})\).

\subsection{Asymptotic distributions and variance comparisons}

We derive limiting distributions for the fitted threshold coefficients and the coverage moments indexed by the coordinates of \(\Phi\), and compare their limiting covariance matrices with those under ordinary quantile regression. Let
\begin{equation*}
  M(\beta)=
  \E\!\left[
  \Phi(X)\{\ind(S\le \Phi(X)^\top\beta)-(1-\alpha)\}
  \right]
\end{equation*}
be the vector of these coverage moments. Assume that \(M(\beta)=0\) has a solution \(\beta_\alpha\). Conditional on the calibration data, \(M(\widehat\beta)\) collects the coverage moments of \(\widehat C_{\mathrm{cond}}\). If \(a^\top\Phi\equiv1\) for some \(a\in\R^d\), then \(a^\top M(\widehat\beta)\) equals training-conditional coverage minus \(1-\alpha\). Let \(\Psi_\alpha(Z)=\Phi(X)\{\ind(S\le \Phi(X)^\top\beta_\alpha)-(1-\alpha)\}\) be the corresponding mean-zero random vector and \(C_\alpha=\E\{\Psi_\alpha(Z)h(Z)^\top\}\) its cross-covariance with \(h(Z)\).

\begin{theorem}[Limiting distributions for threshold coefficients and coverage moments]
  \label{thm:conditional}
  Assume that \(\Sigma_h\) is positive definite, \(\E\{\|h(Z)\|_2^2+\|\Phi(X)\|_2^2\}<\infty\), and \(\P\{S=\Phi(X)^\top\beta_\alpha\}=0\). Suppose \(M(\beta_\alpha+t)=J_\alpha t+o(\|t\|_2)\) as \(t\to0\), where \(J_\alpha\) is symmetric positive definite. Then every minimizer in Eq.~\eqref{eq:conditional-weighted-estimator} satisfies
  \begin{equation*}
    \sqrt n(\widehat\beta-\beta_\alpha)
    =
    -J_\alpha^{-1}
    \frac1{\sqrt n}\sum_{i=1}^n
    \{\Psi_\alpha(Z_i)-C_\alpha\Sigma_h^{-1}h(Z_i)\}
    +o_p(1).
  \end{equation*}
  The vector \(M(\widehat\beta)\) satisfies
  \begin{equation*}
    \sqrt n\,M(\widehat\beta)
    =
    -\frac1{\sqrt n}\sum_{i=1}^n
    \{\Psi_\alpha(Z_i)-C_\alpha\Sigma_h^{-1}h(Z_i)\}
    +o_p(1).
  \end{equation*}
  Consequently, \(\sqrt n(\widehat\beta-\beta_\alpha)\xrightarrow{d}N(0,J_\alpha^{-1}\Var\{\Psi_\alpha(Z)-C_\alpha\Sigma_h^{-1}h(Z)\}J_\alpha^{-1})\). We also have \(\sqrt n\,M(\widehat\beta)\xrightarrow{d}N(0,\Var\{\Psi_\alpha(Z)-C_\alpha\Sigma_h^{-1}h(Z)\})\).
\end{theorem}

The limiting covariance matrix of \(\sqrt n(\widehat\beta-\beta_\alpha)\) is no larger in the positive semidefinite order than the corresponding matrix under ordinary quantile regression, with the difference equal to \(J_\alpha^{-1}C_\alpha\Sigma_h^{-1}C_\alpha^\top J_\alpha^{-1}\). For the coverage moments, the limiting covariance matrix of \(\sqrt n\,M(\widehat\beta)\) is reduced by the positive semidefinite matrix \(C_\alpha\Sigma_h^{-1}C_\alpha^\top\). For a linear combination \(a^\top M(\widehat\beta)\) of coverage moments, the reduction in limiting variance is positive exactly when \(C_\alpha^\top a\ne0\). When \(\Phi(X)\equiv1\), the limiting distribution and variance comparison for training-conditional coverage reduce to those in Theorem~\ref{thm:marginal-stability}.

\begin{remark}[Sufficient condition for the local expansion]
  Suppose \(S\mid X\) admits a conditional density \(f_{S\mid X}\) such that, almost surely, \(u\mapsto f_{S\mid X}(u\mid X)\) is continuous at \(u=\Phi(X)^\top\beta_\alpha\) and \(\sup_{u\in\R}f_{S\mid X}(u\mid X)\le\bar f\) for some constant \(\bar f<\infty\). Then \(\P\{S=\Phi(X)^\top\beta_\alpha\}=0\) and \(M(\beta_\alpha+t)=J_\alpha t+o(\|t\|_2)\), with
  \[
    J_\alpha=
    \E\{\Phi(X)\Phi(X)^\top
    f_{S\mid X}(\Phi(X)^\top\beta_\alpha\mid X)\}.
  \]
  If this matrix is positive definite, \(\beta_\alpha\) is a locally unique solution of \(M(\beta)=0\).
\end{remark}

\section{Simulation Studies}
\label{sec:numerical}

We examine the finite-sample performance of the proposed methods in the following settings: calibration with externally specified moments, semi-supervised calibration using prespecified moment functions or learned moment functions, and conditional calibration. For each data-generating process (DGP) and $(n,p)$ configuration, we generate one training data set and hold it fixed across replications, independently regenerating the calibration and test data and any external-reference or unlabeled data required. The nominal coverage level is $1-\alpha=0.9$, and the finite-sample results in this section are based on 500 replications.

We obtain a conditional mean estimate $\widehat\mu$ from the training data using cubic regression splines~\citep{LeiGsellRinaldoTibshiraniWasserman2018} and use $S=|Y-\widehat\mu(X)|$ as the conformal score in all experiments. Using the same training data and spline construction, we obtain a residual scale estimate $\widehat\rho$ from cross-fitted absolute residuals. Details and additional results are given in Section~S2 of the Supplementary Material.

Throughout the simulations, we consider eight DGPs of the form $Y=\mu(X)+\sigma_s(X)\varepsilon$ with $\mu(X)=\sum_{j=1}^5X_j$, where $X=(X_1,\ldots,X_p)^\top$ has independent $\operatorname{Unif}[-2,2]$ coordinates. The error $\varepsilon$ is standard Gaussian, except under DGP7, where $\varepsilon=\sqrt{3/5}\,t_5$ with $t_5$ following a Student-$t$ distribution with five degrees of freedom. The eight models for the conditional standard deviation $\sigma_s(X)$ are:  DGP1: $\sigma_s(X)=0.45+0.12\mu(X)^2$, DGP2: $\sigma_s(X)=0.5+2.4\operatorname{logit}^{-1}(2(\mu(X)-3))$, DGP3: $\sigma_s(X)=\exp(\mu(X)/3.5)$, DGP4: $\sigma_s(X)=0.45+2.7\operatorname{logit}^{-1}(6(X_1-1.45))$, DGP5: $\sigma_s(X)=0.45+2.7\operatorname{logit}^{-1}(4(X_1X_2-1.35))$, DGP6: $\sigma_s(X)=0.45+0.35|X_1|+2\operatorname{logit}^{-1}(4(X_2-1.15))$, DGP7: $\sigma_s(X)=0.45+2.3\operatorname{logit}^{-1}(4(X_2-1.25))$, and DGP8: $\sigma_s(X)=1$.  The conditional standard deviation depends on $\mu(X)$ in DGP1--DGP3, on $X_1$ in DGP4, on $X_1X_2$ in DGP5, and additively on $X_1$ and $X_2$ in DGP6. DGP7 uses heavy-tailed errors, and DGP8 is homoscedastic.

\subsection{Calibration with Externally Specified Moments}
\label{sec:numerical-known}

\textbf{Design.} The training and calibration data sets each have size $n\in\{100,250,500\}$, and we consider $p\in\{5,10\}$. The independent external-reference and test data sets each contain 5000 observations. The external-reference data are used only to compute empirical means and quantiles supplied as side information. Summaries involving $\widehat\mu$, $\widehat\rho$, or $S$ are computed using the functions fitted on the training data.

Let $\mathcal C_{\mathrm{ext}}(V)$ denote $V$ centred at its empirical mean computed from the external-reference data, componentwise for vector-valued $V$. For scalar $V$, let $Q_V^{\mathrm{ext}}(\cdot)$ denote its empirical quantile function from the same data. We consider seven specifications of the moment function: $h_1=\mathcal C_{\mathrm{ext}}(X_1,X_2,X_1^2+X_2^2,X_1X_2)$ (M1, covariate moments); $h_2=\mathcal C_{\mathrm{ext}}(Y,Y^2)$ (M2, outcome moments); $h_3=\mathcal C_{\mathrm{ext}}(\mathbf 1\{\widehat\mu(X)\le Q_{\widehat\mu}^{\mathrm{ext}}(0.10)\}, \mathbf 1\{\widehat\mu(X)\ge Q_{\widehat\mu}^{\mathrm{ext}}(0.90)\})$ (M3, prediction tails); $h_4=\mathcal C_{\mathrm{ext}}(\widehat\rho(X))$ (M4, residual scale); $h_5=\mathcal C_{\mathrm{ext}}(S)$ (M5, score mean); $h_6=\mathbf 1\{S\le Q_S^{\mathrm{ext}}(0.95)\}-0.95$ (M6, score event at level $0.95$); and $h_7=\mathbf 1\{S\le Q_S^{\mathrm{ext}}(0.9)\}-0.9$ (M7, score event at level $0.9$). Thus M1 and M2 do not depend on the fitted predictor, M3 and M4 evaluate fitted functions on only the covariates in the external-reference data, and M5--M7 use external-reference data to summarize the score distribution. With the true $0.9$ quantile of $S$, M7 attains the maximal reduction in the limiting variance of training-conditional coverage. Additional specifications are reported in Section~S2.1.3 of the Supplementary Material.

\noindent\textbf{Asymptotic variance reduction.} For a fixed score function and a moment function $h$ defined using the true means and quantiles, Theorem~\ref{thm:marginal-stability} gives the relative reduction in the limiting variance of training-conditional coverage as $R_h^2=c_\alpha^\top\Sigma_h^{-1}c_\alpha/\{\alpha(1-\alpha)\}$, compared with standard split conformal prediction (SC). Table~\ref{tab:external-theory} reports these reductions conditional on the fixed training data set of size $n=100$, with $p=5$.

\begin{table}[h]
  \centering
  \caption{Relative reduction in the limiting variance of training-conditional coverage, $R_h^2$, conditional on the fixed training data set of size $n=100$, with $p=5$.}
  \label{tab:external-theory}
  \small
  \linespread{1}\selectfont
  \setlength{\tabcolsep}{5.5pt}
  \begin{tabular}{@{}llrrrrrrrrr@{}}
    \toprule
    & Specification & DGP1 & DGP2 & DGP3 & DGP4 & DGP5 & DGP6 & DGP7 & DGP8 & Average \\
    \midrule
    M1 & Covariate moments & 0.040 & 0.099 & 0.085 & 0.262 & 0.369 & 0.189 & 0.208 & 0.000 & 0.157 \\
    M2 & Outcome moments & 0.181 & 0.226 & 0.305 & 0.075 & 0.163 & 0.152 & 0.058 & 0.008 & 0.146 \\
    M3 & Prediction tails & 0.232 & 0.286 & 0.167 & 0.047 & 0.024 & 0.018 & 0.026 & 0.000 & 0.100 \\
    M4 & Residual scale & 0.081 & 0.350 & 0.232 & 0.326 & 0.268 & 0.198 & 0.238 & 0.000 & 0.212 \\
    M5 & Score mean & 0.608 & 0.633 & 0.598 & 0.627 & 0.676 & 0.613 & 0.604 & 0.489 & 0.606 \\
    M6 & Score event ($0.95$) & 0.474 & 0.474 & 0.474 & 0.474 & 0.474 & 0.474 & 0.474 & 0.474 & 0.474 \\
    M7 & Score event ($0.9$) & 1.000 & 1.000 & 1.000 & 1.000 & 1.000 & 1.000 & 1.000 & 1.000 & 1.000 \\
    \bottomrule
  \end{tabular}
\end{table}

The reductions vary substantially with the relation between $h(Z)$ and $\ind\{S\le q_\alpha\}$. Among M1--M4, M4 has the largest average reduction, with the largest reductions for M1 under DGP4--DGP5 and for M2 and M3 under DGP1--DGP3. The score mean M5 is more strongly aligned with this indicator, with reductions between $0.489$ and $0.676$. For a continuous score distribution, $h(Z)=\ind\{S\le q\}-F(q)$ with $1-\alpha\le F(q)<1$ gives $R_h^2=(1-\alpha)\{1-F(q)\}/\{\alpha F(q)\}$. With $1-\alpha=0.9$, using the true quantiles with $F(q)=0.95$ and $0.9$ yields $R_{h_6}^2=9/19\approx0.474$ and $R_{h_7}^2=1$, respectively.

\noindent\textbf{Evaluation criteria.} Let $\mathcal D_b$ collect the training and calibration data and any external-reference or unlabeled data used in replication $b$, and let $\widehat q_{\alpha,b}$ denote the quantile in that replication. For the resulting prediction set $\widehat C_b$, write $\Delta_b=\P\{Y\in\widehat C_b(X)\mid\mathcal D_b\}-(1-\alpha)$ and $L_b=\mathcal L(\widehat q_{\alpha,b})$, with $\Delta_b$ and $L_b$ estimated from the independent test sample. Across replications, we report mean coverage, coverage variance $\operatorname{Var}_b\{\Delta_b\}$, mean interval length, and mean interval length variance $\operatorname{Var}_b\{L_b\}$, where each variance ratio divides the method's variance by the corresponding variance for SC within the same design. A ratio below one indicates less variation across replications. These empirical variances describe variation in the test-sample estimates across complete replications, with the training data held fixed and the calibration, test, and required external-reference or unlabeled data regenerated each time.

\noindent\textbf{Results.} For $(n,p)=(100,5)$, Table~S.1 in the Supplementary Material shows that mean coverage, averaged over the eight DGPs, is $0.900$ for SC and ranges from $0.901$ to $0.912$ for the proposed method under M1--M7, compared with the nominal level of $0.90$. The corresponding mean interval lengths, averaged over the eight DGPs and expressed relative to SC, range from $0.984$ to $1.040$. Figure~S.1 in the Supplementary Material further shows how marginal coverage varies with the calibration data set size.

Table~\ref{tab:external-main-results} reports the variance ratios for $(n,p)=(100,5)$, while Section~S2.1.2 of the Supplementary Material gives the results for the remaining $(n,p)$ combinations.

\begingroup
\setlength{\intextsep}{6pt}
\begin{table}[h]
  \centering
  \caption{Coverage and mean interval length variance ratios for externally specified moments at $n=100$ and $p=5$. All entries are relative to SC.}
  \label{tab:external-main-results}
  \small
  \linespread{1}\selectfont
  \setlength{\tabcolsep}{3.2pt}
  \begin{tabular}{@{}ll*{9}{r}@{}}
    \toprule
    & & DGP1 & DGP2 & DGP3 & DGP4 & DGP5 & DGP6 & DGP7 & DGP8 & Average \\
    \midrule
    \multicolumn{11}{l}{\textit{Coverage variance ratio}} \\
    M1 & Covariate moments & 1.015 & 0.945 & 0.894 & 0.737 & 0.761 & 0.765 & 0.771 & 1.040 & 0.866 \\
    M2 & Outcome moments & 0.805 & 0.868 & 0.719 & 0.924 & 0.853 & 0.845 & 0.943 & 1.098 & 0.882 \\
    M3 & Prediction tails & 0.850 & 0.749 & 0.832 & 0.950 & 0.952 & 0.952 & 0.932 & 1.064 & 0.910 \\
    M4 & Residual scale & 0.961 & 0.703 & 0.775 & 0.660 & 0.763 & 0.809 & 0.724 & 1.006 & 0.800 \\
    M5 & Score mean & 0.428 & 0.451 & 0.462 & 0.319 & 0.404 & 0.391 & 0.363 & 0.575 & 0.424 \\
    M6 & Score event ($0.95$) & 0.595 & 0.539 & 0.564 & 0.487 & 0.531 & 0.543 & 0.503 & 0.621 & 0.548 \\
    M7 & Score event ($0.9$) & 0.147 & 0.167 & 0.168 & 0.141 & 0.143 & 0.149 & 0.137 & 0.166 & 0.152 \\
    \midrule
    \multicolumn{11}{l}{\textit{Mean interval length variance ratio}} \\
    M1 & Covariate moments & 1.087 & 1.116 & 1.017 & 0.816 & 0.871 & 0.854 & 0.917 & 1.115 & 0.974 \\
    M2 & Outcome moments & 0.912 & 1.005 & 0.835 & 1.115 & 0.963 & 0.904 & 1.054 & 1.139 & 0.991 \\
    M3 & Prediction tails & 0.958 & 0.889 & 0.961 & 1.142 & 1.153 & 1.049 & 1.047 & 1.106 & 1.038 \\
    M4 & Residual scale & 1.057 & 0.792 & 0.943 & 0.788 & 0.879 & 0.939 & 0.832 & 1.037 & 0.908 \\
    M5 & Score mean & 0.496 & 0.463 & 0.440 & 0.425 & 0.399 & 0.448 & 0.455 & 0.654 & 0.473 \\
    M6 & Score event ($0.95$) & 0.514 & 0.401 & 0.440 & 0.364 & 0.413 & 0.409 & 0.471 & 0.532 & 0.443 \\
    M7 & Score event ($0.9$) & 0.192 & 0.235 & 0.213 & 0.271 & 0.205 & 0.240 & 0.221 & 0.198 & 0.222 \\
    \bottomrule
  \end{tabular}
\end{table}

M1--M4 give modest and mechanism-dependent reductions in coverage variance. M5--M7 yield larger reductions. M5--M7 also reduce the mean interval length variance. Figure~\ref{fig:external-projection} compares the empirical relative reductions, calculated as one minus the coverage variance ratio, with the asymptotic reductions in Table~\ref{tab:external-theory}. Their Pearson and Spearman correlations are $0.986$ and $0.982$ over the 56 DGP--specification pairs.

\begin{figure}[H]
  \centering
  \revisedfigure{\textwidth}{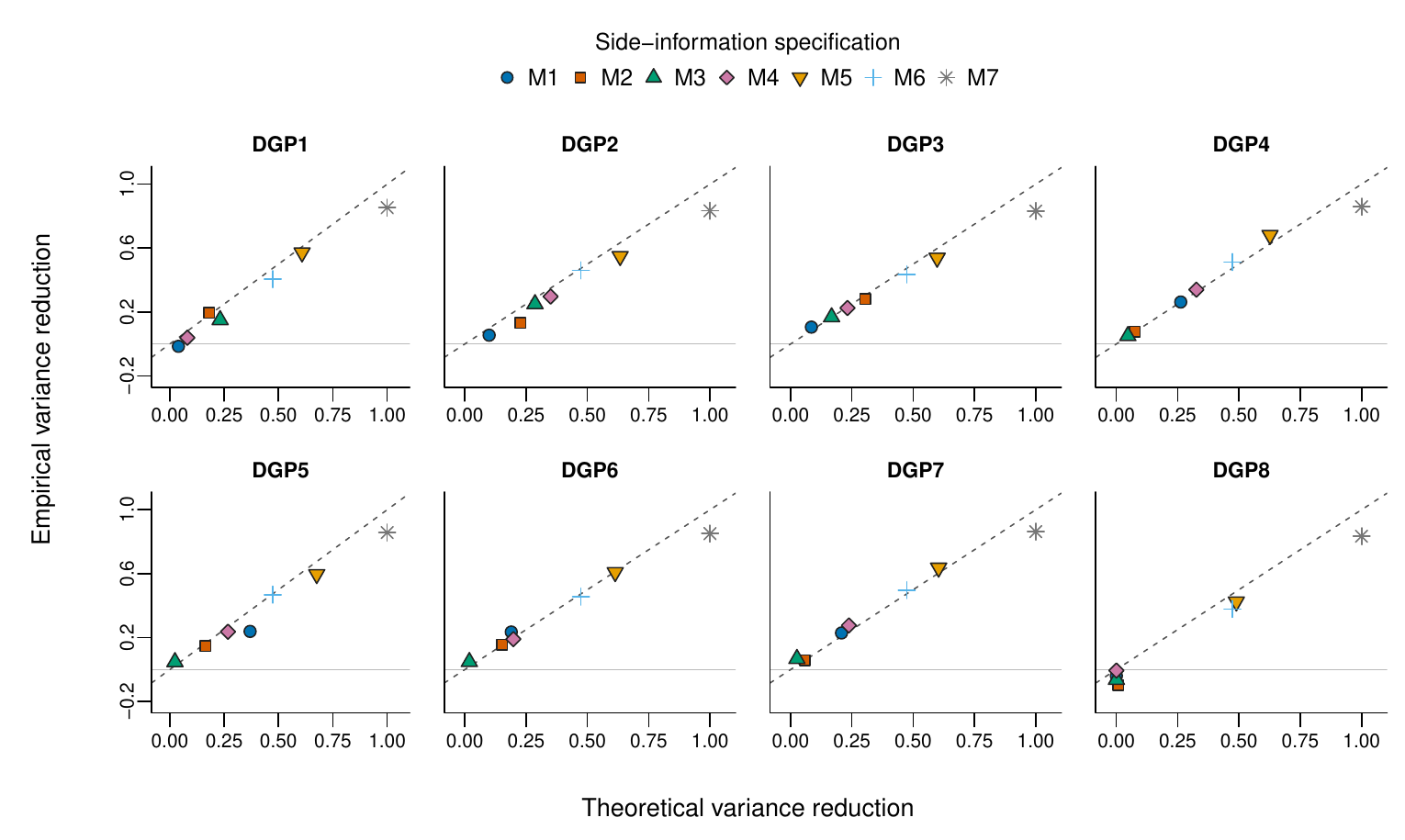}
  \caption{Relative reductions in the limiting variance of training-conditional coverage and in empirical coverage variance for M1--M7 at $n=100$ and $p=5$. Each panel corresponds to one DGP, and the dashed line represents equality.}
  \label{fig:external-projection}
\end{figure}
\endgroup

\subsection{Semi-Supervised Calibration with Prespecified Moment Functions}
\label{sec:numerical-semisupervised}

We next consider prespecified moment functions of the covariates whose means are estimated from the pooled calibration and unlabeled data. The main comparison uses independent training and calibration data sets of size $n=250$ each with $p=5$ covariates, an unlabeled sample of size $m=2500$, and an independent test sample of size 5000. Section~S2.2 of the Supplementary Material reports results for $n\in\{250,500\}$, $p\in\{5,10\}$, and $m/n\in\{1,5,10\}$.

\noindent\textbf{Design.} The training data are also used to obtain estimates $\widehat q_\tau(X)$ of the conditional $\tau$-quantiles of $Y$ given $X$ for $\tau\in\{0.1,0.5,0.9\}$, giving the covariate representation
\[
  A(X)=\bigl\{\widehat\mu(X),\widehat\rho(X),\widehat q_{0.1}(X),\widehat q_{0.5}(X), \widehat q_{0.9}(X)\bigr\}^{\top}.
\]
{\emergencystretch=1em We consider the following seven specifications of the prespecified moment function: SS1:~$h_1(A)=\bigl(\widehat\mu(X),\allowbreak\widehat\mu^2(X)\bigr)^\top$; SS2:~$h_2(A)=\bigl(\widehat\rho(X),\allowbreak\widehat\rho^2(X)\bigr)^\top$; SS3:~$h_3(A)=\bigl(\widehat q_{0.1}(X),\allowbreak \widehat q_{0.5}(X),\allowbreak\widehat q_{0.9}(X)\bigr)^\top$; SS4:~$h_4(A)=\bigl(\widehat\mu(X),\allowbreak\widehat\mu^2(X),\allowbreak \widehat\rho(X),\allowbreak\widehat\rho^2(X)\bigr)^\top$; SS5:~$h_5(A)=\bigl(\widehat\mu(X),\allowbreak\widehat\mu^2(X),\allowbreak \widehat q_{0.1}(X),\allowbreak\widehat q_{0.5}(X),\allowbreak \widehat q_{0.9}(X)\bigr)^\top$; SS6:~$h_6(A)=\bigl(\widehat\rho(X),\allowbreak\widehat\rho^2(X),\allowbreak \widehat q_{0.1}(X),\allowbreak\widehat q_{0.5}(X),\allowbreak \widehat q_{0.9}(X)\bigr)^\top$; SS7:~$h_7(A)=\bigl(\widehat\mu(X),\allowbreak\widehat\mu^2(X),\allowbreak \widehat\rho(X),\allowbreak\widehat\rho^2(X),\allowbreak \widehat q_{0.1}(X),\allowbreak\widehat q_{0.5}(X),\allowbreak \widehat q_{0.9}(X)\bigr)^\top$. Each coordinate is centred at its mean over the pooled calibration and unlabeled data, as in Section~\ref{sec:ss-prespecified}, and we use the same evaluation criteria as in Section~\ref{sec:numerical-known}.\par}

\noindent\textbf{Results.} Tables~S.8--S.11 in the Supplementary Material show that the proposed method's mean coverage, averaged over the eight DGPs, is close to the nominal level of $0.90$ under all seven specifications. Table~\ref{tab:semi-main-results} reports the variance ratios for $n=250$, $p=5$, and $m=2500$. For coverage variance, SS1 is more effective under DGP1--DGP3, where the conditional standard deviation depends on $\mu(X)$, than under DGP4--DGP7. SS2 and SS3 perform better than SS1 under DGP4--DGP7, where the conditional standard deviation depends on individual covariates or their interactions, with heavy-tailed errors in DGP7. The combined specifications SS4--SS7 have average ratios between $0.804$ and $0.812$ across all eight DGPs. The mean interval length variance ratios show a similar pattern.

\begin{table}[h]
  \centering
  \caption{Coverage and mean interval length variance ratios for semi-supervised calibration at $n=250$, $p=5$, and $m=2500$. All entries are relative to SC.}
  \label{tab:semi-main-results}
  \small
  \linespread{1}\selectfont
  \setlength{\tabcolsep}{2.5pt}
  \begin{tabular}{@{}ll*{9}{c}@{}}
    \toprule
    & & DGP1 & DGP2 & DGP3 & DGP4 & DGP5 & DGP6 & DGP7 & DGP8 & Average \\
    \midrule
    \multicolumn{11}{l}{Coverage variance ratio} \\
    SS1 & Prediction moments & 0.828 & 0.737 & 0.824 & 0.956 & 0.953 & 0.935 & 0.953 & 1.007 & 0.899 \\
    SS2 & Residual scale moments & 0.837 & 0.764 & 0.916 & 0.665 & 0.692 & 0.855 & 0.743 & 1.007 & 0.810 \\
    SS3 & Conditional quantiles & 0.846 & 0.794 & 0.855 & 0.678 & 0.682 & 0.875 & 0.796 & 0.992 & 0.815 \\
    SS4 & Prediction and residual scale & 0.822 & 0.740 & 0.864 & 0.677 & 0.730 & 0.850 & 0.765 & 0.996 & 0.806 \\
    SS5 & Prediction and quantiles & 0.817 & 0.739 & 0.825 & 0.687 & 0.696 & 0.871 & 0.781 & 1.016 & 0.804 \\
    SS6 & Residual scale and quantiles & 0.803 & 0.780 & 0.908 & 0.697 & 0.703 & 0.837 & 0.776 & 0.990 & 0.812 \\
    SS7 & All moment functions & 0.827 & 0.776 & 0.877 & 0.683 & 0.715 & 0.843 & 0.773 & 1.002 & 0.812 \\
    \midrule
    \multicolumn{11}{l}{Mean interval length variance ratio} \\
    SS1 & Prediction moments & 0.852 & 0.719 & 0.817 & 1.024 & 1.011 & 0.976 & 0.997 & 1.031 & 0.928 \\
    SS2 & Residual scale moments & 0.883 & 0.747 & 0.941 & 0.691 & 0.729 & 0.875 & 0.752 & 1.046 & 0.833 \\
    SS3 & Conditional quantiles & 0.872 & 0.780 & 0.870 & 0.699 & 0.729 & 0.900 & 0.796 & 1.014 & 0.832 \\
    SS4 & Prediction and residual scale & 0.869 & 0.740 & 0.878 & 0.697 & 0.768 & 0.861 & 0.761 & 1.035 & 0.826 \\
    SS5 & Prediction and quantiles & 0.854 & 0.725 & 0.826 & 0.710 & 0.720 & 0.917 & 0.783 & 1.031 & 0.821 \\
    SS6 & Residual scale and quantiles & 0.830 & 0.775 & 0.950 & 0.739 & 0.736 & 0.868 & 0.773 & 1.023 & 0.837 \\
    SS7 & All moment functions & 0.872 & 0.765 & 0.897 & 0.721 & 0.749 & 0.885 & 0.783 & 1.037 & 0.839 \\
    \bottomrule
  \end{tabular}
\end{table}

Section~S2.2.1 of the Supplementary Material reports a sensitivity analysis with different levels of independent noise in the covariate representation: the empirical variance reductions generally increase as the noise level decreases, while mean coverage remains stable.

\subsection{Semi-Supervised Calibration with Learned Moment Functions}
\label{sec:numerical-estimated-features}

We retain the covariate representation $A$ from Section~\ref{sec:numerical-semisupervised}. Using the calibration data, we learn an approximation to $h_\alpha^*(A)=\P\{S\le q_\alpha\mid A\}-(1-\alpha)$, the optimal scalar moment function identified in Proposition~\ref{prop:ss-basis-optimality}. The main comparison uses independent training and calibration data sets of size $n=250$ each with $p=5$ covariates, an unlabeled sample of size $m=2500$, and an independent test sample of size 5000. Section~S2.3 of the Supplementary Material reports results for $n\in\{250,500\}$, $p\in\{5,10\}$, and $m/n\in\{1,5,10\}$, and provides implementation and tuning details.

\noindent\textbf{Design.} We use SS4, which combines the prediction and residual scale moments from Section~\ref{sec:numerical-semisupervised}, as the benchmark using prespecified moment functions. We compare it with three procedures that learn the moment function by cross-fitted probabilistic classification: ridge logistic regression (CF-Logit), random forests (CF-RF), and gradient-boosted trees (CF-GB).

\noindent\textbf{Results.} Table~S.14 of the Supplementary Material shows that, averaged over the eight DGPs, mean coverage is close to the nominal level of $0.90$ for the four proposed implementations, whose mean interval lengths are $1.007$--$1.013$ times those of SC.

Table~\ref{tab:estimated-feature-main} reports the coverage and mean interval length variance ratios for the main design.\par

\begin{table}[!htb]
  \centering
  \caption{Coverage and mean interval length variance ratios for SS4 and the three procedures using cross-fitted moment functions at $n=250$, $p=5$, and $m=2500$. All entries are relative to SC.}
  \label{tab:estimated-feature-main}
  \small
  \linespread{1}\selectfont
  \begin{tabular}{lccccccccc}
    \toprule
    Method
    & DGP1 & DGP2 & DGP3 & DGP4
    & DGP5 & DGP6 & DGP7 & DGP8 & Average\\
    \midrule
    \multicolumn{10}{l}{Coverage variance ratio}\\
    SS4 & 0.768 & 0.750 & 0.834 & 0.687 & 0.723 & 0.913 & 0.772 & 1.005 & 0.806 \\
    CF-Logit & 0.800 & 0.740 & 0.853 & 0.701 & 0.796 & 0.943 & 0.793 & 1.001 & 0.828 \\
    CF-RF & 0.749 & 0.693 & 0.828 & 0.710 & 0.730 & 0.929 & 0.767 & 1.007 & 0.802 \\
    CF-GB & 0.799 & 0.729 & 0.844 & 0.727 & 0.820 & 0.920 & 0.865 & 1.012 & 0.840 \\
    \midrule
    \multicolumn{10}{l}{Mean interval length variance ratio}\\
    SS4 & 0.796 & 0.806 & 0.935 & 0.793 & 0.748 & 0.943 & 0.814 & 1.034 & 0.858 \\
    CF-Logit & 0.785 & 0.789 & 0.906 & 0.791 & 0.792 & 0.963 & 0.821 & 1.018 & 0.858 \\
    CF-RF & 0.765 & 0.740 & 0.881 & 0.816 & 0.777 & 0.975 & 0.791 & 1.038 & 0.848 \\
    CF-GB & 0.808 & 0.765 & 0.882 & 0.808 & 0.819 & 0.966 & 0.895 & 1.027 & 0.871 \\
    \bottomrule
  \end{tabular}
\end{table}
\noindent All four side-informed procedures reduce coverage variance under DGP1--DGP7, with ratios between $0.687$ and $0.943$, and the procedure with the smallest coverage variance ratio depends on the DGP: SS4 under DGP4--DGP6 and CF-RF under DGP1--DGP3 and DGP7. Under the homoscedastic DGP8 all four ratios are close to one. The mean interval length variance ratios are also below one under DGP1--DGP7.

The asymptotic variance ratios for training-conditional coverage relative to SC in Table~S.13 of the Supplementary Material average $0.754$ for SS4 and $0.769$ for CF-RF, compared with $0.737$ based on the optimal scalar moment function $h_\alpha^*(A)$.

\FloatBarrier
\subsection{Conditional Calibration}
\label{sec:numerical-conditional}

This section considers conditional calibration with summaries of the marginal and group-specific score distributions from external-reference data and covariate moments estimated from the pooled calibration and unlabeled data. The main comparison uses independent training and calibration data sets of size $n=250$ each with $p=5$ covariates, an unlabeled sample of size $m=2500$, and an independent test sample of size 5000, with Marg90 and Group90 additionally using an independent external-reference data set of size 5000. Section~S2.4 of the Supplementary Material reports results for $n\in\{250,500\}$, $p\in\{5,10\}$, and $m/n\in\{1,5,10\}$, and provides implementation details.

\noindent\textbf{Design.} Define $B(X)=\ind(X_1+X_2>0)$ and $\Phi(X)=\{1-B(X),B(X)\}^{\top}$, whose two coordinates indicate membership in the groups $B=0$ and $B=1$. Under the common covariate distribution, $\P\{B(X)=1\}=1/2$. We compare six procedures. SC uses a common threshold for both groups. OQR is ordinary quantile regression at level $1-\alpha=0.9$ with the feature map $\Phi$ and uniform weights $w_i=1/n$ assigned to the calibration data points, and serves as the comparator in Theorem~\ref{thm:conditional}. As benchmarks, Marg90 uses the moment function $\mathcal C_{\mathrm{ext}}(\ind\{S\le Q_S^{\mathrm{ext}}(0.9)\})$, while Group90 uses the two coordinates $\mathcal C_{\mathrm{ext}}(\ind\{B(X)=g\}[\ind\{S\le Q_{S\mid B}^{\mathrm{ext}}(0.9\mid g)\}-0.9])$, $g=0,1$. Here $Q_{S\mid B}^{\mathrm{ext}}(0.9\mid g)$ denotes the empirical $0.9$ quantile of $S$ within group $g$ in the external-reference data. SS2-Marg uses $\{\widehat\rho(X),\widehat\rho^2(X)\}^{\top}$ from Section~\ref{sec:numerical-semisupervised}, while SS2-Group uses the products of these two functions with each group indicator. For both specifications, we centre each coordinate at its mean over the pooled calibration and unlabeled data, compute weights as in Section~\ref{sec:ss-prespecified}, and use them in the weighted quantile regression of Section~\ref{sec:conditional} with feature map $\Phi$.

\noindent\textbf{Evaluation criteria.} Let $n_{\mathrm{test}}$ denote the test sample size, which is 5000 in these simulations. For replication $b$, let $\widehat M_b=n_{\mathrm{test}}^{-1}\sum_{i=1}^{n_{\mathrm{test}}}\Phi(X_{bi}) \{\ind(Y_{bi}\in\widehat C_b(X_{bi}))-0.9\}$ collect the two empirical coverage moments computed from the test data $(X_{bi},Y_{bi})$, and use the sum of absolute empirical coverage moments $\|\widehat M_b\|_1=|\widehat M_{b,0}|+|\widehat M_{b,1}|$. Let $\widehat\Omega$ denote the across-replication covariance matrix of $\widehat M_b$ and $\widehat\Omega_{\rm OQR}$ the corresponding matrix for OQR. For the two coverage moments, we use the normalized trace $\mathrm{tr}(\widehat\Omega_{\rm OQR}^{-1}\widehat\Omega)/2$ as a multivariate extension of the coverage variance ratio relative to OQR. This criterion is invariant to nonsingular reparameterizations of $\Phi$. The variance ratios for the group-specific coverage moments are reported in Table~S.18 of the Supplementary Material. If $\overline L_b$ is the mean interval length in replication $b$, then the mean interval length variance ratio relative to OQR is $\mathrm{Var}_b\{\overline L_b\}/\mathrm{Var}_b\{\overline L_b^{\rm OQR}\}$. The comparisons based on these two criteria in Table~\ref{tab:conditional-main} are restricted to procedures using the same class of group-specific thresholds as OQR.

\noindent\textbf{Results.} Table~\ref{tab:conditional-main} reports the results for $n=250$, $p=5$, and $m=2500$. For DGP2--DGP4 and DGP6--DGP7 SC gives mean $\|\widehat M_b\|_1$ values of $0.060$--$0.077$, which the two group-specific OQR thresholds reduce to $0.021$--$0.023$. Under DGP1, DGP5, and the homoscedastic DGP8 the prespecified groups are only weakly aligned with the score distribution, and SC has mean $\|\widehat M_b\|_1$ values of approximately $0.016$ against approximately $0.023$ for OQR.

\begin{table}[ht]
  \centering
  \caption{Conditional-calibration results at $n=250$, $p=5$, and $m=2500$.}
  \label{tab:conditional-main}
  \small
  \linespread{1}\selectfont
  \setlength{\tabcolsep}{3.6pt}
  \begin{tabular}{lccccccccc}
    \toprule
    Method & DGP1 & DGP2 & DGP3 & DGP4 & DGP5 & DGP6 & DGP7 & DGP8 & Average \\
    \midrule
    \multicolumn{10}{l}{Mean of $\|\widehat M_b\|_1$ across replications} \\
    SC & 0.016 & 0.071 & 0.064 & 0.077 & 0.016 & 0.060 & 0.068 & 0.016 & 0.049 \\
    OQR        & 0.023 & 0.022 & 0.021 & 0.023 & 0.023 & 0.022 & 0.022 & 0.023 & 0.022 \\
    Marg90    & 0.019 & 0.019 & 0.019 & 0.020 & 0.018 & 0.020 & 0.019 & 0.019 & 0.019 \\
    Group90   & 0.011 & 0.011 & 0.010 & 0.011 & 0.011 & 0.011 & 0.011 & 0.010 & 0.011 \\
    SS2-Marg  & 0.021 & 0.022 & 0.021 & 0.022 & 0.022 & 0.022 & 0.021 & 0.023 & 0.022 \\
    SS2-Group & 0.020 & 0.021 & 0.021 & 0.021 & 0.019 & 0.022 & 0.021 & 0.023 & 0.021 \\
    \midrule
    \multicolumn{10}{l}{Coverage variance ratio (normalized trace), relative to OQR} \\
    Marg90    & 0.713 & 0.782 & 0.755 & 0.776 & 0.676 & 0.756 & 0.771 & 0.715 & 0.743 \\
    Group90   & 0.214 & 0.206 & 0.250 & 0.213 & 0.214 & 0.238 & 0.230 & 0.204 & 0.221 \\
    SS2-Marg  & 0.874 & 0.948 & 0.962 & 0.917 & 0.891 & 0.992 & 0.936 & 1.003 & 0.940 \\
    SS2-Group & 0.790 & 0.906 & 0.950 & 0.902 & 0.723 & 0.944 & 0.908 & 1.000 & 0.891 \\
    \midrule
    \multicolumn{10}{l}{Mean interval length variance ratio, relative to OQR} \\
    Marg90    & 0.197 & 0.598 & 0.588 & 0.627 & 0.276 & 0.556 & 0.576 & 0.210 & 0.453 \\
    Group90   & 0.267 & 0.247 & 0.321 & 0.271 & 0.267 & 0.267 & 0.237 & 0.236 & 0.264 \\
    SS2-Marg  & 0.750 & 0.945 & 0.920 & 0.862 & 0.788 & 0.987 & 0.941 & 1.018 & 0.901 \\
    SS2-Group & 0.777 & 0.863 & 0.948 & 0.954 & 0.808 & 0.952 & 0.929 & 1.019 & 0.906 \\
    \bottomrule
  \end{tabular}
\end{table}

Group90 has the smallest coverage variance ratios, averaging $0.221$, consistent with its direct alignment with the two group-specific score events at level $0.9$, while Marg90 averages $0.743$. The semi-supervised residual scale moments give average coverage variance ratios $0.940$ for SS2-Marg and $0.891$ for SS2-Group. The average mean interval length variance ratios are $0.453$ for Marg90 and $0.264$ for Group90, against $0.901$ and $0.906$ for SS2-Marg and SS2-Group.

\FloatBarrier
\section{Real-data Analysis}
\label{sec:applications}

We apply the proposed calibration procedures to three regression data sets to assess whether the variance reductions observed in the simulations persist, while examining mean coverage and mean interval length over repeated random splits. The nominal coverage level is $0.9$ throughout.

\subsection{Data and Study Design}
\label{sec:applications-design}

We consider the Steel Industry Energy Consumption (\url{https://doi.org/10.24432/C52G8C}), Seoul Bike Sharing Demand (\url{https://doi.org/10.24432/C5F62R}), and SGEMM GPU Kernel Performance (\url{https://doi.org/10.24432/C5MK70}) data sets from the UCI Machine Learning Repository. Table~\ref{tab:applications-data} summarizes the data sets and the groups prespecified for conditional calibration before the repeated-split analysis.

\begin{table}[ht]
  \centering
  \caption{The three data sets and prespecified groups used in the empirical applications.}
  \label{tab:applications-data}
  \small
  \linespread{1}\selectfont
  \setlength{\tabcolsep}{4.5pt}
  \begin{tabular}{lrll}
    \toprule
    Data set & $p$ & Response & Group $B=1$\\
    \midrule
    Steel industry energy
    & 18 & Industrial energy use
    & Weekday\\
    Seoul bike sharing
    & 14 & Hourly rental count
    & Winter\\
    SGEMM runtime
    & 14 & Kernel runtime
    & $\mathrm{MWG},\mathrm{NWG}\geq64$\\
    \bottomrule
  \end{tabular}
\end{table}

For each data set, disjoint training and test samples are selected once and held fixed, and the remaining observations are repartitioned over 200 repetitions into mutually disjoint calibration, external-reference, and unlabeled subsets. The training and calibration subsets each have size 250, and the external-reference and unlabeled subsets each contain 2500 observations. The external-reference subset is used only to form the summaries treated as externally supplied, whereas the unlabeled subset is used only through its covariates in the semi-supervised analyses. For the random-forest predictor $\widehat\mu$ fitted once on the training data, the conformal score is $S=|Y-\widehat\mu(X)|$. The same training data are used once to construct the residual-scale and conditional-quantile estimates used in the functions of the covariates. Implementation details are given in Sections~S3.1 and S3.2 of the Supplementary Material.

\subsection{Marginal Calibration}
\label{sec:applications-marginal}

We use the evaluation criteria from Section~\ref{sec:numerical-known}, calculating coverage and mean interval length on the fixed test sample for each of the 200 repeated splits. The reported variance ratios compare their variation across splits with that under standard split conformal prediction within the same data set.

\noindent\textbf{Externally specified moments.} We use M4--M7 from Section~\ref{sec:numerical-known}, denoted Ext-Scale, Ext-ScoreMean, Ext-Score95, and Ext-Score90, with the same specifications for all three data sets. Across standard split conformal prediction and the four specifications, mean coverage ranges from $0.893$ to $0.909$. For the four specifications, mean interval lengths relative to standard split conformal prediction range from $0.992$ to $1.012$ (Table~S.24 in the Supplementary Material).

Table~\ref{tab:applications-external} reports the coverage and mean interval length variance ratios relative to standard split conformal prediction within the same data set. Ext-Scale gives moderate reductions in coverage variance, with ratios from $0.726$ to $0.790$. Ext-ScoreMean performs better than Ext-Scale for all three data sets. Ext-Score90 gives the largest and most consistent reduction in both the coverage variance and the mean interval length variance.

\begin{table}[ht]
  \centering
  \caption{Variance ratios with externally specified moments over 200 repeated splits in the three data sets. Each entry is relative to standard split conformal prediction within the same data set.}
  \label{tab:applications-external}
  \small
  \linespread{1}\selectfont
  \setlength{\tabcolsep}{3.5pt}
  \begin{tabular}{lcccccc}
    \toprule
    & \multicolumn{3}{c}{Coverage variance ratio} & \multicolumn{3}{c}{Mean interval length variance ratio}\\
    \cmidrule(lr){2-4}\cmidrule(lr){5-7}
    Method & Steel & Seoul & SGEMM & Steel & Seoul & SGEMM\\
    \midrule
    Ext-Scale & 0.726 & 0.767 & 0.790 & 0.788 & 0.802 & 0.853\\
    Ext-ScoreMean & 0.344 & 0.357 & 0.486 & 0.331 & 0.376 & 0.422\\
    Ext-Score95 & 0.603 & 0.493 & 0.497 & 0.562 & 0.488 & 0.459\\
    Ext-Score90 & 0.144 & 0.111 & 0.165 & 0.140 & 0.107 & 0.163\\
    \bottomrule
  \end{tabular}
\end{table}

\noindent\textbf{Semi-supervised calibration.} We use the unlabeled data and consider SS4 from Section~\ref{sec:numerical-semisupervised} and CF-RF from Section~\ref{sec:numerical-estimated-features}. For both procedures, the required means are estimated from the pooled calibration and unlabeled covariates, as in the simulation. Table~\ref{tab:applications-semi} reports the corresponding variance ratios.\par

\begin{table}[!htb]
  \centering
  \caption{Semi-supervised marginal-calibration variance ratios over 200 repeated splits in the three data sets. Each entry is relative to standard split conformal prediction within the same data set.}
  \label{tab:applications-semi}
  \small
  \linespread{1}\selectfont
  \setlength{\tabcolsep}{3.5pt}
  \begin{tabular}{lcccccc}
    \toprule
    & \multicolumn{3}{c}{Coverage variance ratio} & \multicolumn{3}{c}{Mean interval length variance ratio}\\
    \cmidrule(lr){2-4}\cmidrule(lr){5-7}
    Method & Steel & Seoul & SGEMM & Steel & Seoul & SGEMM\\
    \midrule
    SS4 & 0.705 & 0.706 & 0.702 & 0.742 & 0.740 & 0.743\\
    CF-RF & 0.749 & 0.796 & 0.720 & 0.795 & 0.835 & 0.692\\
    \bottomrule
  \end{tabular}
\end{table}
\noindent Both procedures reduce the coverage and mean interval length variances in all three data sets, and SS4 has the smaller coverage variance ratio in every data set; for mean interval length variance, SS4 has the smaller ratio for Steel and Seoul and CF-RF for SGEMM. Additional results are reported in Section~S3.4 of the Supplementary Material.

\subsection{Conditional Calibration}
\label{sec:applications-conditional}

We next examine coverage within the prespecified groups. For each data set, $B(X)$ indicates membership in the prespecified group in Table~\ref{tab:applications-data}. We use the six procedures from Section~\ref{sec:numerical-conditional}: Marg90 and Group90 use the external-reference subset, while SS2-Marg and SS2-Group use the unlabeled subset. Table~\ref{tab:applications-conditional} reports the mean of $\|\widehat M_b\|_1$ across splits, the coverage variance ratio, and the mean interval length variance ratio as defined in Section~\ref{sec:numerical-conditional}, with the latter two expressed relative to OQR. Group-specific coverages, variance ratios for the group-specific coverage moments, and relative mean lengths are reported in Section~S3.4 of the Supplementary Material.

\begin{table}[ht]
  \centering
  \caption{Conditional-calibration results over 200 repeated splits in the three data sets. SC denotes standard split conformal prediction. The mean of $\|\widehat M_b\|_1$ across splits, the coverage variance ratio, and the mean interval length variance ratio are shown within each data set. Both ratios are relative to OQR, with the coverage variance ratio computed using the normalized trace. The headings Coverage and Mean length denote these two variance ratios, respectively.}
  \label{tab:applications-conditional}
  \small
  \linespread{1}\selectfont
  \setlength{\tabcolsep}{2.5pt}
  \begin{tabular}{@{}l*{9}{c}@{}}
    \toprule
    & \multicolumn{3}{c}{Steel energy} & \multicolumn{3}{c}{Seoul bike} & \multicolumn{3}{c}{SGEMM runtime}\\
    \cmidrule(lr){2-4}\cmidrule(lr){5-7}\cmidrule(lr){8-10}
    & & \multicolumn{2}{c}{Variance ratios} & & \multicolumn{2}{c}{Variance ratios} & & \multicolumn{2}{c}{Variance ratios}\\
    \cmidrule(lr){3-4}\cmidrule(lr){6-7}\cmidrule(lr){9-10}
    Method & $\|\widehat M_b\|_1$ & Coverage & Mean length & $\|\widehat M_b\|_1$ & Coverage & Mean length & $\|\widehat M_b\|_1$ & Coverage & Mean length\\
    \midrule
    SC & 0.035 & -- & 0.956 & 0.056 & -- & 1.122 & 0.068 & -- & 1.003\\
    OQR & 0.023 & 1.000 & 1.000 & 0.025 & 1.000 & 1.000 & 0.021 & 1.000 & 1.000\\
    Marg90 & 0.018 & 0.675 & 0.413 & 0.019 & 0.616 & 0.289 & 0.018 & 0.737 & 0.553\\
    Group90 & 0.012 & 0.232 & 0.234 & 0.012 & 0.254 & 0.177 & 0.010 & 0.210 & 0.264\\
    SS2-Marg & 0.021 & 0.793 & 0.742 & 0.023 & 0.874 & 0.810 & 0.019 & 0.787 & 0.653\\
    SS2-Group & 0.019 & 0.655 & 0.686 & 0.023 & 0.860 & 0.852 & 0.017 & 0.661 & 0.565\\
    \bottomrule
  \end{tabular}
\end{table}

Standard split conformal prediction gives mean $\|\widehat M_b\|_1$ values between $0.035$ and $0.068$. Allowing separate group thresholds reduces the mean of $\|\widehat M_b\|_1$ in every application. Group90 gives the smallest mean of $\|\widehat M_b\|_1$ and the smallest coverage variance ratio throughout, with the latter ranging from $0.210$ to {$0.254$}, compared with $0.616$ to $0.737$ for Marg90. Both benchmarks also reduce mean interval length variance. For the semi-supervised procedures, the coverage variance ratio is lower for SS2-Group ($0.655$--$0.860$) than for SS2-Marg ($0.787$--$0.874$) in each data set. Their mean interval length variance ratios range from $0.565$ to {$0.852$} and from $0.653$ to {$0.810$}, respectively. For the four side-informed procedures, mean interval lengths relative to OQR remain close to one (Section~S3.4 of the Supplementary Material).

\section{Concluding Remarks}

This work incorporates side information into conformal prediction by using moment restrictions to construct a weighted empirical CDF of the conformal scores, whose quantiles determine the prediction sets. The resulting analysis provides a statistical basis for assessing which summaries are useful for a given prediction problem and how unlabeled data affect the attainable reductions in limiting variance. Finite-sample coverage guarantees and asymptotic variance comparisons provide complementary ways to assess the use of side information.

Future work could study robustness to misspecified moment restrictions when external summaries do not match the distribution generating the calibration data. Quantifying the effect of misspecification on coverage could guide the development of procedures based on approximate moment restrictions. Another direction concerns data-dependent selection from a growing collection of moment restrictions and fitting covariate-dependent thresholds over richer function classes. Regularization and finite-sample theory for these procedures would extend the current fixed-dimensional analysis.

\bibliographystyle{agsm-noin}
\bibliography{refs}

\end{document}


\maketitle

\section{Proofs}
\label{sec:supp-proofs}

As in Section~2.1 of the main paper, probabilities and expectations are conditional on the training data, so the score function \(s\) is treated as fixed. Let \(P\) denote the common distribution of the calibration data points \(Z_1,\ldots,Z_n\) and the test point \(Z_{n+1}\). For any integrable function \(g=g(Z)\), write \(Pg=\E\{g(Z)\}\) and \(P_ng=n^{-1}\sum_{i=1}^n g(Z_i)\). For the weights assigned to the calibration data points, \(\widehat w_1,\ldots,\widehat w_n\), write \(P_{n,w}g=\sum_i\widehat w_i g(Z_i)\). In particular, \(\widehat F_w(q)=P_{n,w}\ind(S\le q)\). Recall that \(S=s(X,Y)\) has distribution function \(F(q)=\P\{S\le q\}\), and set \(h_i=h(Z_i)\).

\subsection{Finite-sample training-conditional coverage (Theorem~1)}

Recall \(\widehat q_\alpha=\inf\{q:\widehat F_w(q)\ge(n+1)(1-\alpha)/n\}\), with \(\inf\varnothing=+\infty\). Set \(F(+\infty)=1\). Let \(F_n(q)=n^{-1}\sum_i\ind(S_i\le q)\), and define \(D_{F_n}=\sup_q|F_n(q)-F(q)|\), \(D_{\widehat F_w}=\sup_q|\widehat F_w(q)-F(q)|\), and \(D_{\widehat w}=\sum_i|\widehat w_i-1/n|\).

\begin{lemma}
  \label{lem:weighted-quantile-inversion}
  Assume the conformal scores \(S_1,\ldots,S_n\) have no ties. Then
  \[
    -D_{\widehat F_w}
    \le
    F(\widehat q_\alpha)-(1-\alpha)
    \le
    D_{\widehat F_w}+\max_i\widehat w_i+\frac1n.
  \]
\end{lemma}

\begin{proof}
  If \(\widehat q_\alpha=+\infty\), then \((n+1)(1-\alpha)/n>1\), so \(\alpha<1/(n+1)<1/n\). Since \(F(+\infty)=1\), we have \(F(\widehat q_\alpha)-(1-\alpha)=\alpha\). Both bounds follow because \(D_{\widehat F_w}\) and \(\max_i\widehat w_i\) are nonnegative.

  Suppose now that \(\widehat q_\alpha\) is finite. The definition gives \(\widehat F_w(\widehat q_\alpha)\ge(n+1)(1-\alpha)/n\), so \(F(\widehat q_\alpha)\ge1-\alpha-D_{\widehat F_w}\). For the upper bound, write \(\widehat F_w(q-)=\lim_{t\uparrow q}\widehat F_w(t)\). Minimality gives \(\widehat F_w(\widehat q_\alpha-)<(n+1)(1-\alpha)/n\). Under the no-tie condition, the jump at \(\widehat q_\alpha\) is at most \(\max_i\widehat w_i\). Comparing \(F\) and \(\widehat F_w\) and using \((1-\alpha)/n\le1/n\) gives the upper bound.
\end{proof}

\begin{lemma}
  \label{lem:cdf-weight-perturbation}
  For any nonnegative weights summing to one, \(D_{\widehat F_w}\le D_{F_n}+D_{\widehat w}\).
\end{lemma}

\begin{proof}
  For each \(q\), the triangle inequality gives \(|\widehat F_w(q)-F(q)|\le |F_n(q)-F(q)|+|\sum_i(\widehat w_i-1/n)\ind(S_i\le q)|\). The second term is at most \(D_{\widehat w}\) because \(0\le\ind(S_i\le q)\le1\). Taking the supremum over \(q\) proves the claim.
\end{proof}

\begin{lemma}
  \label{lem:finite-moment-concentration}
  Assume \(P h=0\), \(\|h(Z)\|_2\le B\) almost surely, and \(\lambda_{\min}(\Sigma_h)\ge\kappa>0\). For every \(\delta\in(0,1)\), with probability at least \(1-2\delta/3\), \(\|P_nh\|_2\le\varepsilon_{\mu,n}(\delta)\) and \(\lambda_{\min}(P_n(hh^\top))\ge\kappa-\varepsilon_{\Sigma,n}(\delta)\).
\end{lemma}

\begin{proof}
  Write \(h^{(j)}\) for coordinate \(j\) of \(h\). Since \(|h^{(j)}(Z)|\le B\) and \(Ph^{(j)}=0\), applying the Hoeffding bound of \citet[Proposition~2.5]{Wainwright2019HDS} coordinatewise and then taking a union bound gives
  \[
    \max_{1\le j\le r}|P_nh^{(j)}|
    \le
    B\left\{\frac{2\log(6r/\delta)}{n}\right\}^{1/2}
  \]
  with probability at least \(1-\delta/3\). On this event, \(\|P_nh\|_2\le\sqrt r\max_j|P_nh^{(j)}|\le\varepsilon_{\mu,n}(\delta)\). Similarly, each \(h^{(j)}(Z)h^{(k)}(Z)\) lies in \([-B^2,B^2]\). Hoeffding's bound followed by a union bound over \(r^2\) coordinate pairs gives
  \[
    \max_{1\le j,k\le r}
    \left|P_n\{h^{(j)}h^{(k)}\}-P\{h^{(j)}h^{(k)}\}\right|
    \le
    B^2\left\{\frac{2\log(6r^2/\delta)}{n}\right\}^{1/2}
  \]
  with probability at least \(1-\delta/3\). On this event, \(\|P_n(hh^\top)-\Sigma_h\|_{\mathrm{op}}\le r\max_{j,k}|P_n\{h^{(j)}h^{(k)}\}-P\{h^{(j)}h^{(k)}\}|\le\varepsilon_{\Sigma,n}(\delta)\), where \(\|\cdot\|_{\mathrm{op}}\) denotes the spectral norm. Weyl's inequality \citep[Eq.~(8.9)]{Wainwright2019HDS} therefore yields \(\lambda_{\min}(P_n(hh^\top))\ge\kappa-\varepsilon_{\Sigma,n}(\delta)\). The intersection of the two events has probability at least \(1-2\delta/3\), which proves the lemma.
\end{proof}

\begin{lemma}
  \label{lem:finite-weight-bound}
  Suppose that \(\|h_i\|_2\le B\), \(\|P_nh\|_2\le\eta\), and \(\lambda_{\min}(P_n(hh^\top))\ge \underline{\kappa}>0\), with \(4B\eta/\underline{\kappa}\le1/2\). Then the optimization in Eq.~(1) has a solution \(\widehat w\) and a Lagrange multiplier \(\widehat\lambda\) such that \(\widehat w_i=1/\{n(1+\widehat\lambda^\top h_i)\}>0\) for every \(i\) and \(P_n\{h/(1+\widehat\lambda^\top h)\}=0\). Moreover, \(\|\widehat\lambda\|_2\le 4\eta/\underline{\kappa}\), \(D_{\widehat w}\le 8B\eta/\underline{\kappa}\), and \(\max_i\widehat w_i\le 2/n\).
\end{lemma}

\begin{proof}
  If \(\eta=0\), then \(P_nh=0\), so \(\widehat\lambda=0\) generates the uniform weights, which solve the optimization and satisfy all the stated bounds. Assume \(\eta>0\) and set \(R=4\eta/\underline{\kappa}\). For \(\|\lambda\|_2\le R\), the condition \(BR\le1/2\) gives \(1+\lambda^\top h_i\in[1/2,3/2]\). Thus the closed ball of radius \(R\) lies in the domain of \(L_n(\lambda)=-P_n\log(1+\lambda^\top h)\), and \(L_n\) attains a minimum on this ball. Its gradient is \(\nabla L_n(\lambda)=-P_n\{h/(1+\lambda^\top h)\}\). On the boundary, \(-a/(1+a)=-a+a^2/(1+a)\) gives
  \[
    \lambda^\top\nabla L_n(\lambda)
    =
    -\lambda^\top P_nh
    +
    P_n\left\{\frac{(\lambda^\top h)^2}{1+\lambda^\top h}\right\}
    \ge
    -R\eta+\frac23\,\underline{\kappa}R^2
    =
    \frac{20\eta^2}{3\underline{\kappa}}
    >0.
  \]
  For a boundary point \(\lambda\), the derivative along the path towards the origin is \(\left.\frac{d}{ds}L_n((1-s)\lambda)\right|_{s=0}=-\lambda^\top\nabla L_n(\lambda)<0\). Thus every boundary point has a descent direction into the ball. No boundary point minimizes \(L_n\), so a minimizer \(\widehat\lambda\) lies in the interior of the ball and satisfies \(P_n\{h/(1+\widehat\lambda^\top h)\}=0\).

  Taking the inner product of \(P_n\{h/(1+\widehat\lambda^\top h)\}=0\) with \(\widehat\lambda\) and using \((1+a)^{-1}=1-a/(1+a)\) gives \(P_n\{1/(1+\widehat\lambda^\top h)\}=1\). Consequently, \(\widehat w_i=1/\{n(1+\widehat\lambda^\top h_i)\}\) are positive, sum to one, and satisfy the moment constraint. The identity \(1/\widehat w_i=n+n\widehat\lambda^\top h_i\) then verifies the stationarity equations for the optimization in Eq.~(1). Strict concavity of the objective therefore identifies these weights as its unique solution. Since \(\widehat\lambda\) lies in the ball, \(\|\widehat\lambda\|_2\le R\). For each \(i\),
  \[
    \left|\widehat w_i-\frac1n\right|
    =
    \frac{|\widehat\lambda^\top h_i|}{n(1+\widehat\lambda^\top h_i)}
    \le
    \frac{2B\|\widehat\lambda\|_2}{n}.
  \]
  Summing this inequality and substituting \(R=4\eta/\underline{\kappa}\) gives \(D_{\widehat w}\le8B\eta/\underline{\kappa}\). The denominator bound also gives \(\max_i\widehat w_i\le2/n\).
\end{proof}

\begin{proof}
  The Dvoretzky--Kiefer--Wolfowitz inequality \citep{Massart1990DKW} gives \(D_{F_n}\le\{\log(6/\delta)/(2n)\}^{1/2}\) with probability at least \(1-\delta/3\). On the event in Lemma~\ref{lem:finite-moment-concentration}, the assumed bounds on \(\varepsilon_{\mu,n}(\delta)\) and \(\varepsilon_{\Sigma,n}(\delta)\) imply \(\|P_nh\|_2\le\varepsilon_{\mu,n}(\delta)\), \(\lambda_{\min}(P_n(hh^\top))\ge\kappa/2\), and \(8B\varepsilon_{\mu,n}(\delta)/\kappa\le1/2\). Lemma~\ref{lem:finite-weight-bound}, with \(\underline{\kappa}=\kappa/2\) and \(\eta=\varepsilon_{\mu,n}(\delta)\), therefore shows that the optimization has a solution with strictly positive weights and gives \(D_{\widehat w}\le16B\varepsilon_{\mu,n}(\delta)/\kappa\) and \(\max_i\widehat w_i\le2/n\).

  On the intersection of these events, Lemmas~\ref{lem:weighted-quantile-inversion} and~\ref{lem:cdf-weight-perturbation} give
  \[
    -D_{F_n}-D_{\widehat w}
    \le
    F(\widehat q_\alpha)-(1-\alpha)
    \le
    D_{F_n}+D_{\widehat w}+\max_i\widehat w_i+\frac1n.
  \]
  Conditional on the calibration data, the score of the independent test point has CDF \(F\), so the training-conditional coverage probability is \(F(\widehat q_\alpha)\). Applying the displayed inequality with the Dvoretzky--Kiefer--Wolfowitz bound and the bounds on \(D_{\widehat w}\) and \(\max_i\widehat w_i\) and taking a union bound proves Theorem~1.
\end{proof}

\subsection{Finite-sample marginal coverage (Corollary~1)}

\begin{proof}

  For the marginal coverage bound, the optimization constraints and the convention \(\widehat w_i=1/n\) when no solution exists ensure that \(\widehat w\) is a probability vector for every realization of the calibration data. Lemma~\ref{lem:weighted-quantile-inversion} therefore applies to the quantile defined on the extended real line. Let
  \[
    \mathcal E_n
    =
    \left\{
    \|P_nh\|_2\le\frac{\kappa}{16B},
    \quad
    \lambda_{\min}(P_n(hh^\top))\ge\frac{\kappa}{2}
    \right\}.
  \]
  The coordinatewise Hoeffding bounds from Lemma~\ref{lem:finite-moment-concentration}, applied at the thresholds \(\kappa/(16B\sqrt r)\) and \(\kappa/(2r)\), followed by Weyl's inequality, give \(\P\{\mathcal E_n^c\}\le p_{\mu,\Sigma,n}\).

  On \(\mathcal E_n\), Lemma~\ref{lem:finite-weight-bound}, with \(\eta=\|P_nh\|_2\) and \(\underline{\kappa}=\kappa/2\), gives \(D_{\widehat w}\le16B\|P_nh\|_2/\kappa\) and \(\max_i\widehat w_i\le2/n\), whereas on \(\mathcal E_n^c\), nonnegativity and \(\sum_i\widehat w_i=1\) give \(D_{\widehat w}\le2\) and \(\max_i\widehat w_i\le1\). Using \(Ph=0\), Jensen's inequality gives \(\E\{\|P_nh\|_2\}\le[\E\{\|P_nh\|_2^2\}]^{1/2}=n^{-1/2}[P\{\|h\|_2^2\}]^{1/2}\le B/\sqrt n\). Splitting the expectations of \(D_{\widehat w}\) and \(\max_i\widehat w_i\) over \(\mathcal E_n\) and \(\mathcal E_n^c\), and integrating the Dvoretzky--Kiefer--Wolfowitz inequality, gives
  \[
    \begin{aligned}
      \E\{D_{\widehat w}\}
      &\le \frac{16B^2}{\kappa\sqrt n}
      +2p_{\mu,\Sigma,n}, \\
      \E\left\{\max_i\widehat w_i\right\}
      &\le \frac2n+p_{\mu,\Sigma,n}, \\
      \E\{D_{F_n}\}
      &\le \sqrt{\frac{\pi}{2n}}.
    \end{aligned}
  \]
  Finally, the tower property gives \(\E\{F(\widehat q_\alpha)\}=\P\{Y_{n+1}\in\widehat C(X_{n+1})\}\). Taking expectations in the bound obtained by combining Lemmas~\ref{lem:weighted-quantile-inversion} and~\ref{lem:cdf-weight-perturbation} proves Corollary~1.
\end{proof}

\subsection{Asymptotic distribution of training-conditional coverage (Theorem~2)}

Let \(\widehat\lambda\) denote a minimizer of the dual objective function \(L_n\) defined in the proof of Lemma~\ref{lem:finite-weight-bound} when one exists, and set \(\widehat\lambda=0\) otherwise.

\begin{lemma}
  \label{lem:lambda-expansion}
  Suppose \(Ph=0\), \(P\{\|h\|_2^2\}<\infty\), and \(\Sigma_h=P(hh^\top)\) is positive definite. With probability tending to one, \(L_n\) has a minimizer \(\widehat\lambda\), the weights satisfy \(\widehat w_i=1/[n\{1+\widehat\lambda^\top h(Z_i)\}]\), and
  \[
    \widehat\lambda=\Sigma_h^{-1}P_nh+o_p(n^{-1/2}),
    \qquad
    \max_{1\le i\le n}|n\widehat w_i-1|=o_p(1).
  \]
  For any scalar-valued or fixed finite-dimensional vector-valued function \(g\) with \(P\{\|g\|_2^2\}<\infty\),
  \[
    P_{n,w}g
    =
    P_ng
    -
    P(gh^\top)\Sigma_h^{-1}P_nh
    +o_p(n^{-1/2}).
  \]
\end{lemma}

\begin{proof}
  The central limit theorem and law of large numbers give the first two relations below:
  \[
    P_nh=O_p(n^{-1/2}),\qquad
    P_n(hh^\top)=\Sigma_h+o_p(1),\qquad
    \max_{1\le i\le n}\|h_i\|_2=o_p(n^{1/2}).
  \]
  For the third relation, the union bound gives, for every \(\epsilon>0\),
  \[
    \P\left\{\max_{i\le n}\|h_i\|_2>\epsilon\sqrt n\right\}
    \le
    \epsilon^{-2}
    \E\!\left\{
    \|h\|_2^2\ind(\|h\|_2^2>\epsilon^2n)
    \right\}
    \to0.
  \]
  Hence, for fixed \(C<\infty\), \(\sup_{\|\lambda\|_2\le Cn^{-1/2}}\max_i|\lambda^\top h_i|=o_p(1)\). On the resulting high-probability event, \(1+\lambda^\top h_i>0\) throughout this ball, and applying \((1+a)^{-1}=1-a+a^2/(1+a)\) pointwise gives
  \[
    P_n\left\{\frac{h}{1+\lambda^\top h}\right\}
    =P_nh-P_n(hh^\top)\lambda
    +P_n\left\{\frac{h(\lambda^\top h)^2}{1+\lambda^\top h}\right\}.
  \]
  The norm of the last term is bounded by
  \[
    \left\|
    P_n\left\{
    \frac{h(\lambda^\top h)^2}{1+\lambda^\top h}
    \right\}
    \right\|_2
    \le
    \frac{\max_i|\lambda^\top h_i|}
    {1-\max_i|\lambda^\top h_i|}
    \|\lambda\|_2P_n\{\|h\|_2^2\}
    =o_p(n^{-1/2})
  \]
  uniformly over \(\|\lambda\|_2\le Cn^{-1/2}\). Consequently,
  \[
    P_n\left\{\frac{h}{1+\lambda^\top h}\right\}
    =
    P_nh-P_n(hh^\top)\lambda+o_p(n^{-1/2}).
  \]
  Let \(\widehat\kappa_n=\lambda_{\min}(P_n(hh^\top))\), set \(R_n=4\|P_nh\|_2/\widehat\kappa_n\) when \(\widehat\kappa_n>0\), and set \(R_n=0\) otherwise. Then \(\widehat\kappa_n\xrightarrow{p}\lambda_{\min}(\Sigma_h)>0\), \(R_n=O_p(n^{-1/2})\), and \(R_n\max_i\|h_i\|_2=o_p(1)\). Therefore, with probability tending to one, Lemma~\ref{lem:finite-weight-bound} yields a multiplier \(\widehat\lambda\) satisfying \(\|\widehat\lambda\|_2\le R_n\) and \(P_n\{h/(1+\widehat\lambda^\top h)\}=0\). Convexity of \(L_n\) shows that \(\widehat\lambda\) minimizes \(L_n\) on its domain. Evaluating the uniform expansion at \(\widehat\lambda\) gives
  \[
    \widehat\lambda=\{P_n(hh^\top)\}^{-1}P_nh+o_p(n^{-1/2}).
  \]
  Since \(\{P_n(hh^\top)\}^{-1}P_nh=\Sigma_h^{-1}P_nh+o_p(n^{-1/2})\), the multiplier expansion follows.

  The representation \(\widehat w_i=1/\{n(1+\widehat\lambda^\top h_i)\}\) from Lemma~\ref{lem:finite-weight-bound} gives
  \[
    \max_i|n\widehat w_i-1|
    \le
    \frac{\max_i|\widehat\lambda^\top h_i|}
    {1-\max_i|\widehat\lambda^\top h_i|}
    =o_p(1).
  \]
  For \(g\) as in the lemma, Cauchy--Schwarz gives \(P_n(\|g\|_2\|h\|_2)=O_p(1)\), so applying \((1+a)^{-1}=1-a+a^2/(1+a)\) pointwise with \(a=\widehat\lambda^\top h\) gives
  \[
    P_{n,w}g=P_ng-P_n(gh^\top)\widehat\lambda+o_p(n^{-1/2}).
  \]
  Since \(P_n(gh^\top)=P(gh^\top)+o_p(1)\), substituting the multiplier expansion proves the result.
\end{proof}

\begin{lemma}
  \label{lem:weighted-threshold-moment-expansion}
  Under the conditions of Theorem~2,
  \[
    \sqrt n\,P_{n,w}\psi_\alpha
    =
    \frac1{\sqrt n}\sum_{i=1}^n
    \{\psi_\alpha(Z_i)-c_\alpha^\top\Sigma_h^{-1}h(Z_i)\}
    +o_p(1).
  \]
\end{lemma}

\begin{proof}
  Apply Lemma~\ref{lem:lambda-expansion} with \(g=\psi_\alpha\), using \(P\psi_\alpha=0\) and \(P(\psi_\alpha h^\top)=c_\alpha^\top\).
\end{proof}

\begin{lemma}
  \label{lem:weighted-quantile-expansion}
  Under the conditions of Theorem~2,
  \[
    \sqrt n(\widehat q_\alpha-q_\alpha)
    =
    -\frac{1}{f(q_\alpha)}
    \frac1{\sqrt n}\sum_{i=1}^n
    \{\psi_\alpha(Z_i)-c_\alpha^\top\Sigma_h^{-1}h(Z_i)\}
    +o_p(1).
  \]
\end{lemma}

\begin{proof}
  For \(t\in\R\), let \(d_{n,t}(Z)=\ind(S\le q_\alpha+t/\sqrt n)-\ind(S\le q_\alpha)\). Set \(A_n=n^{-1/2}\sum_{i=1}^n\{\psi_\alpha(Z_i)-c_\alpha^\top\Sigma_h^{-1}h(Z_i)\}\). Since \(F(q_\alpha)=1-\alpha\) and the weights sum to one, \(P_{n,w}\psi_\alpha=\widehat F_w(q_\alpha)-F(q_\alpha)\). Lemma~\ref{lem:weighted-threshold-moment-expansion} therefore gives
  \begin{equation}
    \label{eq:weighted-fixed-threshold}
    \sqrt n\,P_{n,w}\psi_\alpha
    =\sqrt n\{\widehat F_w(q_\alpha)-F(q_\alpha)\}
    =A_n+o_p(1).
  \end{equation}

  Fix \(M<\infty\). The class \(\{\ind(S\le u):u\in\R\}\) is \(P\)-Donsker by \citet[Example~2.5.4]{VanDerVaartWellner2023WCEP}. Continuity of \(F\) at \(q_\alpha\) implies \(P(S=q_\alpha)=0\) and \(\sup_{|t|\le M}P(d_{n,t}^2)\le P(|S-q_\alpha|\le M n^{-1/2})\to0\). Asymptotic equicontinuity \citep[Corollary~2.3.12]{VanDerVaartWellner2023WCEP} therefore gives
  \begin{equation}
    \label{eq:empirical-local-increment}
    \sup_{|t|\le M}
    \left|\sqrt n(P_n-P)d_{n,t}\right|=o_p(1).
  \end{equation}

  We next prove
  \begin{equation}
    \label{eq:h-local-increment}
    \sup_{|t|\le M}\|P_n(hd_{n,t})\|_2=o_p(1).
  \end{equation}
  For \(0<L<\infty\), let \(h_L^{(j)}=h^{(j)}\ind(|h^{(j)}|\le L)\). The indicators of the nested threshold sets \(\{S\le u\}\) form a VC-subgraph class. Lemma~2.6.20(vi) of \citet{VanDerVaartWellner2023WCEP} then shows that \(\mathcal H_{j,L}=\{h_L^{(j)}\ind(S\le u):u\in\R\}\) is VC-subgraph with envelope \(L\). Writing \(V\) for its VC dimension, Theorem~2.6.7 of \citet{VanDerVaartWellner2023WCEP} gives a constant \(C_V<\infty\), depending only on \(V\), such that, for every probability measure \(Q\) and \(0<\epsilon<1\), \(N(\epsilon L,\mathcal H_{j,L},L_1(Q))\le C_V\epsilon^{-V}\). For fixed \(L\), the bound is uniform in \(Q\), so taking \(Q=P_n\) gives \(n^{-1}\log N(\delta,\mathcal H_{j,L},L_1(P_n))\to0\) for every \(\delta>0\). Theorem~2.4.3 of \citet{VanDerVaartWellner2023WCEP} therefore implies that \(\mathcal H_{j,L}\) is \(P\)-Glivenko--Cantelli. Hence
  \[
    \sup_{|t|\le M}
    |(P_n-P)(h_L^{(j)}d_{n,t})|
    \le
    2\sup_{u\in\R}
    |(P_n-P)\{h_L^{(j)}\ind(S\le u)\}|
    =o_p(1).
  \]
  The truncation remainder \(h^{(j)}-h_L^{(j)}\) satisfies
  \[
    \sup_{|t|\le M}
    |(P_n-P)\{(h^{(j)}-h_L^{(j)})d_{n,t}\}|
    \le
    (P_n+P)\{|h^{(j)}|\ind(|h^{(j)}|>L)\}.
  \]
  For fixed \(L\), the right-hand side is \(2P\{|h^{(j)}|\ind(|h^{(j)}|>L)\}+o_p(1)\). Since \(P\{|h^{(j)}|\}<\infty\) and \(r\) is fixed, first letting \(n\to\infty\) for fixed \(L\) and then \(L\to\infty\) gives
  \[
    \sup_{|t|\le M}\|(P_n-P)(hd_{n,t})\|_2=o_p(1).
  \]
  Dominated convergence also gives \(\sup_{|t|\le M}\|P(hd_{n,t})\|_2\le P\{\|h\|_2\ind(|S-q_\alpha|\le M n^{-1/2})\}=o(1)\). Together with the displayed bound for \((P_n-P)(hd_{n,t})\), this proves Eq.~\eqref{eq:h-local-increment}.

  Lemma~\ref{lem:lambda-expansion} and its proof give \(\widehat\lambda=O_p(n^{-1/2})\), \(\max_i|\widehat\lambda^\top h_i|=o_p(1)\), and \(P_n\{\|h\|_2^2\}=O_p(1)\). Thus, with probability tending to one, substituting \((1+a)^{-1}=1-a+a^2/(1+a)\), with \(a=\widehat\lambda^\top h\), into the representation of \(\widehat w_i\) in Lemma~\ref{lem:lambda-expansion} gives
  \begin{equation}
    \label{eq:weighted-d-expansion}
    P_{n,w}d_{n,t}
    =P_nd_{n,t}-\widehat\lambda^\top P_n(hd_{n,t})+R_{n,t},
  \end{equation}
  where \(R_{n,t}:=P_n\{d_{n,t}(\widehat\lambda^\top h)^2/(1+\widehat\lambda^\top h)\}\). Since \(|d_{n,t}|\le1\), uniformly over \(|t|\le M\),
  \[
    |R_{n,t}|
    \le
    \frac{\|\widehat\lambda\|_2^2P_n\{\|h\|_2^2\}}
    {1-\max_i|\widehat\lambda^\top h_i|}
    =O_p(n^{-1}).
  \]
  Combining Eq.~\eqref{eq:weighted-d-expansion} with \(\sup_{|t|\le M}|R_{n,t}|=O_p(n^{-1})\) and Eqs.~\eqref{eq:empirical-local-increment} and~\eqref{eq:h-local-increment} gives
  \begin{equation}
    \label{eq:weighted-local-increment}
    \begin{aligned}
      \sup_{|t|\le M}
      \left|\sqrt n(P_{n,w}-P)d_{n,t}\right|
      &\le
      \sup_{|t|\le M}
      \left|\sqrt n(P_n-P)d_{n,t}\right| \\
      &\quad+\sqrt n\|\widehat\lambda\|_2
      \sup_{|t|\le M}\|P_n(hd_{n,t})\|_2
      +\sqrt n\sup_{|t|\le M}|R_{n,t}| \\
      &=o_p(1).
    \end{aligned}
  \end{equation}

  Continuous differentiability of \(F\) also gives
  \begin{equation}
    \label{eq:population-local-increment}
    \sup_{|t|\le M}
    \left|\sqrt n\{F(q_\alpha+t/\sqrt n)-F(q_\alpha)\}
    -f(q_\alpha)t\right|=o(1).
  \end{equation}
  Using \(P_{n,w}d_{n,t}=\widehat F_w(q_\alpha+t/\sqrt n)-\widehat F_w(q_\alpha)\), \(Pd_{n,t}=F(q_\alpha+t/\sqrt n)-F(q_\alpha)\), and \(F(q_\alpha)=1-\alpha\), Eqs.~\eqref{eq:weighted-fixed-threshold}, \eqref{eq:weighted-local-increment}, and~\eqref{eq:population-local-increment} yield
  \begin{equation}
    \label{eq:weighted-local-cdf-expansion}
    \sup_{|t|\le M}
    \left|
    \sqrt n\{\widehat F_w(q_\alpha+t/\sqrt n)-(1-\alpha)\}
    -A_n-f(q_\alpha)t
    \right|=o_p(1).
  \end{equation}
  Since \(A_n=O_p(1)\), \(f(q_\alpha)>0\), and \((1-\alpha)/\sqrt n\to0\), applying Eq.~\eqref{eq:weighted-local-cdf-expansion} with \(M=T\) at \(t=\pm T\) gives
  \[
    \lim_{T\to\infty}\liminf_{n\to\infty}
    \P\left\{
    \widehat F_w(q_\alpha-T/\sqrt n)<\frac{n+1}{n}(1-\alpha)
    \le\widehat F_w(q_\alpha+T/\sqrt n)
    \right\}
    =1.
  \]
  When the two inequalities in the preceding display hold, monotonicity and the definition of \(\widehat q_\alpha\) imply \(|\sqrt n(\widehat q_\alpha-q_\alpha)|\le T\). Hence \(t_n:=\sqrt n(\widehat q_\alpha-q_\alpha)=O_p(1)\). Lemma~\ref{lem:lambda-expansion} gives \(\max_i\widehat w_i=\{1+o_p(1)\}/n\). Minimality of \(\widehat q_\alpha\) and the no-tie condition give
  \[
    0\le\widehat F_w(\widehat q_\alpha)-\frac{n+1}{n}(1-\alpha)
    \le\max_i\widehat w_i
    =\frac{1+o_p(1)}{n}.
  \]
  The uniformity in Eq.~\eqref{eq:weighted-local-cdf-expansion} and \(t_n=O_p(1)\) permit evaluation at \(t=t_n\), giving
  \[
    A_n+f(q_\alpha)t_n
    =\sqrt n\left\{\widehat F_w(\widehat q_\alpha)-\frac{n+1}{n}(1-\alpha)\right\}
    +\frac{1-\alpha}{\sqrt n}+o_p(1)
    =o_p(1).
  \]
  Therefore \(t_n=-A_n/f(q_\alpha)+o_p(1)\), which proves the result.
\end{proof}

\begin{proof}
  Conditional on the calibration data, \(\P\{Y_{n+1}\in\widehat C(X_{n+1})\mid Z_1,\ldots,Z_n\}=F(\widehat q_\alpha)\). Lemma~\ref{lem:weighted-quantile-expansion} gives \(\widehat q_\alpha-q_\alpha=O_p(n^{-1/2})\). Differentiability of \(F\) therefore gives
  \[
    \begin{aligned}
      \sqrt n\{F(\widehat q_\alpha)-F(q_\alpha)\}
      &=
      f(q_\alpha)\sqrt n(\widehat q_\alpha-q_\alpha)+o_p(1) \\
      &=
      -\frac1{\sqrt n}\sum_{i=1}^n
      \{\psi_\alpha(Z_i)-c_\alpha^\top\Sigma_h^{-1}h(Z_i)\}
      +o_p(1).
    \end{aligned}
  \]
  Since \(F(q_\alpha)=1-\alpha\), the preceding display proves the coverage expansion in Theorem~2. The summand has mean zero and variance
  \[
    \Var\{\psi_\alpha-c_\alpha^\top\Sigma_h^{-1}h\}
    =
    P\psi_\alpha^2-c_\alpha^\top\Sigma_h^{-1}c_\alpha
    =
    \alpha(1-\alpha)-c_\alpha^\top\Sigma_h^{-1}c_\alpha.
  \]
  The central limit theorem applied to the i.i.d. sum in this expansion, followed by Slutsky's theorem, proves the limiting distribution in Theorem~2.
\end{proof}

\subsection{Asymptotic distribution of average prediction-set size (Corollary~2)}

\begin{proof}
  Lemma~\ref{lem:weighted-quantile-expansion} gives \(\widehat q_\alpha-q_\alpha=O_p(n^{-1/2})\). Differentiability of \(\mathcal L\) at \(q_\alpha\) therefore yields
  \[
    \begin{aligned}
      \sqrt n\{\mathcal L(\widehat q_\alpha)-\mathcal L(q_\alpha)\}
      &=
      \mathcal L'(q_\alpha)\sqrt n(\widehat q_\alpha-q_\alpha)+o_p(1) \\
      &=
      -\frac{\mathcal L'(q_\alpha)}{f(q_\alpha)}
      \frac1{\sqrt n}\sum_{i=1}^n
      \{\psi_\alpha(Z_i)-c_\alpha^\top\Sigma_h^{-1}h(Z_i)\}
      +o_p(1).
    \end{aligned}
  \]
  The central limit theorem and Slutsky's theorem now give Corollary~2.
\end{proof}

\subsection{Asymptotic results with prespecified moment functions (Theorem~3)}

Recall that \(H=h(A)-\E\{h(A)\}\). For functions of \((Z,A)\), write \(Pg=\E\{g(Z,A)\}\) and \(P_ng=n^{-1}\sum_i g(Z_i,A_i)\). Write \(H_i=h(A_i)-\E\{h(A)\}\) and \(H_j^{\mathrm u}=h(A_j^{\mathrm u})-\E\{h(A)\}\), and define
\[
  \bar H_N
  =
  \frac1N\left(\sum_{i=1}^nH_i+\sum_{j=1}^mH_j^{\mathrm u}\right),
  \qquad
  \widetilde H_i=H_i-\bar H_N,
  \qquad
  \Delta_{h,n}
  =
  P_n\widetilde H
  =
  \bar h_n-\bar h_N.
\]
For a scalar function \(g=g(Z,A)\), write \(P_{n,w}^{\mathrm{ss}}g=\sum_i\widehat w_i^{\,\mathrm{ss}}g(Z_i,A_i)\).

\begin{lemma}
  \label{lem:ss-weight-expansion}
  Under the conditions of Theorem~3, with probability tending to one, the optimization in Eq.~(1), with \(h(Z_i)\) replaced by \(\widetilde H_i\), has a solution with strictly positive weights. For a corresponding multiplier \(\widehat\lambda_{\mathrm{ss}}\), the weights satisfy \(\widehat w_i^{\,\mathrm{ss}}=1/\{n(1+\widehat\lambda_{\mathrm{ss}}^\top\widetilde H_i)\}\), \(\widehat\lambda_{\mathrm{ss}}=\Sigma_A^{-1}\Delta_{h,n}+o_p(n^{-1/2})\), and \(\max_{1\le i\le n}|n\widehat w_i^{\,\mathrm{ss}}-1|=o_p(1)\). For every fixed bounded scalar function \(g=g(Z,A)\), \(P_{n,w}^{\mathrm{ss}}g=P_ng-P(gH^\top)\Sigma_A^{-1}\Delta_{h,n}+o_p(n^{-1/2})\).
\end{lemma}

\begin{proof}
  The calibration data and unlabeled data are independent, so \(\E\{\Delta_{h,n}\}=0\). Moreover, \(\E\{\Delta_{h,n}\Delta_{h,n}^\top\}=m/(nN)\Sigma_A\), which gives \(\Delta_{h,n}=O_p(n^{-1/2})\). Chebyshev's inequality, the law of large numbers, and the tail argument in the proof of Lemma~\ref{lem:lambda-expansion} give, respectively:
  \[
    \bar H_N=O_p(N^{-1/2}),\qquad
    P_n(\widetilde H\widetilde H^\top)=\Sigma_A+o_p(1),
    \qquad
    \max_{1\le i\le n}\|\widetilde H_i\|_2=o_p(n^{1/2}).
  \]
  Applying the proof of Lemma~\ref{lem:lambda-expansion} with \(h_i\), \(P_nh\), and \(P_n(hh^\top)\) replaced by \(\widetilde H_i\), \(\Delta_{h,n}\), and \(P_n(\widetilde H\widetilde H^\top)\), respectively, proves existence of the solution, the stated expansion of \(\widehat\lambda_{\mathrm{ss}}\), and \(\max_i|n\widehat w_i^{\,\mathrm{ss}}-1|=o_p(1)\). For bounded \(g\), \(P_n(g\widetilde H^\top)=P_n(gH^\top)-(P_ng)\bar H_N^\top=P(gH^\top)+o_p(1)\). Substituting the formula for \(\widehat w_i^{\,\mathrm{ss}}\) and the multiplier expansion into \(P_{n,w}^{\mathrm{ss}}g\), as in the last paragraph of that proof, gives its stated expansion.
\end{proof}

\begin{lemma}
  \label{lem:ss-weighted-threshold-moment-expansion}
  Under the conditions of Theorem~3,
  \[
    \sqrt n\,P_{n,w}^{\mathrm{ss}}\psi_\alpha
    =
    \frac1{\sqrt n}\sum_{i=1}^n\psi_\alpha(Z_i)
    -c_{\alpha,A}^\top\Sigma_A^{-1}\sqrt n\,\Delta_{h,n}
    +o_p(1).
  \]
\end{lemma}

\begin{proof}
  Apply Lemma~\ref{lem:ss-weight-expansion} with \(g=\psi_\alpha\), using \(P\psi_\alpha=0\) and \(P(\psi_\alpha H^\top)=c_{\alpha,A}^\top\).
\end{proof}

\begin{lemma}
  \label{lem:ss-weighted-quantile-expansion}
  Under the conditions of Theorem~3,
  \[
    \sqrt n(\widehat q_\alpha^{\,\mathrm{ss}}-q_\alpha)
    =
    -\frac1{f(q_\alpha)}\left\{
    \frac1{\sqrt n}\sum_{i=1}^n\psi_\alpha(Z_i)
    -c_{\alpha,A}^\top\Sigma_A^{-1}\sqrt n\,\Delta_{h,n}
    \right\}
    +o_p(1).
  \]
\end{lemma}

\begin{proof}
  For \(t\in\R\), let \(d_{n,t}(Z)=\ind(S\le q_\alpha+t/\sqrt n)-\ind(S\le q_\alpha)\), and set \(A_n^{\mathrm{ss}}=n^{-1/2}\sum_{i=1}^n\psi_\alpha(Z_i)-c_{\alpha,A}^\top\Sigma_A^{-1}\sqrt n\,\Delta_{h,n}\). Since \(F(q_\alpha)=1-\alpha\) and the weights sum to one, \(P_{n,w}^{\mathrm{ss}}\psi_\alpha=\widehat F_{\mathrm{ss}}(q_\alpha)-F(q_\alpha)\). Lemma~\ref{lem:ss-weighted-threshold-moment-expansion} therefore gives
  \begin{equation}
    \label{eq:ss-weighted-fixed-threshold}
    \sqrt n\,P_{n,w}^{\mathrm{ss}}\psi_\alpha
    =\sqrt n\{\widehat F_{\mathrm{ss}}(q_\alpha)-F(q_\alpha)\}
    =A_n^{\mathrm{ss}}+o_p(1).
  \end{equation}

  Fix \(M<\infty\). The proof of Eq.~\eqref{eq:h-local-increment}, applied to \(H\), gives \(\sup_{|t|\le M}\|P_n(Hd_{n,t})\|_2=o_p(1)\). Since \(P_n(\widetilde H d_{n,t})=P_n(Hd_{n,t})-\bar H_NP_nd_{n,t}\), \(|P_nd_{n,t}|\le1\), and \(\bar H_N=o_p(1)\),
  \[
    \sup_{|t|\le M}\|P_n(\widetilde H d_{n,t})\|_2=o_p(1).
  \]
  Lemma~\ref{lem:ss-weight-expansion}, Eq.~\eqref{eq:weighted-d-expansion} and the subsequent bound on its remainder, applied with \(h\), \(\widehat\lambda\), and \(P_{n,w}\) replaced by \(\widetilde H\), \(\widehat\lambda_{\mathrm{ss}}\), and \(P_{n,w}^{\mathrm{ss}}\), together with Eq.~\eqref{eq:empirical-local-increment}, give
  \begin{equation}
    \label{eq:ss-weighted-local-increment}
    \sup_{|t|\le M}
    \left|\sqrt n(P_{n,w}^{\mathrm{ss}}-P)d_{n,t}\right|
    =o_p(1).
  \end{equation}

  Using \(P_{n,w}^{\mathrm{ss}}d_{n,t}=\widehat F_{\mathrm{ss}}(q_\alpha+t/\sqrt n)-\widehat F_{\mathrm{ss}}(q_\alpha)\) and \(Pd_{n,t}=F(q_\alpha+t/\sqrt n)-F(q_\alpha)\), Eqs.~\eqref{eq:ss-weighted-fixed-threshold}, \eqref{eq:ss-weighted-local-increment}, and~\eqref{eq:population-local-increment} give
  \[
    \sup_{|t|\le M}
    \left|
    \sqrt n\{\widehat F_{\mathrm{ss}}(q_\alpha+t/\sqrt n)-(1-\alpha)\}
    -A_n^{\mathrm{ss}}-f(q_\alpha)t
    \right|=o_p(1).
  \]
  The central limit theorem and \(\Delta_{h,n}=O_p(n^{-1/2})\) give \(A_n^{\mathrm{ss}}=O_p(1)\). Because the adjusted quantile level differs from \(1-\alpha\) by \((1-\alpha)/n\), repeating the argument after Eq.~\eqref{eq:weighted-local-cdf-expansion} based on monotonicity and the adjusted quantile definition gives \(\sqrt n(\widehat q_\alpha^{\,\mathrm{ss}}-q_\alpha)=O_p(1)\). Lemma~\ref{lem:ss-weight-expansion} gives \(\max_i\widehat w_i^{\,\mathrm{ss}}=\{1+o_p(1)\}/n\). Minimality and the no-tie condition give
  \[
    0\le
    \widehat F_{\mathrm{ss}}(\widehat q_\alpha^{\,\mathrm{ss}})
    -\frac{n+1}{n}(1-\alpha)
    \le
    \max_i\widehat w_i^{\,\mathrm{ss}}
    =\frac{1+o_p(1)}{n}.
  \]
  Evaluating the preceding uniform expansion at \(\sqrt n(\widehat q_\alpha^{\,\mathrm{ss}}-q_\alpha)\) therefore gives
  \[
    \sqrt n(\widehat q_\alpha^{\,\mathrm{ss}}-q_\alpha)
    =-\frac1{f(q_\alpha)}
    \left\{A_n^{\mathrm{ss}}-\frac{1-\alpha}{\sqrt n}\right\}
    +o_p(1)
    =-\frac{A_n^{\mathrm{ss}}}{f(q_\alpha)}+o_p(1).
  \]
\end{proof}

\begin{proof}
  Conditional on \(\mathcal D_n^{\mathrm{ss}}\), \(\P\{Y_{n+1}\in\widehat C_{\mathrm{ss}}(X_{n+1})\mid\mathcal D_n^{\mathrm{ss}}\}=F(\widehat q_\alpha^{\,\mathrm{ss}})\). Lemma~\ref{lem:ss-weighted-quantile-expansion} and differentiability of \(F\) at \(q_\alpha\) give
  \[
    \sqrt n\{F(\widehat q_\alpha^{\,\mathrm{ss}})-F(q_\alpha)\}
    =
    -\frac1{\sqrt n}\sum_{i=1}^n\psi_\alpha(Z_i)
    +c_{\alpha,A}^\top\Sigma_A^{-1}\sqrt n\,\Delta_{h,n}
    +o_p(1).
  \]
  Since \(F(q_\alpha)=1-\alpha\) and \(\Delta_{h,n}=\bar h_n-\bar h_N\), this proves the coverage expansion in Theorem~3.

  For the joint limit, set
  \[
    B_n=\frac1{\sqrt n}\sum_{i=1}^n\psi_\alpha(Z_i),
    \qquad
    D_n=\sqrt n\,\Delta_{h,n}=
    \frac{m}{N}\frac1{\sqrt n}\sum_{i=1}^nH_i
    -
    \frac{\sqrt n}{N}\sum_{j=1}^mH_j^{\mathrm u}.
  \]
  Independence of the calibration data and unlabeled data gives
  \[
    \Var\{B_n\}=\alpha(1-\alpha),\qquad
    \Cov\{B_n,D_n\}=\frac mN c_{\alpha,A}^\top,\qquad
    \Var\{D_n\}=\frac mN\Sigma_A.
  \]
  When \(\gamma<1\), the multivariate central limit theorem applies to \(n^{-1/2}\sum_{i=1}^n(\psi_\alpha(Z_i),H_i^\top)^\top\) and \(m^{-1/2}\sum_{j=1}^mH_j^{\mathrm u}\). When \(\gamma=1\), \(\Var\{D_n\}=m\Sigma_A/N\to0\), so \(D_n\xrightarrow{p}0\), while the central limit theorem applies to \(B_n\). Independence and Slutsky's theorem therefore give
  \[
    \begin{pmatrix}B_n\\D_n\end{pmatrix}
    \xrightarrow{d}
    N\left(
    0,
    \begin{pmatrix}
      \alpha(1-\alpha) & (1-\gamma)c_{\alpha,A}^\top\\
      (1-\gamma)c_{\alpha,A} & (1-\gamma)\Sigma_A
    \end{pmatrix}
    \right).
  \]
  Applying the linear map \((b,d)\mapsto-b+c_{\alpha,A}^\top\Sigma_A^{-1}d\) to this limit gives a centred normal limit with variance \(\alpha(1-\alpha)-(1-\gamma)c_{\alpha,A}^\top\Sigma_A^{-1}c_{\alpha,A}\). Together with the coverage expansion just established, this proves the limiting distribution in Theorem~3.
\end{proof}

\subsection{Semi-supervised average prediction-set size (Corollary~3)}

\begin{proof}
  Lemma~\ref{lem:ss-weighted-quantile-expansion} gives \(\widehat q_\alpha^{\,\mathrm{ss}}-q_\alpha=O_p(n^{-1/2})\). Differentiability of \(\mathcal L\) at \(q_\alpha\) yields
  \[
    \sqrt n\{\mathcal L(\widehat q_\alpha^{\,\mathrm{ss}})-\mathcal L(q_\alpha)\}
    =
    \mathcal L'(q_\alpha)\sqrt n(\widehat q_\alpha^{\,\mathrm{ss}}-q_\alpha)+o_p(1).
  \]
  Substituting the expansion in Lemma~\ref{lem:ss-weighted-quantile-expansion} into the preceding display gives the root-\(n\) expansion. The joint limit in the proof of Theorem~3 then gives Corollary~3.
\end{proof}

\subsection{Nested linear spans and optimality (Proposition~1)}

\begin{proof}
  By the defining property of conditional expectation, \(\psi_\alpha(Z)-h^*_\alpha(A)\) is orthogonal to every square-integrable function of \(A\). For \(j=1,2\), let \(\Pi_j\) denote the orthogonal projection in \(L_2(P)\) onto \(\mathcal V_j\). Hence \(\Pi_j\psi_\alpha=\Pi_jh^*_\alpha\), and Remark~2, applied with \(h_j(A)-\E\{h_j(A)\}\) in place of \(h(Z)\), identifies \(c_{\alpha,A,j}^\top\Sigma_{A,j}^{-1}c_{\alpha,A,j}\) with \(\|\Pi_jh^*_\alpha\|_{P,2}^2\). Since \(\E\{h^*_\alpha(A)\}=\E\{\psi_\alpha(Z)\}=0\), \(\|h^*_\alpha\|_{P,2}^2=\Var\{h^*_\alpha(A)\}\). Since \(\mathcal V_1\subseteq\mathcal V_2\), the projection theorem yields
  \[
    \begin{aligned}
      c_{\alpha,A,1}^\top\Sigma_{A,1}^{-1}c_{\alpha,A,1}
      &=\|\Pi_1h^*_\alpha\|_{P,2}^2
      \le \|\Pi_2h^*_\alpha\|_{P,2}^2\\
      &=c_{\alpha,A,2}^\top\Sigma_{A,2}^{-1}c_{\alpha,A,2}
      \le \|h^*_\alpha\|_{P,2}^2
      =\Var\{h^*_\alpha(A)\}.
    \end{aligned}
  \]
  Each \(\mathcal V_j\) is finite dimensional and hence closed, so \(\|\Pi_jh^*_\alpha\|_{P,2}=\|h^*_\alpha\|_{P,2}\) holds exactly when \(h^*_\alpha(A)\in\mathcal V_j\).
\end{proof}

\subsection{Asymptotic distributions with a learned moment function (Theorem~4)}

\begin{lemma}
  \label{lem:ss-crossfit-replacement}
  Under the conditions of Theorem~4, let \(H^\dagger=h^\dagger(A)-\E\{h^\dagger(A)\}\), \(\Sigma_\dagger=\E\{(H^\dagger)^2\}\), and \(c_{\alpha,\dagger}=\E\{\psi_\alpha(Z)H^\dagger\}\). Let \(\bar h_n^\dagger=n^{-1}\sum_{i=1}^n h^\dagger(A_i)\) and \(\bar h_N^\dagger=N^{-1}\{\sum_{i=1}^n h^\dagger(A_i)+\sum_{j=1}^m h^\dagger(A_j^{\mathrm u})\}\). Then
  \[
    \sqrt n\,P_n\widehat H
    =
    \sqrt n(\bar h_n^\dagger-\bar h_N^\dagger)+o_p(1),
  \]
  \[
    P_n(\widehat H^2)=\Sigma_\dagger+o_p(1),
    \qquad
    P_n(\psi_\alpha\widehat H)=c_{\alpha,\dagger}+o_p(1),
    \qquad
    \max_{1\le i\le n}|\widehat H_i|=o_p(n^{1/2}).
  \]
\end{lemma}

\begin{proof}
  Fix the data-independent fold assignments and the randomization used to fit the classifiers and write \(n_k=|\mathcal I_k|\), \(m_k=|\mathcal J_k|\), and \(N_k=n_k+m_k\). For \(i\in\mathcal I_k\), define
  \[
    \widetilde H_i^\dagger
    =
    h^\dagger(A_i)
    -
    \frac1{N_k}
    \left\{
    \sum_{\ell\in\mathcal I_k}h^\dagger(A_\ell)
    +
    \sum_{\ell\in\mathcal J_k}h^\dagger(A_\ell^{\mathrm u})
    \right\}.
  \]
  Set \(R_i=\widehat H_i-\widetilde H_i^\dagger\) and \(\delta_k=\widehat h^{(-k)}-h^\dagger\). With \(P\delta_k=\E\{\delta_k(A)\mid\widehat h^{(-k)}\}\), substituting the definitions of \(\widehat H_i\) and \(\widetilde H_i^\dagger\) into \(R_i\) and centring each sum at \(P\delta_k\) gives
  \[
    \frac1{\sqrt n}\sum_{i\in\mathcal I_k}R_i
    =
    \frac1{\sqrt n}\left[
    \frac{m_k}{N_k}\sum_{i\in\mathcal I_k}\{\delta_k(A_i)-P\delta_k\}
    -
    \frac{n_k}{N_k}\sum_{j\in\mathcal J_k}\{\delta_k(A_j^{\mathrm u})-P\delta_k\}
    \right].
  \]
  Conditional on \(\widehat h^{(-k)}\), the calibration data points indexed by \(\mathcal I_k\) and the unlabeled observations indexed by \(\mathcal J_k\) are independent, which gives
  \[
    \E\left\{
    \left|\frac1{\sqrt n}\sum_{i\in\mathcal I_k}R_i\right|^2
    \mathrel{\Big|}\widehat h^{(-k)}
    \right\}
    \le
    \frac{n_km_k}{nN_k}
    \E\{|\delta_k(A)|^2\mid\widehat h^{(-k)}\}.
  \]
  Jensen's inequality likewise gives
  \[
    \E\left\{
    \frac1n\sum_{i\in\mathcal I_k}|R_i|^2
    \mathrel{\Big|}\widehat h^{(-k)}
    \right\}
    \le
    \frac{4n_k}{n}
    \E\{|\delta_k(A)|^2\mid\widehat h^{(-k)}\}.
  \]
  Since \(K\) is fixed, the conditional mean-square convergence assumption in Theorem~4 and conditional Markov's inequality give the first two bounds below. The inequality \(\max_i|R_i|/\sqrt n\le[P_n\{|R|^2\}]^{1/2}\) gives the third:
  \[
    \sqrt n\,P_nR=o_p(1),\qquad
    P_n\{|R|^2\}=o_p(1),\qquad
    \max_{1\le i\le n}|R_i|=o_p(n^{1/2}).
  \]

  Write \(T_{\mathcal I_k}^\dagger=\sum_{i\in\mathcal I_k}[h^\dagger(A_i)-\E\{h^\dagger(A)\}]\) and \(T_{\mathcal J_k}^{\mathrm u,\dagger}=\sum_{j\in\mathcal J_k}[h^\dagger(A_j^{\mathrm u})-\E\{h^\dagger(A)\}]\), where an empty sum is zero. Direct calculation gives
  \[
    \begin{aligned}
      P_n\widetilde H^\dagger
      &=\sum_{k=1}^K\left\{
      \frac{m_k}{nN_k}T_{\mathcal I_k}^\dagger
      -\frac{n_k}{nN_k}T_{\mathcal J_k}^{\mathrm u,\dagger}
      \right\},\\
      \bar h_n^\dagger-\bar h_N^\dagger
      &=\sum_{k=1}^K\left\{
      \frac{m}{nN}T_{\mathcal I_k}^\dagger
      -\frac1N T_{\mathcal J_k}^{\mathrm u,\dagger}
      \right\}.
    \end{aligned}
  \]
  If \(n_k=n/K\) and \(m_k=m/K\) for every \(k\), the two right-hand sides are identical. For the nearly equal-sized folds specified in the construction, \(n_k=n/K+O(1)\) and \(m_k=m/K+O(1)\), and each coefficient in the first equality differs from its counterpart in the second by \(O((nN)^{-1})\). Since \(K\) is fixed, \(T_{\mathcal I_k}^\dagger=O_p(n^{1/2})\) and \(T_{\mathcal J_k}^{\mathrm u,\dagger}=O_p((m+1)^{1/2})\), so
  \[
    \sqrt n\,P_n\widetilde H^\dagger
    =
    \sqrt n(\bar h_n^\dagger-\bar h_N^\dagger)+o_p(1).
  \]

  The bounds for \(T_{\mathcal I_k}^\dagger\) and \(T_{\mathcal J_k}^{\mathrm u,\dagger}\) in the preceding paragraph imply \(N_k^{-1}(T_{\mathcal I_k}^\dagger+T_{\mathcal J_k}^{\mathrm u,\dagger})=o_p(1)\) for every \(k\). Using the definition of \(\widetilde H_i^\dagger\), the law of large numbers then gives the second-moment and cross-moment limits below. Applying the tail argument from the proof of Lemma~\ref{lem:lambda-expansion} to \(H^\dagger\) and using the preceding bound gives the third relation:
  \[
    P_n((\widetilde H^\dagger)^2)
    =\Sigma_\dagger+o_p(1),
    \qquad
    P_n(\psi_\alpha\widetilde H^\dagger)
    =c_{\alpha,\dagger}+o_p(1),
    \qquad
    \max_{1\le i\le n}|\widetilde H_i^\dagger|=o_p(n^{1/2}).
  \]

  Finally, \(\widehat H_i=\widetilde H_i^\dagger+R_i\). The bound \(\sqrt n\,P_nR=o_p(1)\) proves the asserted mean expansion, and the triangle inequality gives the asserted bound on \(\max_i|\widehat H_i|\). Cauchy--Schwarz and \(P_n\{|R|^2\}=o_p(1)\) give \(P_n(\widetilde H^\dagger R)=o_p(1)\), \(P_n(R^2)=o_p(1)\), and \(P_n(\psi_\alpha R)=o_p(1)\). Expanding \(P_n(\widehat H^2)\) and \(P_n(\psi_\alpha\widehat H)\) proves the remaining two limits.
\end{proof}

\begin{proof}
  Set \(d_{n,t}=\ind(S\le q_\alpha+t/\sqrt n)-\ind(S\le q_\alpha)\), and use the decomposition \(\widehat H_i=\widetilde H_i^\dagger+R_i\) from the proof of Lemma~\ref{lem:ss-crossfit-replacement}. For fixed \(M<\infty\), applying the argument for Eq.~\eqref{eq:h-local-increment} foldwise to \(H^\dagger\) and using that its pooled average over \(\mathcal I_k\) and \(\mathcal J_k\) is \(o_p(1)\) for every \(k\) give \(\sup_{|t|\le M}|P_n(\widetilde H^\dagger d_{n,t})|=o_p(1)\). Since \(P_n\{|R|^2\}=o_p(1)\), Cauchy--Schwarz yields
  \[
    \sup_{|t|\le M}|P_n(\widehat H d_{n,t})|\le\sup_{|t|\le M}|P_n(\widetilde H^\dagger d_{n,t})|+[P_n\{|R|^2\}]^{1/2}=o_p(1).
  \]
  The mean, second-moment, and maximum bounds in Lemma~\ref{lem:ss-crossfit-replacement} verify the conditions used in the proof of Lemma~\ref{lem:ss-weight-expansion}, with \(\widehat H\) and \(P_n\widehat H\) in place of \(\widetilde H\) and \(\Delta_{h,n}\). The cross-moment convergence in Lemma~\ref{lem:ss-crossfit-replacement} then gives the corresponding expansion for the weighted empirical mean of \(\psi_\alpha\), where the weights are obtained from \(\widehat H_i\), with \(\Sigma_\dagger\) and \(c_{\alpha,\dagger}\) in place of \(\Sigma_A\) and \(c_{\alpha,A}\). The uniform bound in the preceding display supplies the local empirical-process condition used in the proof of Lemma~\ref{lem:ss-weighted-quantile-expansion}. Substituting \(\sqrt n\,P_n\widehat H=\sqrt n(\bar h_n^\dagger-\bar h_N^\dagger)+o_p(1)\) into that quantile expansion, then applying the Taylor expansion and joint central limit theorem from the proof of Theorem~3, gives the stated expansion and limiting distribution for training-conditional coverage. When \(\mathcal L\) is differentiable at \(q_\alpha\), the proof of Corollary~3 gives the stated limiting distribution for average prediction-set size. If \(h^\dagger=h^*_\alpha\), then \(\E\{h^*_\alpha(A)\}=0\) and iterated expectation gives \(\E\{\psi_\alpha(Z)h^*_\alpha(A)\}=\E[\{h^*_\alpha(A)\}^2]\). Thus \(c_{\alpha,\dagger}=\Sigma_\dagger=\Var\{h^*_\alpha(A)\}\), and hence \(c_{\alpha,\dagger}^2/\Sigma_\dagger=\Var\{h^*_\alpha(A)\}\), which gives the stated variance.
\end{proof}

\subsection{Finite-sample bounds for coverage moments (Theorem~5)}

For \(f\in\mathcal F_K^+\), \(\beta\in\R^d\), and \(\diamond\in\{<,\le\}\), define \(g^\diamond_{f,\beta}(Z)=f(X)\{\ind(S\mathbin{\diamond}\Phi(X)^\top\beta)-(1-\alpha)\}\). Let \(\mathcal G_K=\{g^\diamond_{f,\beta}:f\in\mathcal F_K^+,\ \beta\in\R^d,\ \diamond\in\{<,\le\}\}\).

\begin{lemma}
  \label{lem:conditional-uniform-deviation}
  There is a universal constant \(C\) such that, for every \(K\ge1\) and \(\eta\in(0,1)\),
  \[
    \P\!\left\{
    \sup_{g\in\mathcal G_K}|(P_n-P)g|
    \le
    CK\sqrt{\frac{d+\log(1/\eta)}{n}}
    \right\}
    \ge1-\eta.
  \]
\end{lemma}

\begin{proof}
  For \(f(x)=a^\top\Phi(x)\), the subgraph of \(g^\le_{f,\beta}\) is
  \[
    \begin{aligned}
      \{(z,u):u<g^\le_{f,\beta}(z)\}
      &=
      [\{S\le\Phi(X)^\top\beta\}
      \cap\{u<\alpha a^\top\Phi(X)\}]\\
      &\quad\cup
      [\{S>\Phi(X)^\top\beta\}
      \cap\{u<-(1-\alpha)a^\top\Phi(X)\}].
    \end{aligned}
  \]
  For \(g^<_{f,\beta}\), replace the two score events in this display by \(\{S<\Phi(X)^\top\beta\}\) and \(\{S\ge\Phi(X)^\top\beta\}\). The resulting subgraphs for \(g^\le_{f,\beta}\) and \(g^<_{f,\beta}\) are fixed Boolean combinations of halfspaces in \((\Phi(X),S,u)\). Problem~2.6.14 and Problem~2.6.23 of \citet{VanDerVaartWellner2023WCEP} therefore show that \(\mathcal G_K\) is VC-subgraph with dimension \(V\le c_0d\) for a universal constant \(c_0\). Since \(0\le f\le K\), the class has envelope \(K\). Theorem~2.6.7 of \citet{VanDerVaartWellner2023WCEP} gives, for every discrete probability measure \(Q\) and \(0<\epsilon<1\),
  \[
    N(\epsilon K,\mathcal G_K,L_2(Q))
    \le C_0V(16e)^V\epsilon^{-2V}.
  \]
  Hence its uniform entropy integral satisfies
  \[
    \begin{aligned}
      \sup_Q\int_0^1
      \sqrt{1+\log N(\epsilon K,\mathcal G_K,L_2(Q))}\,d\epsilon
      &\le
      C_1\sqrt V\int_0^1
      \sqrt{1+\log(1/\epsilon)}\,d\epsilon\\
      &\le C_2\sqrt d.
    \end{aligned}
  \]
  Theorem~2.14.1 of \citet{VanDerVaartWellner2023WCEP} now gives
  \[
    \E\!\left\{
    \sup_{g\in\mathcal G_K}
    |\sqrt n(P_n-P)g|
    \right\}
    \le CK\sqrt d.
  \]

  Let \(T=\sup_{g\in\mathcal G_K}|(P_n-P)g|\). Every \(g\in\mathcal G_K\) takes values in \([-(1-\alpha)K,\alpha K]\), so replacing one observation changes \(T\) by at most \(K/n\). The bounded-difference inequality \citep[e.g., Proposition~2.15.3 of][]{VanDerVaartWellner2023WCEP} therefore gives
  \[
    \P[T>\E\{T\}+t]
    \le
    \exp\!\left(-\frac{2nt^2}{K^2}\right).
  \]
  Taking \(t=K\{\log(1/\eta)/(2n)\}^{1/2}\), using \(\E\{T\}\le CK\sqrt{d/n}\), and enlarging \(C\) proves the lemma.
\end{proof}

\begin{proof}
  Set \(r_i=S_i-\Phi(X_i)^\top\widehat\beta\). Fix \(f(x)=a^\top\Phi(x)\in\mathcal F_K^+\) and set \(L_f(t)=\sum_i\widehat w_i\rho_{1-\alpha}(r_i-tf(X_i))\). Along the path \(\widehat\beta+ta\), \(L_f(t)\) is the objective in Eq.~(3), so convexity and optimality of \(\widehat\beta\) give \(0\in\partial L_f(0)\). As in the one-dimensional subgradient calculation in the proof of Theorem~3 of \citet[Section~D, Supplementary Material]{GibbsCherianCandes2025}, the subdifferential of the \(i\)th summand at zero is
  \[
    \partial\!\left[
    t\mapsto
    \widehat w_i\rho_{1-\alpha}(r_i-tf(X_i))
    \right](0)
    =
    \begin{cases}
      \{\widehat w_i\alpha f(X_i)\}, & r_i<0,\\
      \widehat w_i f(X_i)[\alpha-1,\alpha], & r_i=0,\\
      \{-\widehat w_i(1-\alpha)f(X_i)\}, & r_i>0.
    \end{cases}
  \]
  Because \(\widehat w_i f(X_i)\ge0\), the sums of the lower and upper endpoints are the left and right derivatives of \(L_f\), respectively. Thus \(0\in\partial L_f(0)\) gives
  \[
    L'_{f,-}(0)
    =
    P_{n,w}g^<_{f,\widehat\beta}
    \le0
    \le
    P_{n,w}g^\le_{f,\widehat\beta}
    =
    L'_{f,+}(0).
  \]

  For every realized value of \(\widehat\beta\), \(g^\diamond_{f,\widehat\beta}\in\mathcal G_K\). To compare \(P_{n,w}g^\diamond_{f,\widehat\beta}\) with \(Pg^\diamond_{f,\widehat\beta}\), let \(\Delta_{n,K}=\sup_{g\in\mathcal G_K}|(P_n-P)g|\) and \(D_{\widehat w}=\sum_i|\widehat w_i-1/n|\). Since \(|g^\diamond_{f,\beta}|\le K\),
  \[
    |(P_{n,w}-P_n)g^\diamond_{f,\beta}|
    =
    \left|\sum_{i=1}^n
    \left(\widehat w_i-\frac1n\right)
    g^\diamond_{f,\beta}(Z_i)\right|
    \le KD_{\widehat w}.
  \]
  Moreover, \(Pg^\diamond_{f,\widehat\beta}=P_{n,w}g^\diamond_{f,\widehat\beta}+(P-P_n)g^\diamond_{f,\widehat\beta}+(P_n-P_{n,w})g^\diamond_{f,\widehat\beta}\). Combining this identity with \(P_{n,w}g^\le_{f,\widehat\beta}\ge0\), \(P_{n,w}g^<_{f,\widehat\beta}\le0\), the definition of \(\Delta_{n,K}\), and \(|(P_{n,w}-P_n)g^\diamond_{f,\beta}|\le KD_{\widehat w}\) gives, simultaneously for every \(f\in\mathcal F_K^+\),
  \begin{equation}
    \label{eq:conditional-finite-reduction}
    \begin{aligned}
      Pg^\le_{f,\widehat\beta}
      &\ge-\Delta_{n,K}-KD_{\widehat w},\\
      Pg^<_{f,\widehat\beta}
      &\le \Delta_{n,K}+KD_{\widehat w}.
    \end{aligned}
  \end{equation}

  Conditional on \(Z_1,\ldots,Z_n\), \(\widehat\beta\) is fixed and \(Z_{n+1}\) is an independent draw from \(P\). The definition of \(\widehat C_{\rm cond}\) therefore gives
  \begin{equation}
    \label{eq:conditional-test-moment}
    \begin{aligned}
      Pg^\le_{f,\widehat\beta}
      &=
      \E\!\left[
      f(X_{n+1})
      \{\ind(Y_{n+1}\in\widehat C_{\rm cond}(X_{n+1}))-(1-\alpha)\}
      \,\middle|\,Z_1,\ldots,Z_n
      \right].
    \end{aligned}
  \end{equation}

  Lemma~\ref{lem:conditional-uniform-deviation} with \(\eta=\delta/3\) gives \(\Delta_{n,K}\le CK\{d+\log(3/\delta)\}^{1/2}/\sqrt n\) with probability at least \(1-\delta/3\). Lemma~\ref{lem:finite-moment-concentration} and the assumption \(\varepsilon_{\Sigma,n}(\delta)\le\kappa/2\) give \(\|P_nh\|_2\le\varepsilon_{\mu,n}(\delta)\) and \(\lambda_{\min}(P_n(hh^\top))\ge\kappa-\varepsilon_{\Sigma,n}(\delta)\ge\kappa/2\) with probability at least \(1-2\delta/3\).

  On this event, take \(\eta=\varepsilon_{\mu,n}(\delta)\) and \(\underline\kappa=\kappa/2\). Then \(4B\eta/\underline\kappa=8B\varepsilon_{\mu,n}(\delta)/\kappa\le1/2\), so Lemma~\ref{lem:finite-weight-bound} gives \(D_{\widehat w}\le8B\eta/\underline\kappa=16B\varepsilon_{\mu,n}(\delta)/\kappa\). A union bound makes the bounds for \(\Delta_{n,K}\) and \(D_{\widehat w}\) simultaneous with probability at least \(1-\delta\). The first inequality in Eq.~\eqref{eq:conditional-finite-reduction} then gives, simultaneously for every \(f\in\mathcal F_K^+\),
  \[
    Pg^\le_{f,\widehat\beta}
    \ge
    -CK\sqrt{\frac{d+\log(3/\delta)}{n}}
    -\frac{16KB}{\kappa}\varepsilon_{\mu,n}(\delta)
    =
    -\varepsilon^{\mathrm{cond}}_{n,K}(\delta).
  \]
  Eq.~\eqref{eq:conditional-test-moment} now gives the stated lower bound conditional on the calibration data.

  Suppose now that the conditional CDF of \(S\) given \(X\) is continuous almost surely. Conditional on the calibration data, \(\widehat\beta\) is fixed and the test point is independent, so
  \[
    \begin{aligned}
      &\P\{S=\Phi(X)^\top\widehat\beta\mid Z_1,\ldots,Z_n\}\\
      &\quad=
      \E[
      \P\{S=\Phi(X)^\top\widehat\beta
      \mid X,Z_1,\ldots,Z_n\}
      \mid Z_1,\ldots,Z_n]
      =0.
    \end{aligned}
  \]
  Hence \(Pg^\le_{f,\widehat\beta}=Pg^<_{f,\widehat\beta}\), and the two inequalities in Eq.~\eqref{eq:conditional-finite-reduction} give \(\sup_{f\in\mathcal F_K^+}|Pg^\le_{f,\widehat\beta}|\le\varepsilon^{\mathrm{cond}}_{n,K}(\delta)\). Together with Eq.~\eqref{eq:conditional-test-moment}, this proves the two-sided bound conditional on the calibration data.

  For the marginal bounds, the proof of Lemma~\ref{lem:conditional-uniform-deviation} gives \(\E\{\Delta_{n,K}\}\le CK\sqrt{d/n}\), while the calculation in the proof of Corollary~1 gives \(\E\{D_{\widehat w}\}\le16B^2/(\kappa\sqrt n)+2p_{\mu,\Sigma,n}\). For every realization of the calibration data, the subgradient inequalities derived from \(0\in\partial L_f(0)\) hold for the resulting weights, including the uniform weights used when the optimization in Eq.~(1) has no solution. Hence Eq.~\eqref{eq:conditional-finite-reduction} holds for every such realization. Taking expectations therefore gives
  \[
    \begin{aligned}
      \inf_{f\in\mathcal F_K^+}\E\{Pg^\le_{f,\widehat\beta}\}
      &\ge-\E\{\Delta_{n,K}\}-K\E\{D_{\widehat w}\}
      \ge-\bar\varepsilon^{\mathrm{cond}}_{n,K},\\
      \sup_{f\in\mathcal F_K^+}
      \left|\E\{Pg^\le_{f,\widehat\beta}\}\right|
      &\le\E\{\Delta_{n,K}\}+K\E\{D_{\widehat w}\}
      \le\bar\varepsilon^{\mathrm{cond}}_{n,K}
      \quad\text{under the stated continuity condition}.
    \end{aligned}
  \]
  Eq.~\eqref{eq:conditional-test-moment} and the tower property yield the two marginal bounds in Theorem~5.
\end{proof}

\subsection{Asymptotic distributions of threshold coefficients and coverage moments (Theorem~6)}

\begin{lemma}
  \label{lem:conditional-weighted-moment-expansion}
  Under the conditions of Theorem~6,
  \[
    \sqrt n\,P_{n,w}\Psi_\alpha
    =
    \frac1{\sqrt n}\sum_{i=1}^n
    \{\Psi_\alpha(Z_i)-C_\alpha\Sigma_h^{-1}h(Z_i)\}
    +o_p(1).
  \]
\end{lemma}

\begin{proof}
  Apply Lemma~\ref{lem:lambda-expansion} with \(g=\Psi_\alpha\), using \(\|\Psi_\alpha(Z)\|_2\le\|\Phi(X)\|_2\), \(P\Psi_\alpha=M(\beta_\alpha)=0\), and \(P(\Psi_\alpha h^\top)=C_\alpha\).
\end{proof}

\begin{proof}
  Let \(U_i=S_i-\Phi(X_i)^\top\beta_\alpha\) and, for \(t\in\R^d\), define
  \[
    Q_n(t)=n\sum_{i=1}^n\widehat w_i
    \left\{
    \rho_{1-\alpha}(U_i-\Phi(X_i)^\top t/\sqrt n)
    -\rho_{1-\alpha}(U_i)
    \right\}.
  \]
  Knight's identity \citep{Knight1998} gives
  \begin{equation}
    \label{eq:conditional-knight-identity}
    \rho_{1-\alpha}(U_i-\Phi(X_i)^\top t/\sqrt n)-\rho_{1-\alpha}(U_i)
    =\frac{\Phi(X_i)^\top t}{\sqrt n}\{\ind(U_i\le0)-(1-\alpha)\}+R_{ni}(t).
  \end{equation}
  Here \(R_{ni}(t)=\int_0^{\Phi(X_i)^\top t/\sqrt n}\{\ind(U_i\le u)-\ind(U_i\le0)\}\,du\ge0\). Since \(\Phi(X_i)\{\ind(U_i\le0)-(1-\alpha)\}=\Psi_\alpha(Z_i)\), summing gives
  \begin{equation}
    \label{eq:conditional-local-objective}
    Q_n(t)
    =t^\top\sqrt n\,P_{n,w}\Psi_\alpha
    +n\sum_{i=1}^n\widehat w_iR_{ni}(t).
  \end{equation}

  We first derive an expansion of \(n\sum_i\widehat w_iR_{ni}(t)\) that is uniform over \(\|t\|_2\le T\) for every fixed \(T<\infty\). Absolute continuity of the check loss, Fubini's theorem, and the local expansion of \(M\) give, for \(v\to0\),
  \[
    \begin{aligned}
      &P\{\rho_{1-\alpha}(S-\Phi(X)^\top(\beta_\alpha+v))
      -\rho_{1-\alpha}(S-\Phi(X)^\top\beta_\alpha)\}\\
      &\qquad=\int_0^1v^\top M(\beta_\alpha+sv)\,ds
      =\frac12v^\top J_\alpha v+o(\|v\|_2^2).
    \end{aligned}
  \]
  At \(v=t/\sqrt n\), the expected sum of the first term on the right-hand side of Eq.~\eqref{eq:conditional-knight-identity} is \(\sqrt n\,t^\top M(\beta_\alpha)=0\). Taking expectations after summing Knight's identity and using the preceding expansion of the expected check loss therefore gives \(\E\{\sum_iR_{ni}(t)\}=t^\top J_\alpha t/2+o(1)\). The integral definition gives \(R_{ni}(t)=0\) unless \(\sqrt n|U_i|\le|\Phi(X_i)^\top t|\), and \(R_{ni}(t)\le n^{-1/2}|\Phi(X_i)^\top t|\). Independence therefore bounds \(\Var\{\sum_iR_{ni}(t)\}\) by the expectation of \((\Phi(X)^\top t)^2\) on that event. This expectation tends to zero by dominated convergence, using \(\P\{S=\Phi(X)^\top\beta_\alpha\}=0\) and \(P\{\|\Phi(X)\|_2^2\}<\infty\). Hence
  \[
    \sum_iR_{ni}(t)=\frac12t^\top J_\alpha t+o_p(1)=O_p(1).
  \]
  Lemma~\ref{lem:lambda-expansion} gives \(\max_i|n\widehat w_i-1|=o_p(1)\). Since \(R_{ni}(t)\ge0\),
  \[
    \left|
    n\sum_{i=1}^n\widehat w_iR_{ni}(t)
    -\sum_{i=1}^nR_{ni}(t)
    \right|
    \le
    \max_i|n\widehat w_i-1|
    \sum_{i=1}^nR_{ni}(t)
    =o_p(1).
  \]
  Therefore, for each fixed \(t\),
  \[
    n\sum_i\widehat w_iR_{ni}(t)=\frac12t^\top J_\alpha t+o_p(1).
  \]
  To obtain the uniform expansion, Eq.~\eqref{eq:conditional-knight-identity} shows that \(t\mapsto R_{ni}(t)\), and hence \(t\mapsto n\sum_i\widehat w_iR_{ni}(t)\), is finite and convex. The pointwise limit \(t\mapsto t^\top J_\alpha t/2\) is continuous, so the convexity lemma of \citet{Pollard1991LAD} gives, for every fixed \(T<\infty\),
  \begin{equation}
    \label{eq:conditional-weighted-remainder-uniform}
    \sup_{\|t\|_2\le T}
    \left|
    n\sum_i\widehat w_iR_{ni}(t)
    -\frac12t^\top J_\alpha t
    \right|=o_p(1).
  \end{equation}

  Write \(A_n=n^{-1/2}\sum_{i=1}^n\{\Psi_\alpha(Z_i)-C_\alpha\Sigma_h^{-1}h(Z_i)\}\). Lemma~\ref{lem:conditional-weighted-moment-expansion} gives \(\sqrt n\,P_{n,w}\Psi_\alpha=A_n+o_p(1)\). For every fixed \(T<\infty\), combining this relation with Eqs.~\eqref{eq:conditional-local-objective} and~\eqref{eq:conditional-weighted-remainder-uniform} gives
  \begin{equation}
    \label{eq:conditional-objective-quadratic}
    \sup_{\|t\|_2\le T}
    \left|
    Q_n(t)-t^\top A_n
    -\frac12t^\top J_\alpha t
    \right|=o_p(1).
  \end{equation}

  For any minimizer \(\widehat\beta\) in Eq.~(3), the vector \(t_n=\sqrt n(\widehat\beta-\beta_\alpha)\) minimizes \(Q_n\). Since \(A_n=O_p(1)\) and \(J_\alpha\) is positive definite, Eq.~\eqref{eq:conditional-objective-quadratic} gives
  \[
    \inf_{\|t\|_2=T}Q_n(t)
    \ge
    -T\|A_n\|_2
    +\frac12\lambda_{\min}(J_\alpha)T^2
    +o_p(1).
  \]
  Given \(\varepsilon>0\), choose a fixed \(T\) large enough that \(\inf_{\|t\|_2=T}Q_n(t)>0\) with probability at least \(1-\varepsilon\) for all sufficiently large \(n\). If a minimizer satisfied \(\|t_n\|_2>T\), then \(s=T/\|t_n\|_2\in(0,1)\) and convexity would give \(Q_n(st_n)\le(1-s)Q_n(0)+sQ_n(t_n)\le0\), contradicting \(\inf_{\|t\|_2=T}Q_n(t)>0\). Thus, on the same event, \(t_n\) lies in this ball for every minimizer and is therefore \(O_p(1)\).

  Since \(J_\alpha^{-1}A_n=O_p(1)\), enlarging \(T\) if necessary ensures that \(\|J_\alpha^{-1}A_n\|_2\le T\) and \(\|t_n\|_2\le T\) for every minimizer with probability arbitrarily close to one. When both points lie in this ball, Eq.~\eqref{eq:conditional-objective-quadratic}, optimality of \(t_n\), and completing the square give, simultaneously for every minimizer,
  \[
    \frac{\lambda_{\min}(J_\alpha)}2
    \|t_n+J_\alpha^{-1}A_n\|_2^2
    \le
    2\sup_{\|t\|_2\le T}
    \left|
    Q_n(t)-t^\top A_n
    -\frac12t^\top J_\alpha t
    \right|
    =o_p(1).
  \]
  Hence \(t_n=-J_\alpha^{-1}A_n+o_p(1)\), which, by the definitions of \(t_n\) and \(A_n\), proves the coefficient expansion in Theorem~6.

  Because \(t_n=O_p(1)\), the assumed expansion of \(M\) at \(\widehat\beta=\beta_\alpha+t_n/\sqrt n\) gives \(\sqrt n\,M(\widehat\beta)=J_\alpha t_n+o_p(1)=-A_n+o_p(1)\), proving the coverage-moment expansion in Theorem~6. The multivariate central limit theorem applied to \(A_n\), followed by Slutsky's theorem, gives the asserted limits for \(\sqrt n(\widehat\beta-\beta_\alpha)\) and \(\sqrt n\,M(\widehat\beta)\).
\end{proof}

\FloatBarrier

\section{Simulation Designs and Additional Results}
\label{sec:supp-sim}

This section gives implementation details and additional numerical results for the simulation experiments in Section~5 of the main paper. SC denotes standard split conformal prediction. Unless available analytically, the asymptotic variance and covariance benchmarks are computed using independent samples of size 50,000 under one fixed training fit per DGP--$(n,p)$ setting, with 5000 population replications. Marginal calibration uses the adjusted quantile level $(n+1)(1-\alpha)/n$ in Sections~2 and~3 of the main paper, whereas conditional calibration uses the nominal check-loss level $1-\alpha$ in Section~4. The finite-sample experiments use 500 Monte Carlo replications, except for the marginal-coverage experiment in Section~\ref{sec:supp-marginal-coverage}, which uses 2000 independent replications with one fixed training set and independently regenerated calibration, external-reference, and test data at each common sample size.

\paragraph{Prediction and residual scale estimation.}
We estimate the conditional mean and conditional mean absolute residual using cubic regression splines \citep{LeiGsellRinaldoTibshiraniWasserman2018}.  On the single fixed training data set for each DGP--$(n,p)$ setting, the conditional mean is fitted once using the additive model $\widehat\mu(x)=\widehat a_\mu+\sum_{j=1}^p\widehat f_j(x_j)$. Each component uses a seven-dimensional cubic regression-spline basis. Its quadratic roughness penalty is normalized by the mean of its positive eigenvalues, and its smoothing parameter is selected from $\{10^{-4},\ldots,10^4\}$ on a nine-point logarithmic grid by five-fold cross-validation of the corresponding marginal spline. The resulting component-specific penalties are used in the final additive fit.

To estimate the residual scale without using in-sample residuals, we partition the fixed training observations into five fixed folds.  For an observation $i$ in fold $k(i)$, let $\widehat\mu^{(-k(i))}$ denote the additive cubic regression-spline fit obtained without that fold, and define the cross-fitted absolute residual $\widetilde R_i=\left|Y_i-\widehat\mu^{(-k(i))}(X_i)\right|$. Let $W_i=\left\{\widehat\mu^{(-k(i))}(X_i),\,X_{i1},\,X_{i2},\,X_{i1}X_{i2}\right\}^{\top}$. We regress $\widetilde R_i$ on the four coordinates of $W_i$ using the same additive cubic regression-spline construction,
\[
  \widehat\rho(x)=\max\left\{0,\widehat a_\rho+\sum_{\ell=1}^4\widehat g_\ell(W_\ell(x))\right\}, \qquad
  W(x)=\left\{\widehat\mu(x),x_1,x_2,x_1x_2\right\}^{\top},
\]
with component-specific smoothing parameters again selected by five-fold cross-validation. Negative scale predictions are truncated at zero. The fitted-mean coordinate accommodates DGP1--DGP3, while the remaining coordinates allow the residual scale function to represent the covariate effects and interactions in DGP4--DGP7.

\FloatBarrier

\subsection{Calibration with Externally Specified Moments}
\label{sec:supp-known}

This section reports additional results for the experiment with externally specified moments in Section~5.1 of the main paper. We first assess finite-sample marginal coverage using analytic or numerically estimated means and quantiles and then examine several side-information specifications not included in the main comparison.

\FloatBarrier

\subsubsection{Marginal Coverage}
\label{sec:supp-marginal-coverage}

\paragraph{Design.}
We use DGP1--DGP8 and the side-information specifications M1--M7 from Section~5.1 of the main paper, set $p=5$, and take $1-\alpha=0.9$. The common training and calibration data sizes are
\[
  n\in\{24,44,84,164,324,644\}.
\]
For each method, marginal coverage is estimated by averaging $\ind(S^{\mathrm{test}}\leq\widehat q)$ over the independent test observations and Monte Carlo replications.

The means for M1 are analytic, and the second moment for M2 is computed from $\E\{Y^2\}=20/3+\E\{\sigma_s^2(X)\}$ using analytic expressions or one-dimensional numerical integration. For M3--M7, the required moments and quantiles are estimated from an independent external-reference data set of 50,000 observations $(X,Y)$ drawn from the corresponding DGP in each calibration replication.

M1--M7 use the weighted empirical CDF and adjusted quantile level from Section~2 of the main paper,
\[
  \widehat F_w(q)=\sum_{i=1}^n \widehat w_i\ind(S_i\leq q),
  \qquad
  \widehat q_\alpha=\inf\left\{q:\widehat F_w(q)\geq\frac{n+1}{n}(1-\alpha)\right\}.
\]

\begin{figure}[htp]
  \centering
  \linespread{1}\selectfont
  \revisedfigure{0.99\textwidth}{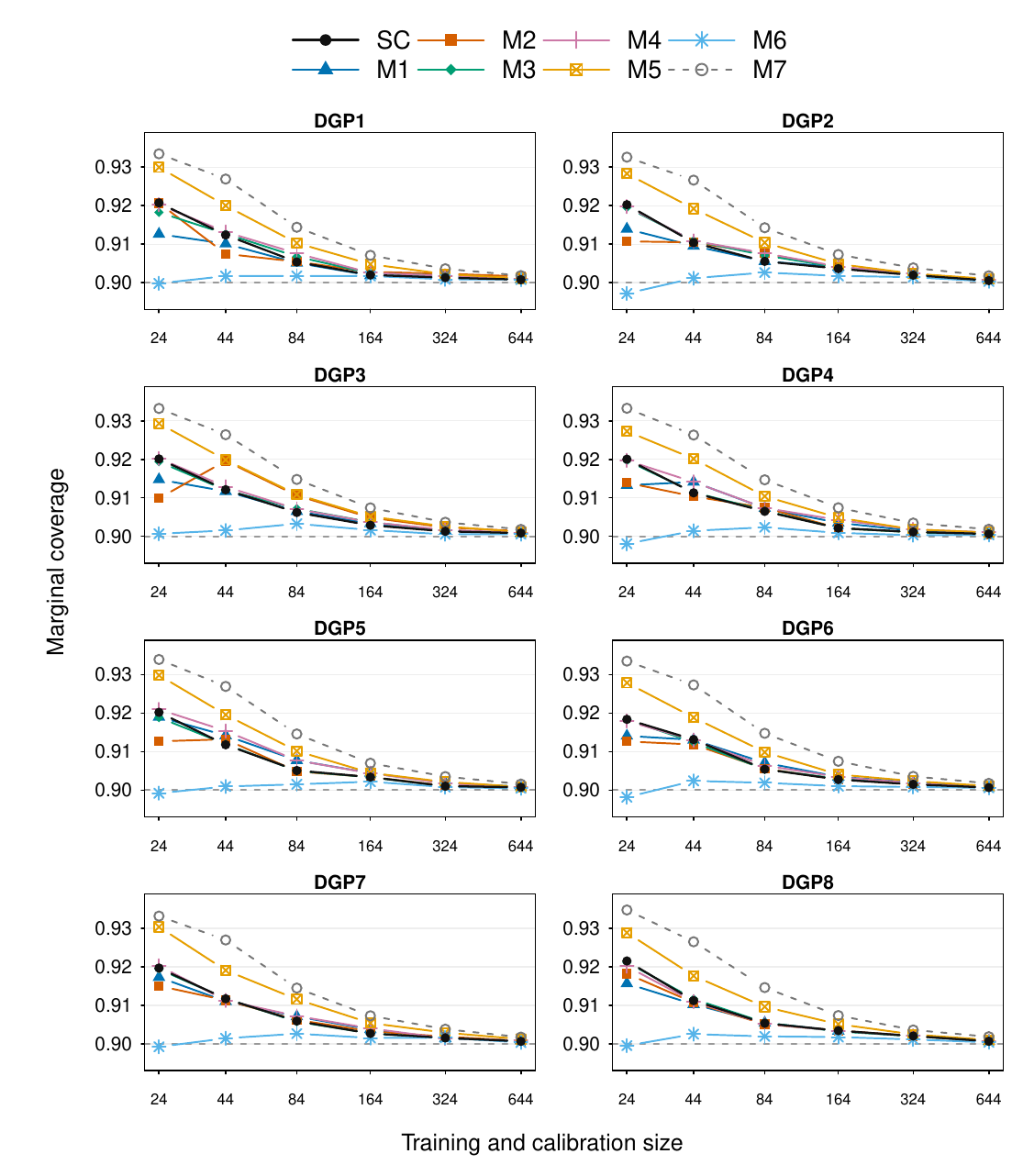}
  \caption{Marginal coverage with analytic or numerically estimated means and quantiles. Each panel corresponds to one DGP and compares SC with M1--M7. The dashed horizontal line marks the nominal level $0.9$.}
  \label{fig:supp-marginal-coverage}
\end{figure}

\paragraph{Results.}
Figure~\ref{fig:supp-marginal-coverage} reports the resulting marginal coverage estimates. At $n=24$, DGP-averaged coverage is $0.9201$ for SC, close to its exact value $0.9200$. The corresponding averages for M1--M4 range from $0.9142$ to $0.9199$, while those for M5, M6, and M7 are $0.9290$, $0.8990$, and $0.9335$, respectively. Across all DGP--method pairs at this sample size, the coverage range from $0.8972$ to $0.9348$ and occur on both sides of $0.9$.
At $n=644$, every estimate lies between $0.9003$ and $0.9018$. Hence the deviations are small at the largest common training and calibration data size considered.

\FloatBarrier

\subsubsection{Additional Sample-Size and Dimension Settings}
\label{sec:supp-known-grid}

\begin{figure}[!htbp]
  \centering
  \linespread{1}\selectfont
  \revisedfigure{0.98\textwidth}{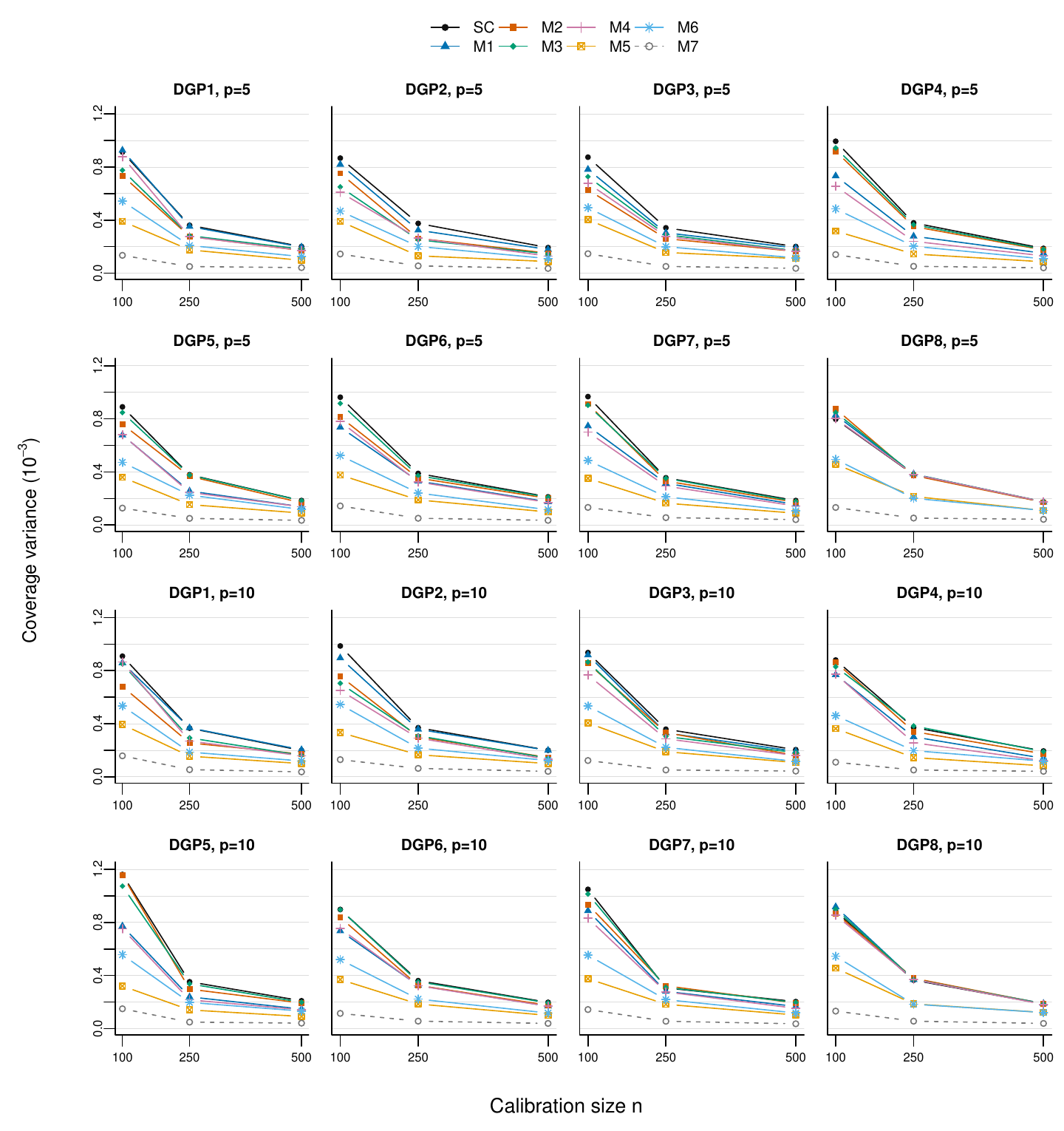}
  \caption{Coverage variance for SC and M1--M7 under DGP1--DGP8.}
  \label{fig:supp-external-variance-grid}
\end{figure}

We use the coverage variance and mean interval length variance defined in Section~5.1 of the main paper. Each variance ratio divides the method's variance by that of SC, and each relative reduction equals one minus the corresponding variance ratio. Figure~\ref{fig:supp-external-variance-grid} displays the coverage variances for all $(n,p)$ combinations. The upper two rows correspond to $p=5$ and the lower two rows to $p=10$. Absolute variability decreases substantially with the calibration data size for every method and DGP. The ordering is also broadly stable across the two dimensions: M5--M7 generally have the smallest variances, while the performance of M1--M4 depends more strongly on the data-generating mechanism.

Tables~\ref{tab:supp-known-n100}--\ref{tab:supp-known-n500} report the full numerical results across all $(n,p)$ settings. Each table combines the $p=5$ and $p=10$ cases at a common calibration data size. Entries for coverage variance and mean interval length variance are ratios relative to SC using the same data in each replication. The final column reports mean coverage in the coverage variance block and mean interval length relative to SC in the mean interval length block.

\begin{table}[!htbp]
  \centering
  \linespread{1}\selectfont
  \setlength{\belowcaptionskip}{6pt}
  \caption{Externally specified-moment results at $n=100$ for $p\in\{5,10\}$.}
  \label{tab:supp-known-n100}
  \label{tab:supp-known-n100-p5}
  \label{tab:supp-known-n100-p10}
  \linespread{1}\fontsize{8.5}{9.5}\selectfont
  \setlength{\tabcolsep}{2.2pt}
  \renewcommand{\arraystretch}{1.2}
  \begin{tabular}{@{}cllrrrrrrrrr@{}}
    \toprule
    $p$ & & Specification & DGP1 & DGP2 & DGP3 & DGP4 & DGP5 & DGP6 & DGP7 & DGP8 & Mean \\
    \midrule
    \multicolumn{12}{@{}l}{\textit{Coverage variance ratio relative to SC (mean coverage in the final column)}} \\
    \addlinespace[1pt]
    5 & SC & -- & 1.000 & 1.000 & 1.000 & 1.000 & 1.000 & 1.000 & 1.000 & 1.000 & 0.900 \\
    & M1 & Covariate moments & 1.015 & 0.945 & 0.894 & 0.737 & 0.761 & 0.765 & 0.771 & 1.040 & 0.905 \\
    & M2 & Outcome moments & 0.805 & 0.868 & 0.719 & 0.924 & 0.853 & 0.845 & 0.943 & 1.098 & 0.905 \\
    & M3 & Prediction tails & 0.850 & 0.749 & 0.832 & 0.950 & 0.952 & 0.952 & 0.932 & 1.064 & 0.904 \\
    & M4 & Residual scale & 0.961 & 0.703 & 0.775 & 0.660 & 0.763 & 0.809 & 0.724 & 1.006 & 0.905 \\
    & M5 & Score mean & 0.428 & 0.451 & 0.462 & 0.319 & 0.404 & 0.391 & 0.363 & 0.575 & 0.908 \\
    & M6 & Score event ($0.95$) & 0.595 & 0.539 & 0.564 & 0.487 & 0.531 & 0.543 & 0.503 & 0.621 & 0.901 \\
    & M7 & Score event ($0.9$) & 0.147 & 0.167 & 0.168 & 0.141 & 0.143 & 0.149 & 0.137 & 0.166 & 0.912 \\
    \addlinespace[3pt]
    10 & SC & -- & 1.000 & 1.000 & 1.000 & 1.000 & 1.000 & 1.000 & 1.000 & 1.000 & 0.900 \\
    & M1 & Covariate moments & 0.945 & 0.908 & 0.980 & 0.874 & 0.665 & 0.820 & 0.846 & 1.038 & 0.905 \\
    & M2 & Outcome moments & 0.749 & 0.769 & 0.913 & 0.982 & 0.995 & 0.933 & 0.890 & 0.983 & 0.905 \\
    & M3 & Prediction tails & 0.935 & 0.714 & 0.924 & 0.942 & 0.925 & 0.999 & 0.967 & 1.018 & 0.904 \\
    & M4 & Residual scale & 0.952 & 0.661 & 0.819 & 0.879 & 0.648 & 0.840 & 0.792 & 0.967 & 0.905 \\
    & M5 & Score mean & 0.434 & 0.339 & 0.433 & 0.413 & 0.276 & 0.412 & 0.358 & 0.516 & 0.908 \\
    & M6 & Score event ($0.95$) & 0.588 & 0.553 & 0.570 & 0.524 & 0.479 & 0.579 & 0.528 & 0.617 & 0.901 \\
    & M7 & Score event ($0.9$) & 0.174 & 0.131 & 0.131 & 0.125 & 0.129 & 0.128 & 0.137 & 0.149 & 0.912 \\
    \midrule
    \multicolumn{12}{@{}l}{\textit{Mean interval length variance ratio relative to SC (mean length relative to SC in the final column)}} \\
    \addlinespace[1pt]
    5 & SC & -- & 1.000 & 1.000 & 1.000 & 1.000 & 1.000 & 1.000 & 1.000 & 1.000 & 1.000 \\
    & M1 & Covariate moments & 1.087 & 1.116 & 1.017 & 0.816 & 0.871 & 0.854 & 0.917 & 1.115 & 1.027 \\
    & M2 & Outcome moments & 0.912 & 1.005 & 0.835 & 1.115 & 0.963 & 0.904 & 1.054 & 1.139 & 1.028 \\
    & M3 & Prediction tails & 0.958 & 0.889 & 0.961 & 1.142 & 1.153 & 1.049 & 1.047 & 1.106 & 1.025 \\
    & M4 & Residual scale & 1.057 & 0.792 & 0.943 & 0.788 & 0.879 & 0.939 & 0.832 & 1.037 & 1.027 \\
    & M5 & Score mean & 0.496 & 0.463 & 0.440 & 0.425 & 0.399 & 0.448 & 0.455 & 0.654 & 1.026 \\
    & M6 & Score event ($0.95$) & 0.514 & 0.401 & 0.440 & 0.364 & 0.413 & 0.409 & 0.471 & 0.532 & 0.984 \\
    & M7 & Score event ($0.9$) & 0.192 & 0.235 & 0.213 & 0.271 & 0.205 & 0.240 & 0.221 & 0.198 & 1.040 \\
    \addlinespace[3pt]
    10 & SC & -- & 1.000 & 1.000 & 1.000 & 1.000 & 1.000 & 1.000 & 1.000 & 1.000 & 1.000 \\
    & M1 & Covariate moments & 1.046 & 1.132 & 1.066 & 1.132 & 0.832 & 0.935 & 1.117 & 1.094 & 1.024 \\
    & M2 & Outcome moments & 0.815 & 0.891 & 1.041 & 1.142 & 1.222 & 1.074 & 1.071 & 1.071 & 1.022 \\
    & M3 & Prediction tails & 0.972 & 0.869 & 0.950 & 1.071 & 1.102 & 1.085 & 1.108 & 1.085 & 1.021 \\
    & M4 & Residual scale & 1.041 & 0.882 & 0.905 & 1.064 & 0.773 & 0.949 & 0.963 & 1.031 & 1.022 \\
    & M5 & Score mean & 0.464 & 0.390 & 0.519 & 0.550 & 0.331 & 0.444 & 0.460 & 0.590 & 1.025 \\
    & M6 & Score event ($0.95$) & 0.459 & 0.438 & 0.485 & 0.486 & 0.374 & 0.513 & 0.463 & 0.557 & 0.987 \\
    & M7 & Score event ($0.9$) & 0.290 & 0.223 & 0.211 & 0.206 & 0.290 & 0.182 & 0.287 & 0.190 & 1.034 \\
    \bottomrule
  \end{tabular}
\end{table}

The conclusions from the $n=100$, $p=5$ comparison are broadly unchanged across all scenarios. M5--M7 remain the strongest specifications, M4 gives moderate reductions under the heterogeneous designs, and M1--M3 are more mechanism-dependent. Increasing $p$ from 5 to 10 has little effect on this ordering, and the mean interval length variances show the same broad pattern.

\begin{table}[!htbp]
  \centering
  \linespread{1}\selectfont
  \setlength{\belowcaptionskip}{6pt}
  \caption{Externally specified-moment results at $n=250$ for $p\in\{5,10\}$.}
  \label{tab:supp-known-n250}
  \label{tab:supp-known-n250-p5}
  \label{tab:supp-known-n250-p10}
  \linespread{1}\fontsize{8.5}{9.5}\selectfont
  \setlength{\tabcolsep}{2.2pt}
  \renewcommand{\arraystretch}{1.2}
  \begin{tabular}{@{}cllrrrrrrrrr@{}}
    \toprule
    $p$ & & Specification & DGP1 & DGP2 & DGP3 & DGP4 & DGP5 & DGP6 & DGP7 & DGP8 & Mean \\
    \midrule
    \multicolumn{12}{@{}l}{\textit{Coverage variance ratio relative to SC (mean coverage in the final column)}} \\
    \addlinespace[1pt]
    5 & SC & -- & 1.000 & 1.000 & 1.000 & 1.000 & 1.000 & 1.000 & 1.000 & 1.000 & 0.900 \\
    & M1 & Covariate moments & 0.975 & 0.864 & 0.890 & 0.734 & 0.675 & 0.851 & 0.883 & 1.014 & 0.902 \\
    & M2 & Outcome moments & 0.767 & 0.700 & 0.762 & 0.927 & 0.970 & 0.909 & 0.930 & 0.984 & 0.902 \\
    & M3 & Prediction tails & 0.782 & 0.671 & 0.849 & 0.964 & 0.993 & 0.957 & 0.988 & 1.009 & 0.902 \\
    & M4 & Residual scale & 0.763 & 0.703 & 0.811 & 0.633 & 0.649 & 0.834 & 0.825 & 1.006 & 0.902 \\
    & M5 & Score mean & 0.486 & 0.348 & 0.457 & 0.382 & 0.405 & 0.486 & 0.461 & 0.569 & 0.903 \\
    & M6 & Score event ($0.95$) & 0.578 & 0.536 & 0.584 & 0.535 & 0.594 & 0.620 & 0.597 & 0.530 & 0.902 \\
    & M7 & Score event ($0.9$) & 0.139 & 0.146 & 0.149 & 0.136 & 0.132 & 0.130 & 0.156 & 0.138 & 0.905 \\
    \addlinespace[3pt]
    10 & SC & -- & 1.000 & 1.000 & 1.000 & 1.000 & 1.000 & 1.000 & 1.000 & 1.000 & 0.901 \\
    & M1 & Covariate moments & 1.006 & 0.968 & 0.922 & 0.810 & 0.678 & 0.900 & 0.914 & 1.011 & 0.903 \\
    & M2 & Outcome moments & 0.696 & 0.812 & 0.925 & 0.919 & 0.843 & 0.902 & 1.050 & 1.039 & 0.903 \\
    & M3 & Prediction tails & 0.800 & 0.837 & 0.848 & 1.033 & 0.956 & 0.979 & 1.013 & 1.018 & 0.902 \\
    & M4 & Residual scale & 0.737 & 0.787 & 0.799 & 0.692 & 0.615 & 0.895 & 0.900 & 1.015 & 0.902 \\
    & M5 & Score mean & 0.424 & 0.449 & 0.527 & 0.391 & 0.399 & 0.514 & 0.597 & 0.510 & 0.903 \\
    & M6 & Score event ($0.95$) & 0.508 & 0.584 & 0.619 & 0.531 & 0.559 & 0.616 & 0.711 & 0.504 & 0.902 \\
    & M7 & Score event ($0.9$) & 0.147 & 0.171 & 0.143 & 0.138 & 0.139 & 0.156 & 0.181 & 0.154 & 0.905 \\
    \midrule
    \multicolumn{12}{@{}l}{\textit{Mean interval length variance ratio relative to SC (mean length relative to SC in the final column)}} \\
    \addlinespace[1pt]
    5 & SC & -- & 1.000 & 1.000 & 1.000 & 1.000 & 1.000 & 1.000 & 1.000 & 1.000 & 1.000 \\
    & M1 & Covariate moments & 0.995 & 0.908 & 0.910 & 0.848 & 0.737 & 0.931 & 0.882 & 1.047 & 1.010 \\
    & M2 & Outcome moments & 0.798 & 0.721 & 0.775 & 0.982 & 1.015 & 0.950 & 0.928 & 1.013 & 1.011 \\
    & M3 & Prediction tails & 0.790 & 0.683 & 0.871 & 1.009 & 1.038 & 0.999 & 1.020 & 1.034 & 1.008 \\
    & M4 & Residual scale & 0.786 & 0.701 & 0.823 & 0.707 & 0.702 & 0.906 & 0.824 & 1.013 & 1.009 \\
    & M5 & Score mean & 0.462 & 0.332 & 0.455 & 0.370 & 0.394 & 0.511 & 0.468 & 0.574 & 1.008 \\
    & M6 & Score event ($0.95$) & 0.543 & 0.505 & 0.538 & 0.520 & 0.569 & 0.595 & 0.538 & 0.524 & 1.003 \\
    & M7 & Score event ($0.9$) & 0.105 & 0.118 & 0.121 & 0.125 & 0.123 & 0.111 & 0.128 & 0.110 & 1.013 \\
    \addlinespace[3pt]
    10 & SC & -- & 1.000 & 1.000 & 1.000 & 1.000 & 1.000 & 1.000 & 1.000 & 1.000 & 1.000 \\
    & M1 & Covariate moments & 1.067 & 1.017 & 0.926 & 0.825 & 0.751 & 0.951 & 1.001 & 1.044 & 1.008 \\
    & M2 & Outcome moments & 0.785 & 0.922 & 0.997 & 0.980 & 0.937 & 0.942 & 1.072 & 1.075 & 1.011 \\
    & M3 & Prediction tails & 0.830 & 0.897 & 0.835 & 1.078 & 1.069 & 1.012 & 1.088 & 1.048 & 1.009 \\
    & M4 & Residual scale & 0.772 & 0.835 & 0.796 & 0.692 & 0.666 & 0.940 & 0.911 & 1.029 & 1.006 \\
    & M5 & Score mean & 0.435 & 0.431 & 0.520 & 0.353 & 0.386 & 0.508 & 0.559 & 0.539 & 1.005 \\
    & M6 & Score event ($0.95$) & 0.489 & 0.576 & 0.599 & 0.486 & 0.531 & 0.598 & 0.610 & 0.503 & 1.000 \\
    & M7 & Score event ($0.9$) & 0.114 & 0.144 & 0.123 & 0.092 & 0.129 & 0.130 & 0.144 & 0.123 & 1.009 \\
    \bottomrule
  \end{tabular}
\end{table}

Averaged over the eight DGPs, mean coverage is between $0.901$ and $0.908$ for M1--M6. For M7, mean coverage decreases from $0.912$ at $n=100$ to $0.905$ at $n=250$ and $0.902$ at $n=500$. For M7, the excess above the nominal coverage level is largest at the smallest calibration data size and diminishes as $n$ increases. Across all $(n,p)$ configurations and M1--M7, relative mean lengths, averaged over DGPs, range from $0.984$ to $1.040$. Absolute coverage variance decreases with $n$ for all specifications, while their broad ranking remains stable.

\begin{table}[!htbp]
  \centering
  \linespread{1}\selectfont
  \setlength{\belowcaptionskip}{6pt}
  \caption{Externally specified-moment results at $n=500$ for $p\in\{5,10\}$.}
  \label{tab:supp-known-n500}
  \label{tab:supp-known-n500-p5}
  \label{tab:supp-known-n500-p10}
  \linespread{1}\fontsize{8.5}{9.5}\selectfont
  \setlength{\tabcolsep}{2.2pt}
  \renewcommand{\arraystretch}{1.2}
  \begin{tabular}{@{}cllrrrrrrrrr@{}}
    \toprule
    $p$ & & Specification & DGP1 & DGP2 & DGP3 & DGP4 & DGP5 & DGP6 & DGP7 & DGP8 & Mean \\
    \midrule
    \multicolumn{12}{@{}l}{\textit{Coverage variance ratio relative to SC (mean coverage in the final column)}} \\
    \addlinespace[1pt]
    5 & SC & -- & 1.000 & 1.000 & 1.000 & 1.000 & 1.000 & 1.000 & 1.000 & 1.000 & 0.900 \\
    & M1 & Covariate moments & 0.981 & 0.927 & 0.946 & 0.802 & 0.746 & 0.824 & 0.841 & 0.986 & 0.901 \\
    & M2 & Outcome moments & 0.851 & 0.783 & 0.816 & 0.938 & 0.908 & 0.948 & 0.930 & 0.983 & 0.901 \\
    & M3 & Prediction tails & 0.906 & 0.752 & 0.844 & 0.962 & 0.992 & 0.986 & 0.966 & 1.005 & 0.901 \\
    & M4 & Residual scale & 0.857 & 0.666 & 0.825 & 0.699 & 0.743 & 0.801 & 0.777 & 1.001 & 0.901 \\
    & M5 & Score mean & 0.483 & 0.456 & 0.551 & 0.460 & 0.488 & 0.463 & 0.483 & 0.623 & 0.902 \\
    & M6 & Score event ($0.95$) & 0.622 & 0.571 & 0.575 & 0.577 & 0.637 & 0.542 & 0.574 & 0.626 & 0.901 \\
    & M7 & Score event ($0.9$) & 0.209 & 0.185 & 0.180 & 0.215 & 0.189 & 0.165 & 0.220 & 0.244 & 0.902 \\
    \addlinespace[3pt]
    10 & SC & -- & 1.000 & 1.000 & 1.000 & 1.000 & 1.000 & 1.000 & 1.000 & 1.000 & 0.900 \\
    & M1 & Covariate moments & 1.023 & 0.995 & 0.938 & 0.716 & 0.702 & 0.858 & 0.821 & 1.019 & 0.901 \\
    & M2 & Outcome moments & 0.861 & 0.726 & 0.796 & 0.887 & 0.919 & 0.889 & 0.949 & 1.006 & 0.901 \\
    & M3 & Prediction tails & 0.814 & 0.701 & 0.893 & 0.975 & 0.946 & 0.982 & 0.969 & 1.007 & 0.901 \\
    & M4 & Residual scale & 0.793 & 0.657 & 0.782 & 0.623 & 0.675 & 0.851 & 0.752 & 0.983 & 0.901 \\
    & M5 & Score mean & 0.501 & 0.499 & 0.528 & 0.425 & 0.437 & 0.504 & 0.505 & 0.654 & 0.901 \\
    & M6 & Score event ($0.95$) & 0.600 & 0.613 & 0.567 & 0.604 & 0.630 & 0.593 & 0.581 & 0.652 & 0.901 \\
    & M7 & Score event ($0.9$) & 0.186 & 0.204 & 0.212 & 0.212 & 0.197 & 0.197 & 0.179 & 0.211 & 0.902 \\
    \midrule
    \multicolumn{12}{@{}l}{\textit{Mean interval length variance ratio relative to SC (mean length relative to SC in the final column)}} \\
    \addlinespace[1pt]
    5 & SC & -- & 1.000 & 1.000 & 1.000 & 1.000 & 1.000 & 1.000 & 1.000 & 1.000 & 1.000 \\
    & M1 & Covariate moments & 0.990 & 0.923 & 0.959 & 0.780 & 0.772 & 0.816 & 0.863 & 1.012 & 1.003 \\
    & M2 & Outcome moments & 0.834 & 0.807 & 0.850 & 0.945 & 0.949 & 0.962 & 0.940 & 1.002 & 1.005 \\
    & M3 & Prediction tails & 0.932 & 0.744 & 0.866 & 0.977 & 1.028 & 1.010 & 0.971 & 1.018 & 1.005 \\
    & M4 & Residual scale & 0.885 & 0.645 & 0.829 & 0.654 & 0.739 & 0.785 & 0.773 & 1.027 & 1.002 \\
    & M5 & Score mean & 0.447 & 0.380 & 0.519 & 0.411 & 0.456 & 0.442 & 0.428 & 0.599 & 1.002 \\
    & M6 & Score event ($0.95$) & 0.545 & 0.498 & 0.528 & 0.533 & 0.605 & 0.495 & 0.511 & 0.600 & 0.999 \\
    & M7 & Score event ($0.9$) & 0.114 & 0.109 & 0.110 & 0.144 & 0.141 & 0.101 & 0.132 & 0.150 & 1.003 \\
    \addlinespace[3pt]
    10 & SC & -- & 1.000 & 1.000 & 1.000 & 1.000 & 1.000 & 1.000 & 1.000 & 1.000 & 1.000 \\
    & M1 & Covariate moments & 1.035 & 1.041 & 0.974 & 0.741 & 0.669 & 0.827 & 0.785 & 1.040 & 1.003 \\
    & M2 & Outcome moments & 0.890 & 0.727 & 0.800 & 0.896 & 0.931 & 0.923 & 0.967 & 1.014 & 1.005 \\
    & M3 & Prediction tails & 0.818 & 0.690 & 0.905 & 1.006 & 0.975 & 1.004 & 0.990 & 1.025 & 1.005 \\
    & M4 & Residual scale & 0.775 & 0.672 & 0.804 & 0.621 & 0.638 & 0.823 & 0.726 & 0.985 & 1.002 \\
    & M5 & Score mean & 0.462 & 0.439 & 0.482 & 0.388 & 0.390 & 0.457 & 0.479 & 0.617 & 1.000 \\
    & M6 & Score event ($0.95$) & 0.519 & 0.544 & 0.516 & 0.555 & 0.585 & 0.563 & 0.566 & 0.604 & 0.997 \\
    & M7 & Score event ($0.9$) & 0.112 & 0.123 & 0.130 & 0.130 & 0.128 & 0.127 & 0.111 & 0.142 & 1.002 \\
    \bottomrule
  \end{tabular}
\end{table}

\FloatBarrier

\subsubsection{Additional Side-Information Specifications}
\label{sec:supp-known-additional}

We consider additional forms of side information that complement the main analysis. The covariate and outcome moments are examined componentwise, and we also include a joint event, bounded summaries based on the outcome and estimated scale, and oracle summaries constructed from the true conditional scale. Table~\ref{tab:supp-known-additional-library} lists the resulting specifications.

\begin{table}[!htbp]
  \centering
  \linespread{1}\selectfont
  \setlength{\belowcaptionskip}{6pt}
  \caption{Additional side-information specifications for the externally specified-moment experiment.}
  \label{tab:supp-known-additional-library}
  \linespread{1}\fontsize{8.5}{9.5}\selectfont
  \setlength{\tabcolsep}{3.2pt}
  \renewcommand{\arraystretch}{1}
  \begin{tabular}{ll}
    \toprule
    Specification & Moment function\\
    \midrule
    X-Linear &
    $\mathcal C_{\mathrm{ext}}(X_1,X_2)$\\
    X-Quadratic &
    $\mathcal C_{\mathrm{ext}}(X_1^2+X_2^2,X_1X_2)$\\
    Y-Mean &
    $\mathcal C_{\mathrm{ext}}(Y)$\\
    Y-Second &
    $\mathcal C_{\mathrm{ext}}(Y^2)$\\
    Joint event &
    $\mathcal C_{\mathrm{ext}}(\ind(X_1>0)\ind(Y>0))$\\
    $\widehat\rho$-Tail &
    $\mathcal C_{\mathrm{ext}}
    \bigl(\ind\{\widehat\rho(X)>
    Q_{\widehat\rho}^{\mathrm{ext}}(.75)\}\bigr)$\\
    Y-Range &
    $\mathcal C_{\mathrm{ext}}
    \bigl(\ind\{Y<Q_Y^{\mathrm{ext}}(.10)
    \text{ or }Y>Q_Y^{\mathrm{ext}}(.90)\}\bigr)$\\
    True-Scale &
    $\mathcal C_{\mathrm{ext}}(\sigma_s(X))$\\
    True-Scale-Tail &
    $\mathcal C_{\mathrm{ext}}
    \bigl(\ind\{\sigma_s(X)>Q_{\sigma_s}^{\mathrm{ext}}(.75)\}\bigr)$\\
    \bottomrule
  \end{tabular}
\end{table}

The analysis uses $n=100$ and $p=5$, matching the main design. Table~\ref{tab:supp-known-additional-theory} reports the relative reductions in the limiting variance of training-conditional coverage. Table~\ref{tab:supp-known-additional-results} reports the finite-sample coverage variance ratios and mean interval length variance ratios, together with mean coverage and mean interval length relative to SC.

\begin{table}[!htbp]
  \centering
  \linespread{1}\selectfont
  \setlength{\belowcaptionskip}{6pt}
  \caption{Relative reductions in the limiting variance of training-conditional coverage for the additional moment specifications at $n=100$ and $p=5$. M1, M2, and M4 are included for comparison.}
  \label{tab:supp-known-additional-theory}
  \linespread{1}\fontsize{8.5}{9.5}\selectfont
  \setlength{\tabcolsep}{3.2pt}
  \renewcommand{\arraystretch}{1}
  \begin{tabular}{@{}lrrrrrrrrr@{}}
    \toprule
    Specification & DGP1 & DGP2 & DGP3 & DGP4 & DGP5 & DGP6 & DGP7 & DGP8 & Average \\
    \midrule
    M1 & 0.040 & 0.099 & 0.085 & 0.262 & 0.369 & 0.189 & 0.208 & 0.000 & 0.157 \\
    X-Linear & 0.000 & 0.086 & 0.076 & 0.174 & 0.000 & 0.135 & 0.166 & 0.000 & 0.080 \\
    X-Quadratic & 0.040 & 0.014 & 0.009 & 0.088 & 0.369 & 0.055 & 0.042 & 0.000 & 0.077 \\
    \midrule
    M2 & 0.181 & 0.226 & 0.305 & 0.075 & 0.163 & 0.152 & 0.058 & 0.008 & 0.146 \\
    Y-Mean & 0.000 & 0.160 & 0.223 & 0.017 & 0.000 & 0.036 & 0.014 & 0.001 & 0.056 \\
    Y-Second & 0.180 & 0.132 & 0.219 & 0.067 & 0.163 & 0.134 & 0.051 & 0.008 & 0.119 \\
    Y-Range & 0.052 & 0.043 & 0.138 & 0.027 & 0.067 & 0.069 & 0.018 & 0.004 & 0.052 \\
    Joint event & 0.001 & 0.069 & 0.048 & 0.031 & 0.002 & 0.001 & 0.000 & 0.000 & 0.019 \\
    \midrule
    M4 & 0.081 & 0.350 & 0.232 & 0.326 & 0.268 & 0.198 & 0.238 & 0.000 & 0.212 \\
    $\widehat\rho$-Tail & 0.056 & 0.271 & 0.174 & 0.248 & 0.218 & 0.184 & 0.222 & 0.000 & 0.172 \\
    True-Scale & 0.288 & 0.384 & 0.247 & 0.366 & 0.446 & 0.211 & 0.268 & 0.000 & 0.276 \\
    True-Scale-Tail & 0.225 & 0.271 & 0.194 & 0.251 & 0.318 & 0.189 & 0.233 & 0.000 & 0.210 \\
    \bottomrule
  \end{tabular}
\end{table}

\begin{table}[!htbp]
  \centering
  \linespread{1}\selectfont
  \setlength{\belowcaptionskip}{6pt}
  \caption{Finite-sample variance ratios for coverage and mean interval length relative to SC for the additional moment specifications at $n=100$ and $p=5$.}
  \label{tab:supp-known-additional-results}
  \linespread{1}\fontsize{8.5}{9.5}\selectfont
  \setlength{\tabcolsep}{3.2pt}
  \renewcommand{\arraystretch}{1}
  \begin{tabular}{@{}lrrrrrrrrr@{}}
    \toprule
    Specification & DGP1 & DGP2 & DGP3 & DGP4 & DGP5 & DGP6 & DGP7 & DGP8 & Mean \\
    \midrule
    \multicolumn{10}{l}{\textit{Coverage variance ratio relative to SC (final column: mean coverage)}} \\
    SC & 1.000 & 1.000 & 1.000 & 1.000 & 1.000 & 1.000 & 1.000 & 1.000 & 0.900 \\
    M1 & 1.015 & 0.945 & 0.894 & 0.737 & 0.761 & 0.765 & 0.771 & 1.040 & 0.905 \\
    X-Linear & 1.009 & 0.877 & 0.864 & 0.771 & 1.031 & 0.844 & 0.815 & 1.028 & 0.905 \\
    X-Quadratic & 1.003 & 1.026 & 1.006 & 0.896 & 0.704 & 0.916 & 0.957 & 0.988 & 0.905 \\
    \midrule
    M2 & 0.805 & 0.868 & 0.719 & 0.924 & 0.853 & 0.845 & 0.943 & 1.098 & 0.905 \\
    Y-Mean & 1.004 & 0.872 & 0.769 & 0.930 & 0.983 & 0.972 & 0.948 & 1.056 & 0.904 \\
    Y-Second & 0.832 & 0.908 & 0.789 & 0.968 & 0.839 & 0.885 & 0.942 & 1.086 & 0.905 \\
    Y-Range & 0.932 & 0.964 & 0.882 & 1.001 & 0.890 & 0.925 & 0.999 & 1.030 & 0.904 \\
    Joint event & 0.989 & 0.939 & 0.890 & 0.929 & 1.018 & 1.016 & 0.957 & 1.008 & 0.903 \\
    \midrule
    M4 & 0.961 & 0.703 & 0.775 & 0.660 & 0.763 & 0.809 & 0.724 & 1.006 & 0.905 \\
    $\widehat\rho$-Tail & 1.009 & 0.742 & 0.784 & 0.711 & 0.837 & 0.837 & 0.760 & 1.036 & 0.905 \\
    True-Scale & 0.767 & 0.641 & 0.751 & 0.641 & 0.659 & 0.780 & 0.731 & 1.000 & 0.905 \\
    True-Scale-Tail & 0.806 & 0.745 & 0.813 & 0.690 & 0.745 & 0.862 & 0.745 & 1.000 & 0.905 \\
    \midrule
    \multicolumn{10}{l}{\textit{Mean interval length variance ratio relative to SC (final column: relative mean length)}} \\
    SC & 1.000 & 1.000 & 1.000 & 1.000 & 1.000 & 1.000 & 1.000 & 1.000 & 1.000 \\
    M1 & 1.087 & 1.116 & 1.017 & 0.816 & 0.871 & 0.854 & 0.917 & 1.115 & 1.027 \\
    X-Linear & 1.128 & 1.032 & 0.946 & 0.902 & 1.169 & 0.929 & 0.944 & 1.098 & 1.025 \\
    X-Quadratic & 1.150 & 1.283 & 1.190 & 1.025 & 0.817 & 1.013 & 1.097 & 1.045 & 1.028 \\
    \midrule
    M2 & 0.912 & 1.005 & 0.835 & 1.115 & 0.963 & 0.904 & 1.054 & 1.139 & 1.028 \\
    Y-Mean & 1.099 & 1.095 & 0.860 & 1.144 & 1.095 & 1.048 & 1.057 & 1.086 & 1.022 \\
    Y-Second & 0.948 & 1.011 & 0.847 & 1.029 & 0.944 & 0.936 & 1.073 & 1.118 & 1.026 \\
    Y-Range & 1.031 & 1.125 & 1.019 & 1.082 & 0.984 & 0.968 & 1.067 & 1.062 & 1.022 \\
    Joint event & 1.055 & 1.109 & 1.014 & 1.107 & 1.116 & 1.064 & 1.033 & 1.063 & 1.020 \\
    \midrule
    M4 & 1.057 & 0.792 & 0.943 & 0.788 & 0.879 & 0.939 & 0.832 & 1.037 & 1.027 \\
    $\widehat\rho$-Tail & 1.134 & 0.896 & 0.863 & 0.839 & 0.980 & 0.922 & 0.877 & 1.066 & 1.026 \\
    True-Scale & 0.883 & 0.775 & 0.867 & 0.722 & 0.766 & 0.871 & 0.828 & 1.000 & 1.024 \\
    True-Scale-Tail & 0.866 & 0.897 & 0.927 & 0.803 & 0.881 & 0.953 & 0.847 & 1.000 & 1.024 \\
    \bottomrule
  \end{tabular}
\end{table}

The two components of M1 are informative under different mechanisms. X-Linear accounts for most of the asymptotic variance reduction under DGP2--DGP4 and DGP6--DGP7, whereas X-Quadratic is more important under DGP1 and DGP5. Neither component alone gives a consistent finite-sample improvement across the eight DGPs: their average coverage variance ratios are $0.905$ and $0.937$, compared with $0.866$ for M1. For M2, most of the asymptotic variance reduction is associated with Y-Second. Its average asymptotic variance reduction is $0.119$, compared with $0.056$ for Y-Mean and $0.146$ for M2. The corresponding average coverage variance ratios are $0.906$, $0.942$, and $0.882$. Y-Range and the joint event give small average reductions in coverage variance, with ratios $0.953$ and $0.968$, respectively.

The residual scale specification M4 is more informative than its upper-quartile version. Their average asymptotic variance reductions are $0.212$ and $0.172$, and their finite-sample coverage variance ratios are $0.800$ and $0.840$, respectively. True-Scale has average asymptotic variance reduction $0.276$ and average coverage variance ratio $0.746$, illustrating the gain available from observing the true conditional scale. True-Scale-Tail has average coverage variance ratio $0.801$, showing a smaller reduction after thresholding. Under DGP8, the true scale is constant, and both oracle specifications reduce to SC.

\FloatBarrier

\subsection{Semi-Supervised Calibration with Prespecified Moment Functions}
\label{sec:supp-ss}

This section reports the asymptotic variance reductions, finite-sample results for the additional sample configurations, and a sensitivity analysis for the semi-supervised experiment in Section~5.2 of the main paper.

\paragraph{Conditional-quantile estimation.}
The three conditional quantile functions are fitted once on the fixed independent training data by additive quantile regression using the same seven-dimensional cubic regression-spline basis for each coordinate of $W=\{\widehat\mu(X),X_1,X_2,X_1X_2\}^{\top}$.
No calibration or unlabeled data enter these fits.

\paragraph{Asymptotic variance reduction.}
For a fixed covariate representation $A$ and a prespecified moment function $h$, define
\[
  R_h^2=\frac{c_{\alpha,A}^{\top}\Sigma_A^{-1}c_{\alpha,A}}{\alpha(1-\alpha)},
\]
where $\Sigma_A=\Var\{h(A)\}$ and $c_{\alpha,A}=\E\left[\left\{\ind(S\le q_\alpha)-(1-\alpha)\right\}\left\{h(A)-\E (h(A))\right\}\right]$. Theorem~3 shows that estimating the mean of $h(A)$ from the pooled calibration and unlabeled data gives the relative reduction in limiting variance $R_{h,\mathrm{ss}}^2=\{m/(n+m)\}R_h^2$. The multiplier $m/(n+m)$ equals $1/2$, $5/6$, and $10/11$ when $m/n=1,5,$ and $10$, respectively.

Table~\ref{tab:supp-semi-theory} reports $R_{h,\mathrm{ss}}^2$ for the main setting $m/n=10$. The criterion is evaluated conditional on the fitted mean, scale, and quantile functions using independent samples of size 50,000 and then averaged over 5000 population draws under the same fixed training fit.

\begin{table}[!htbp]
  \centering
  \linespread{1}\selectfont
  \setlength{\belowcaptionskip}{6pt}
  \caption{Relative reductions in the limiting variance of training-conditional coverage for the prespecified moment functions at $n=250$, $p=5$, and $m/n=10$. The reported criterion is $R_{h,\mathrm{ss}}^2=\{m/(n+m)\}R_h^2$.}
  \label{tab:supp-semi-theory}
  \linespread{1}\fontsize{8.5}{9.5}\selectfont
  \setlength{\tabcolsep}{2.2pt}
  \renewcommand{\arraystretch}{1}
  \begin{tabular}{llccccccccc}
    \toprule
    & Specification & DGP1 & DGP2 & DGP3 & DGP4 & DGP5 & DGP6 & DGP7 & DGP8 & Avg. \\
    \midrule
    SS1 & Prediction moments & 0.259 & 0.304 & 0.227 & 0.040 & 0.052 & 0.029 & 0.024 & 0.000 & 0.117 \\
    SS2 & Residual scale moments & 0.234 & 0.296 & 0.185 & 0.381 & 0.362 & 0.169 & 0.237 & 0.000 & 0.233 \\
    SS3 & Conditional quantiles & 0.228 & 0.289 & 0.197 & 0.374 & 0.381 & 0.151 & 0.214 & 0.000 & 0.229 \\
    SS4 & Prediction and residual scale & 0.260 & 0.319 & 0.228 & 0.383 & 0.369 & 0.171 & 0.238 & 0.000 & 0.246 \\
    SS5 & Prediction and quantiles & 0.260 & 0.312 & 0.229 & 0.376 & 0.384 & 0.153 & 0.218 & 0.000 & 0.242 \\
    SS6 & Residual scale and quantiles & 0.247 & 0.321 & 0.198 & 0.386 & 0.385 & 0.170 & 0.241 & 0.000 & 0.244 \\
    SS7 & All moment functions & 0.261 & 0.329 & 0.229 & 0.386 & 0.388 & 0.171 & 0.242 & 0.000 & 0.251 \\
    \bottomrule
  \end{tabular}
\end{table}

SS4--SS7 contain the functions in the corresponding smaller specifications, so their asymptotic variance reductions are at least as large. In particular,
\[
  \begin{aligned}
    R_{h_4,\mathrm{ss}}^2&\ge\max\{R_{h_1,\mathrm{ss}}^2,R_{h_2,\mathrm{ss}}^2\},&
    R_{h_5,\mathrm{ss}}^2&\ge\max\{R_{h_1,\mathrm{ss}}^2,R_{h_3,\mathrm{ss}}^2\},\\
    R_{h_6,\mathrm{ss}}^2&\ge\max\{R_{h_2,\mathrm{ss}}^2,R_{h_3,\mathrm{ss}}^2\},&
    R_{h_7,\mathrm{ss}}^2&\ge\max_{1\le j\le6}R_{h_j,\mathrm{ss}}^2.
  \end{aligned}
\]

The individual specifications show two distinct patterns. SS1 has average asymptotic variance reduction $0.117$ and is most informative under DGP1--DGP3. SS2 and SS3 have substantially larger average reductions, $0.233$ and $0.229$, driven mainly by DGP4--DGP7. Combining the functions increases the asymptotic variance reduction, but the increments are modest: the averages for SS4--SS7 range from $0.242$ to $0.251$. SS7 has the largest asymptotic variance reduction for every DGP, as implied by the nested projection spaces. All reductions are negligible under the homoscedastic DGP8.

\paragraph{Finite-sample results.}
Tables~\ref{tab:supp-semi-n250-p5-coverage}--\ref{tab:supp-semi-n500-p10-length} give the complete numerical results. For each $(n,p)$ combination, coverage variance and mean interval length variance are reported for $m/n=1,5,$ and $10$, with the two criteria combined in one table. Variance entries are ratios relative to SC using the same data in each replication. The final column reports mean coverage in the coverage variance block and mean interval length relative to SC in the mean interval length block.

\begin{table}[!htbp]
  \centering
  \linespread{1}\selectfont
  \setlength{\belowcaptionskip}{6pt}
  \caption{Coverage variance and mean interval length variance results for semi-supervised calibration at $n=250$ and $p=5$. Variance entries are ratios relative to SC.}
  \label{tab:supp-semi-n250-p5-coverage}
  \label{tab:supp-semi-n250-p5-length}
  \linespread{1}\fontsize{8.5}{9.5}\selectfont
  \setlength{\tabcolsep}{3.2pt}
  \renewcommand{\arraystretch}{1}
  \begin{tabular*}{\textwidth}{@{\extracolsep{\fill}}clrrrrrrrrr@{}}
    \toprule
    $m/n$ & Method & DGP1 & DGP2 & DGP3 & DGP4 & DGP5 & DGP6 & DGP7 & DGP8 & Mean \\
    \midrule
    \multicolumn{11}{@{}l}{\makebox[0pt][l]{\textit{Coverage variance ratio relative to SC (mean coverage in the final column)}}} \\
    \addlinespace[1pt]
    1 & SC & 1.000 & 1.000 & 1.000 & 1.000 & 1.000 & 1.000 & 1.000 & 1.000 & 0.900 \\
    & SS1 & 0.894 & 0.815 & 0.890 & 0.949 & 0.961 & 0.947 & 0.966 & 0.993 & 0.902 \\
    & SS2 & 0.880 & 0.848 & 0.921 & 0.788 & 0.838 & 0.891 & 0.864 & 1.008 & 0.902 \\
    & SS3 & 0.866 & 0.861 & 0.909 & 0.794 & 0.806 & 0.934 & 0.885 & 0.975 & 0.902 \\
    & SS4 & 0.883 & 0.814 & 0.896 & 0.786 & 0.851 & 0.897 & 0.864 & 1.001 & 0.902 \\
    & SS5 & 0.891 & 0.833 & 0.902 & 0.804 & 0.831 & 0.924 & 0.887 & 0.978 & 0.902 \\
    & SS6 & 0.873 & 0.835 & 0.913 & 0.793 & 0.843 & 0.899 & 0.888 & 0.986 & 0.902 \\
    & SS7 & 0.893 & 0.851 & 0.911 & 0.789 & 0.865 & 0.904 & 0.880 & 0.992 & 0.902 \\
    \addlinespace[1.5pt]
    5 & SC & 1.000 & 1.000 & 1.000 & 1.000 & 1.000 & 1.000 & 1.000 & 1.000 & 0.900 \\
    & SS1 & 0.803 & 0.747 & 0.838 & 0.960 & 0.971 & 0.937 & 0.946 & 0.992 & 0.902 \\
    & SS2 & 0.817 & 0.748 & 0.899 & 0.708 & 0.718 & 0.846 & 0.788 & 1.010 & 0.902 \\
    & SS3 & 0.812 & 0.784 & 0.863 & 0.710 & 0.705 & 0.879 & 0.806 & 0.984 & 0.902 \\
    & SS4 & 0.790 & 0.738 & 0.848 & 0.717 & 0.744 & 0.857 & 0.811 & 0.996 & 0.902 \\
    & SS5 & 0.776 & 0.765 & 0.830 & 0.724 & 0.728 & 0.875 & 0.801 & 1.007 & 0.902 \\
    & SS6 & 0.787 & 0.773 & 0.921 & 0.719 & 0.731 & 0.844 & 0.817 & 0.997 & 0.902 \\
    & SS7 & 0.807 & 0.781 & 0.875 & 0.730 & 0.750 & 0.843 & 0.817 & 1.000 & 0.902 \\
    \addlinespace[1.5pt]
    10 & SC & 1.000 & 1.000 & 1.000 & 1.000 & 1.000 & 1.000 & 1.000 & 1.000 & 0.900 \\
    & SS1 & 0.828 & 0.737 & 0.824 & 0.956 & 0.953 & 0.935 & 0.953 & 1.007 & 0.902 \\
    & SS2 & 0.837 & 0.764 & 0.916 & 0.665 & 0.692 & 0.855 & 0.743 & 1.007 & 0.902 \\
    & SS3 & 0.846 & 0.794 & 0.855 & 0.678 & 0.682 & 0.875 & 0.796 & 0.992 & 0.902 \\
    & SS4 & 0.822 & 0.740 & 0.864 & 0.677 & 0.730 & 0.850 & 0.765 & 0.996 & 0.903 \\
    & SS5 & 0.817 & 0.739 & 0.825 & 0.687 & 0.696 & 0.871 & 0.781 & 1.016 & 0.902 \\
    & SS6 & 0.803 & 0.780 & 0.908 & 0.697 & 0.703 & 0.837 & 0.776 & 0.990 & 0.902 \\
    & SS7 & 0.827 & 0.776 & 0.877 & 0.683 & 0.715 & 0.843 & 0.773 & 1.002 & 0.902 \\
    \midrule
    \multicolumn{11}{@{}l}{\makebox[0pt][l]{\textit{Mean interval length variance ratio relative to SC (mean length relative to SC in the final column)}}} \\
    \addlinespace[1pt]
    1 & SC & 1.000 & 1.000 & 1.000 & 1.000 & 1.000 & 1.000 & 1.000 & 1.000 & 1.000 \\
    & SS1 & 0.917 & 0.808 & 0.923 & 1.019 & 1.017 & 0.987 & 1.007 & 1.010 & 1.011 \\
    & SS2 & 0.899 & 0.845 & 0.960 & 0.833 & 0.900 & 0.929 & 0.861 & 1.038 & 1.010 \\
    & SS3 & 0.887 & 0.868 & 0.949 & 0.837 & 0.863 & 0.963 & 0.876 & 0.997 & 1.009 \\
    & SS4 & 0.919 & 0.806 & 0.929 & 0.838 & 0.912 & 0.932 & 0.865 & 1.033 & 1.010 \\
    & SS5 & 0.915 & 0.822 & 0.948 & 0.847 & 0.877 & 0.950 & 0.886 & 1.004 & 1.009 \\
    & SS6 & 0.905 & 0.824 & 0.962 & 0.834 & 0.896 & 0.946 & 0.884 & 1.009 & 1.009 \\
    & SS7 & 0.933 & 0.854 & 0.953 & 0.836 & 0.899 & 0.944 & 0.875 & 1.021 & 1.010 \\
    \addlinespace[1.5pt]
    5 & SC & 1.000 & 1.000 & 1.000 & 1.000 & 1.000 & 1.000 & 1.000 & 1.000 & 1.000 \\
    & SS1 & 0.827 & 0.723 & 0.850 & 1.037 & 1.027 & 0.981 & 0.995 & 1.014 & 1.010 \\
    & SS2 & 0.840 & 0.710 & 0.914 & 0.738 & 0.751 & 0.870 & 0.795 & 1.046 & 1.010 \\
    & SS3 & 0.835 & 0.756 & 0.884 & 0.745 & 0.743 & 0.921 & 0.824 & 1.008 & 1.009 \\
    & SS4 & 0.812 & 0.701 & 0.862 & 0.747 & 0.773 & 0.883 & 0.816 & 1.036 & 1.010 \\
    & SS5 & 0.798 & 0.735 & 0.833 & 0.753 & 0.769 & 0.927 & 0.817 & 1.033 & 1.009 \\
    & SS6 & 0.814 & 0.749 & 0.943 & 0.758 & 0.764 & 0.889 & 0.819 & 1.031 & 1.009 \\
    & SS7 & 0.834 & 0.777 & 0.878 & 0.768 & 0.781 & 0.900 & 0.832 & 1.039 & 1.010 \\
    \addlinespace[1.5pt]
    10 & SC & 1.000 & 1.000 & 1.000 & 1.000 & 1.000 & 1.000 & 1.000 & 1.000 & 1.000 \\
    & SS1 & 0.852 & 0.719 & 0.817 & 1.024 & 1.011 & 0.976 & 0.997 & 1.031 & 1.011 \\
    & SS2 & 0.883 & 0.747 & 0.941 & 0.691 & 0.729 & 0.875 & 0.752 & 1.046 & 1.010 \\
    & SS3 & 0.872 & 0.780 & 0.870 & 0.699 & 0.729 & 0.900 & 0.796 & 1.014 & 1.009 \\
    & SS4 & 0.869 & 0.740 & 0.878 & 0.697 & 0.768 & 0.861 & 0.761 & 1.035 & 1.011 \\
    & SS5 & 0.854 & 0.725 & 0.826 & 0.710 & 0.720 & 0.917 & 0.783 & 1.031 & 1.010 \\
    & SS6 & 0.830 & 0.775 & 0.950 & 0.739 & 0.736 & 0.868 & 0.773 & 1.023 & 1.010 \\
    & SS7 & 0.872 & 0.765 & 0.897 & 0.721 & 0.749 & 0.885 & 0.783 & 1.037 & 1.010 \\
    \bottomrule
  \end{tabular*}
\end{table}

\begin{table}[!htbp]
  \centering
  \linespread{1}\selectfont
  \setlength{\belowcaptionskip}{6pt}
  \caption{Coverage variance and mean interval length variance results for semi-supervised calibration at $n=250$ and $p=10$. Variance entries are ratios relative to SC.}
  \label{tab:supp-semi-n250-p10-coverage}
  \label{tab:supp-semi-n250-p10-length}
  \linespread{1}\fontsize{8.5}{9.5}\selectfont
  \setlength{\tabcolsep}{3.2pt}
  \renewcommand{\arraystretch}{1}
  \begin{tabular*}{\textwidth}{@{\extracolsep{\fill}}clrrrrrrrrr@{}}
    \toprule
    $m/n$ & Method & DGP1 & DGP2 & DGP3 & DGP4 & DGP5 & DGP6 & DGP7 & DGP8 & Mean \\
    \midrule
    \multicolumn{11}{@{}l}{\makebox[0pt][l]{\textit{Coverage variance ratio relative to SC (mean coverage in the final column)}}} \\
    \addlinespace[1pt]
    1 & SC & 1.000 & 1.000 & 1.000 & 1.000 & 1.000 & 1.000 & 1.000 & 1.000 & 0.901 \\
    & SS1 & 0.877 & 0.842 & 0.935 & 1.023 & 0.979 & 0.970 & 0.984 & 1.000 & 0.902 \\
    & SS2 & 0.893 & 0.870 & 0.938 & 0.868 & 0.859 & 0.908 & 0.871 & 1.006 & 0.902 \\
    & SS3 & 0.936 & 0.831 & 0.947 & 0.866 & 0.857 & 0.951 & 0.872 & 1.007 & 0.902 \\
    & SS4 & 0.880 & 0.855 & 0.933 & 0.852 & 0.852 & 0.884 & 0.845 & 0.998 & 0.902 \\
    & SS5 & 0.883 & 0.819 & 0.916 & 0.848 & 0.863 & 0.932 & 0.872 & 1.013 & 0.902 \\
    & SS6 & 0.865 & 0.854 & 0.935 & 0.864 & 0.855 & 0.896 & 0.867 & 1.001 & 0.902 \\
    & SS7 & 0.869 & 0.848 & 0.947 & 0.868 & 0.872 & 0.879 & 0.868 & 1.006 & 0.902 \\
    \addlinespace[1.5pt]
    5 & SC & 1.000 & 1.000 & 1.000 & 1.000 & 1.000 & 1.000 & 1.000 & 1.000 & 0.901 \\
    & SS1 & 0.807 & 0.723 & 0.890 & 1.020 & 0.977 & 0.985 & 0.992 & 1.005 & 0.902 \\
    & SS2 & 0.821 & 0.750 & 0.926 & 0.808 & 0.672 & 0.815 & 0.821 & 0.984 & 0.902 \\
    & SS3 & 0.901 & 0.733 & 0.917 & 0.826 & 0.684 & 0.878 & 0.875 & 0.987 & 0.902 \\
    & SS4 & 0.814 & 0.735 & 0.947 & 0.795 & 0.677 & 0.820 & 0.844 & 0.988 & 0.902 \\
    & SS5 & 0.832 & 0.721 & 0.902 & 0.826 & 0.719 & 0.887 & 0.861 & 1.018 & 0.902 \\
    & SS6 & 0.816 & 0.730 & 0.918 & 0.811 & 0.691 & 0.821 & 0.847 & 0.988 & 0.902 \\
    & SS7 & 0.829 & 0.730 & 0.937 & 0.805 & 0.684 & 0.825 & 0.848 & 1.006 & 0.902 \\
    \addlinespace[1.5pt]
    10 & SC & 1.000 & 1.000 & 1.000 & 1.000 & 1.000 & 1.000 & 1.000 & 1.000 & 0.901 \\
    & SS1 & 0.815 & 0.680 & 0.864 & 1.013 & 0.967 & 1.000 & 1.002 & 0.993 & 0.902 \\
    & SS2 & 0.830 & 0.719 & 0.882 & 0.780 & 0.635 & 0.796 & 0.816 & 0.982 & 0.902 \\
    & SS3 & 0.923 & 0.719 & 0.889 & 0.792 & 0.649 & 0.871 & 0.867 & 0.994 & 0.902 \\
    & SS4 & 0.830 & 0.706 & 0.893 & 0.772 & 0.637 & 0.821 & 0.834 & 1.010 & 0.902 \\
    & SS5 & 0.837 & 0.678 & 0.859 & 0.798 & 0.660 & 0.876 & 0.827 & 1.023 & 0.902 \\
    & SS6 & 0.828 & 0.708 & 0.884 & 0.780 & 0.634 & 0.815 & 0.821 & 1.005 & 0.902 \\
    & SS7 & 0.823 & 0.696 & 0.882 & 0.785 & 0.667 & 0.834 & 0.824 & 1.022 & 0.902 \\
    \midrule
    \multicolumn{11}{@{}l}{\makebox[0pt][l]{\textit{Mean interval length variance ratio relative to SC (mean length relative to SC in the final column)}}} \\
    \addlinespace[1pt]
    1 & SC & 1.000 & 1.000 & 1.000 & 1.000 & 1.000 & 1.000 & 1.000 & 1.000 & 1.000 \\
    & SS1 & 0.891 & 0.843 & 0.951 & 1.092 & 1.028 & 1.008 & 1.043 & 1.030 & 1.009 \\
    & SS2 & 0.941 & 0.886 & 0.949 & 0.893 & 0.868 & 0.929 & 0.863 & 1.018 & 1.007 \\
    & SS3 & 0.962 & 0.866 & 0.955 & 0.882 & 0.868 & 0.987 & 0.882 & 1.021 & 1.008 \\
    & SS4 & 0.914 & 0.861 & 0.937 & 0.866 & 0.866 & 0.904 & 0.840 & 1.026 & 1.007 \\
    & SS5 & 0.907 & 0.852 & 0.926 & 0.852 & 0.883 & 0.954 & 0.889 & 1.036 & 1.008 \\
    & SS6 & 0.915 & 0.877 & 0.933 & 0.875 & 0.866 & 0.923 & 0.875 & 1.017 & 1.008 \\
    & SS7 & 0.913 & 0.877 & 0.942 & 0.884 & 0.877 & 0.900 & 0.872 & 1.024 & 1.008 \\
    \addlinespace[1.5pt]
    5 & SC & 1.000 & 1.000 & 1.000 & 1.000 & 1.000 & 1.000 & 1.000 & 1.000 & 1.000 \\
    & SS1 & 0.798 & 0.722 & 0.891 & 1.086 & 1.014 & 1.037 & 1.054 & 1.030 & 1.008 \\
    & SS2 & 0.837 & 0.755 & 0.915 & 0.805 & 0.671 & 0.842 & 0.826 & 1.005 & 1.007 \\
    & SS3 & 0.902 & 0.755 & 0.921 & 0.804 & 0.708 & 0.906 & 0.935 & 1.004 & 1.007 \\
    & SS4 & 0.818 & 0.751 & 0.927 & 0.793 & 0.682 & 0.856 & 0.861 & 1.015 & 1.007 \\
    & SS5 & 0.829 & 0.750 & 0.904 & 0.819 & 0.732 & 0.904 & 0.914 & 1.040 & 1.007 \\
    & SS6 & 0.828 & 0.764 & 0.899 & 0.794 & 0.686 & 0.853 & 0.877 & 1.007 & 1.007 \\
    & SS7 & 0.839 & 0.762 & 0.914 & 0.800 & 0.672 & 0.850 & 0.884 & 1.025 & 1.007 \\
    \addlinespace[1.5pt]
    10 & SC & 1.000 & 1.000 & 1.000 & 1.000 & 1.000 & 1.000 & 1.000 & 1.000 & 1.000 \\
    & SS1 & 0.805 & 0.668 & 0.860 & 1.066 & 1.019 & 1.039 & 1.058 & 1.022 & 1.009 \\
    & SS2 & 0.842 & 0.727 & 0.875 & 0.760 & 0.626 & 0.820 & 0.810 & 1.002 & 1.007 \\
    & SS3 & 0.925 & 0.734 & 0.883 & 0.756 & 0.655 & 0.903 & 0.913 & 1.011 & 1.007 \\
    & SS4 & 0.825 & 0.696 & 0.874 & 0.752 & 0.628 & 0.859 & 0.828 & 1.034 & 1.006 \\
    & SS5 & 0.835 & 0.677 & 0.851 & 0.777 & 0.678 & 0.900 & 0.886 & 1.044 & 1.007 \\
    & SS6 & 0.848 & 0.719 & 0.868 & 0.753 & 0.620 & 0.848 & 0.838 & 1.017 & 1.007 \\
    & SS7 & 0.835 & 0.697 & 0.861 & 0.769 & 0.652 & 0.872 & 0.837 & 1.044 & 1.007 \\
    \bottomrule
  \end{tabular*}
\end{table}

\begin{table}[!htbp]
  \centering
  \linespread{1}\selectfont
  \setlength{\belowcaptionskip}{6pt}
  \caption{Coverage variance and mean interval length variance results for semi-supervised calibration at $n=500$ and $p=5$. Variance entries are ratios relative to SC.}
  \label{tab:supp-semi-n500-p5-coverage}
  \label{tab:supp-semi-n500-p5-length}
  \linespread{1}\fontsize{8.5}{9.5}\selectfont
  \setlength{\tabcolsep}{3.2pt}
  \renewcommand{\arraystretch}{1}
  \begin{tabular*}{\textwidth}{@{\extracolsep{\fill}}clrrrrrrrrr@{}}
    \toprule
    $m/n$ & Method & DGP1 & DGP2 & DGP3 & DGP4 & DGP5 & DGP6 & DGP7 & DGP8 & Mean \\
    \midrule
    \multicolumn{11}{@{}l}{\makebox[0pt][l]{\textit{Coverage variance ratio relative to SC (mean coverage in the final column)}}} \\
    \addlinespace[1pt]
    1 & SC & 1.000 & 1.000 & 1.000 & 1.000 & 1.000 & 1.000 & 1.000 & 1.000 & 0.900 \\
    & SS1 & 0.881 & 0.832 & 0.867 & 0.971 & 0.986 & 0.979 & 1.035 & 0.994 & 0.901 \\
    & SS2 & 0.901 & 0.815 & 0.869 & 0.821 & 0.788 & 0.922 & 0.932 & 1.001 & 0.901 \\
    & SS3 & 0.901 & 0.823 & 0.874 & 0.819 & 0.789 & 0.925 & 0.904 & 1.009 & 0.901 \\
    & SS4 & 0.896 & 0.806 & 0.882 & 0.819 & 0.791 & 0.923 & 0.919 & 0.986 & 0.901 \\
    & SS5 & 0.899 & 0.815 & 0.879 & 0.824 & 0.781 & 0.920 & 0.901 & 0.992 & 0.901 \\
    & SS6 & 0.900 & 0.818 & 0.872 & 0.832 & 0.782 & 0.937 & 0.915 & 0.985 & 0.901 \\
    & SS7 & 0.903 & 0.803 & 0.881 & 0.823 & 0.787 & 0.923 & 0.912 & 0.976 & 0.901 \\
    \addlinespace[1.5pt]
    5 & SC & 1.000 & 1.000 & 1.000 & 1.000 & 1.000 & 1.000 & 1.000 & 1.000 & 0.900 \\
    & SS1 & 0.793 & 0.796 & 0.826 & 0.946 & 0.964 & 0.949 & 1.013 & 1.012 & 0.901 \\
    & SS2 & 0.799 & 0.764 & 0.817 & 0.683 & 0.654 & 0.896 & 0.777 & 1.007 & 0.901 \\
    & SS3 & 0.792 & 0.767 & 0.844 & 0.693 & 0.656 & 0.885 & 0.758 & 1.005 & 0.901 \\
    & SS4 & 0.799 & 0.751 & 0.812 & 0.679 & 0.663 & 0.899 & 0.781 & 1.008 & 0.901 \\
    & SS5 & 0.797 & 0.772 & 0.848 & 0.696 & 0.666 & 0.882 & 0.776 & 0.979 & 0.901 \\
    & SS6 & 0.793 & 0.768 & 0.823 & 0.691 & 0.647 & 0.908 & 0.767 & 0.985 & 0.901 \\
    & SS7 & 0.803 & 0.751 & 0.826 & 0.702 & 0.658 & 0.906 & 0.775 & 0.983 & 0.901 \\
    \addlinespace[1.5pt]
    10 & SC & 1.000 & 1.000 & 1.000 & 1.000 & 1.000 & 1.000 & 1.000 & 1.000 & 0.900 \\
    & SS1 & 0.782 & 0.787 & 0.830 & 0.947 & 0.972 & 0.966 & 1.013 & 1.006 & 0.901 \\
    & SS2 & 0.762 & 0.756 & 0.810 & 0.633 & 0.633 & 0.892 & 0.763 & 1.001 & 0.901 \\
    & SS3 & 0.790 & 0.779 & 0.850 & 0.651 & 0.624 & 0.900 & 0.742 & 0.996 & 0.901 \\
    & SS4 & 0.780 & 0.739 & 0.823 & 0.637 & 0.641 & 0.892 & 0.758 & 1.002 & 0.901 \\
    & SS5 & 0.789 & 0.775 & 0.825 & 0.664 & 0.619 & 0.889 & 0.761 & 0.979 & 0.901 \\
    & SS6 & 0.786 & 0.773 & 0.822 & 0.656 & 0.614 & 0.899 & 0.746 & 0.986 & 0.901 \\
    & SS7 & 0.803 & 0.769 & 0.837 & 0.656 & 0.625 & 0.908 & 0.753 & 0.979 & 0.901 \\
    \midrule
    \multicolumn{11}{@{}l}{\makebox[0pt][l]{\textit{Mean interval length variance ratio relative to SC (mean length relative to SC in the final column)}}} \\
    \addlinespace[1pt]
    1 & SC & 1.000 & 1.000 & 1.000 & 1.000 & 1.000 & 1.000 & 1.000 & 1.000 & 1.000 \\
    & SS1 & 0.883 & 0.780 & 0.889 & 1.007 & 1.012 & 0.992 & 1.051 & 1.004 & 1.005 \\
    & SS2 & 0.905 & 0.749 & 0.898 & 0.829 & 0.797 & 0.915 & 0.954 & 1.012 & 1.003 \\
    & SS3 & 0.892 & 0.770 & 0.894 & 0.822 & 0.784 & 0.901 & 0.937 & 1.026 & 1.004 \\
    & SS4 & 0.898 & 0.740 & 0.909 & 0.829 & 0.803 & 0.917 & 0.941 & 1.002 & 1.003 \\
    & SS5 & 0.898 & 0.764 & 0.892 & 0.821 & 0.792 & 0.902 & 0.927 & 1.005 & 1.004 \\
    & SS6 & 0.898 & 0.746 & 0.897 & 0.833 & 0.785 & 0.930 & 0.939 & 1.007 & 1.004 \\
    & SS7 & 0.906 & 0.732 & 0.904 & 0.826 & 0.794 & 0.911 & 0.937 & 0.993 & 1.004 \\
    \addlinespace[1.5pt]
    5 & SC & 1.000 & 1.000 & 1.000 & 1.000 & 1.000 & 1.000 & 1.000 & 1.000 & 1.000 \\
    & SS1 & 0.786 & 0.731 & 0.851 & 0.976 & 0.990 & 0.955 & 1.029 & 1.022 & 1.005 \\
    & SS2 & 0.809 & 0.679 & 0.838 & 0.686 & 0.645 & 0.871 & 0.782 & 1.020 & 1.004 \\
    & SS3 & 0.795 & 0.705 & 0.877 & 0.693 & 0.647 & 0.858 & 0.769 & 1.018 & 1.004 \\
    & SS4 & 0.801 & 0.670 & 0.823 & 0.682 & 0.653 & 0.877 & 0.786 & 1.021 & 1.004 \\
    & SS5 & 0.791 & 0.707 & 0.874 & 0.699 & 0.650 & 0.856 & 0.780 & 0.997 & 1.004 \\
    & SS6 & 0.797 & 0.682 & 0.842 & 0.697 & 0.642 & 0.883 & 0.774 & 0.999 & 1.004 \\
    & SS7 & 0.802 & 0.669 & 0.844 & 0.701 & 0.646 & 0.878 & 0.779 & 1.002 & 1.004 \\
    \addlinespace[1.5pt]
    10 & SC & 1.000 & 1.000 & 1.000 & 1.000 & 1.000 & 1.000 & 1.000 & 1.000 & 1.000 \\
    & SS1 & 0.762 & 0.739 & 0.842 & 0.978 & 1.009 & 0.968 & 1.034 & 1.014 & 1.005 \\
    & SS2 & 0.759 & 0.664 & 0.825 & 0.635 & 0.619 & 0.861 & 0.765 & 1.018 & 1.004 \\
    & SS3 & 0.766 & 0.704 & 0.871 & 0.648 & 0.615 & 0.874 & 0.744 & 1.013 & 1.004 \\
    & SS4 & 0.779 & 0.648 & 0.834 & 0.635 & 0.644 & 0.858 & 0.759 & 1.017 & 1.004 \\
    & SS5 & 0.770 & 0.713 & 0.829 & 0.655 & 0.607 & 0.866 & 0.766 & 0.996 & 1.004 \\
    & SS6 & 0.775 & 0.685 & 0.838 & 0.649 & 0.609 & 0.866 & 0.749 & 1.003 & 1.004 \\
    & SS7 & 0.797 & 0.678 & 0.842 & 0.656 & 0.616 & 0.875 & 0.757 & 0.999 & 1.004 \\
    \bottomrule
  \end{tabular*}
\end{table}

\begin{table}[!htbp]
  \centering
  \linespread{1}\selectfont
  \setlength{\belowcaptionskip}{6pt}
  \caption{Coverage variance and mean interval length variance results for semi-supervised calibration at $n=500$ and $p=10$. Variance entries are ratios relative to SC.}
  \label{tab:supp-semi-n500-p10-coverage}
  \label{tab:supp-semi-n500-p10-length}
  \linespread{1}\fontsize{8.5}{9.5}\selectfont
  \setlength{\tabcolsep}{3.2pt}
  \renewcommand{\arraystretch}{1}
  \begin{tabular*}{\textwidth}{@{\extracolsep{\fill}}clrrrrrrrrr@{}}
    \toprule
    $m/n$ & Method & DGP1 & DGP2 & DGP3 & DGP4 & DGP5 & DGP6 & DGP7 & DGP8 & Mean \\
    \midrule
    \multicolumn{11}{@{}l}{\makebox[0pt][l]{\textit{Coverage variance ratio relative to SC (mean coverage in the final column)}}} \\
    \addlinespace[1pt]
    1 & SC & 1.000 & 1.000 & 1.000 & 1.000 & 1.000 & 1.000 & 1.000 & 1.000 & 0.900 \\
    & SS1 & 0.895 & 0.870 & 0.936 & 0.984 & 0.949 & 1.005 & 0.989 & 0.993 & 0.900 \\
    & SS2 & 0.885 & 0.863 & 0.946 & 0.830 & 0.711 & 0.959 & 0.914 & 1.004 & 0.901 \\
    & SS3 & 0.903 & 0.870 & 0.943 & 0.816 & 0.725 & 0.949 & 0.898 & 0.990 & 0.901 \\
    & SS4 & 0.890 & 0.869 & 0.940 & 0.833 & 0.702 & 0.956 & 0.900 & 0.988 & 0.901 \\
    & SS5 & 0.902 & 0.858 & 0.942 & 0.825 & 0.716 & 0.940 & 0.896 & 0.989 & 0.901 \\
    & SS6 & 0.905 & 0.855 & 0.950 & 0.836 & 0.705 & 0.943 & 0.893 & 0.992 & 0.901 \\
    & SS7 & 0.904 & 0.866 & 0.946 & 0.837 & 0.699 & 0.950 & 0.900 & 0.993 & 0.901 \\
    \addlinespace[1.5pt]
    5 & SC & 1.000 & 1.000 & 1.000 & 1.000 & 1.000 & 1.000 & 1.000 & 1.000 & 0.900 \\
    & SS1 & 0.834 & 0.767 & 0.827 & 0.972 & 0.926 & 0.994 & 1.002 & 0.991 & 0.901 \\
    & SS2 & 0.811 & 0.722 & 0.843 & 0.707 & 0.598 & 0.863 & 0.852 & 1.007 & 0.901 \\
    & SS3 & 0.811 & 0.732 & 0.843 & 0.706 & 0.622 & 0.877 & 0.848 & 0.990 & 0.901 \\
    & SS4 & 0.829 & 0.727 & 0.841 & 0.704 & 0.601 & 0.870 & 0.854 & 1.007 & 0.901 \\
    & SS5 & 0.814 & 0.728 & 0.820 & 0.696 & 0.626 & 0.880 & 0.857 & 0.989 & 0.901 \\
    & SS6 & 0.808 & 0.738 & 0.837 & 0.680 & 0.610 & 0.867 & 0.843 & 0.993 & 0.901 \\
    & SS7 & 0.822 & 0.744 & 0.827 & 0.682 & 0.615 & 0.877 & 0.843 & 1.000 & 0.901 \\
    \addlinespace[1.5pt]
    10 & SC & 1.000 & 1.000 & 1.000 & 1.000 & 1.000 & 1.000 & 1.000 & 1.000 & 0.900 \\
    & SS1 & 0.800 & 0.735 & 0.791 & 0.954 & 0.929 & 0.988 & 1.002 & 0.989 & 0.901 \\
    & SS2 & 0.782 & 0.706 & 0.811 & 0.689 & 0.599 & 0.871 & 0.828 & 1.006 & 0.901 \\
    & SS3 & 0.789 & 0.703 & 0.819 & 0.681 & 0.613 & 0.880 & 0.817 & 0.984 & 0.901 \\
    & SS4 & 0.782 & 0.703 & 0.799 & 0.681 & 0.590 & 0.890 & 0.824 & 1.004 & 0.901 \\
    & SS5 & 0.792 & 0.720 & 0.802 & 0.686 & 0.623 & 0.887 & 0.826 & 0.991 & 0.901 \\
    & SS6 & 0.794 & 0.709 & 0.804 & 0.684 & 0.611 & 0.887 & 0.817 & 0.993 & 0.901 \\
    & SS7 & 0.796 & 0.705 & 0.794 & 0.674 & 0.606 & 0.886 & 0.825 & 0.996 & 0.901 \\
    \midrule
    \multicolumn{11}{@{}l}{\makebox[0pt][l]{\textit{Mean interval length variance ratio relative to SC (mean length relative to SC in the final column)}}} \\
    \addlinespace[1pt]
    1 & SC & 1.000 & 1.000 & 1.000 & 1.000 & 1.000 & 1.000 & 1.000 & 1.000 & 1.000 \\
    & SS1 & 0.911 & 0.842 & 0.938 & 1.027 & 0.970 & 1.032 & 1.011 & 1.002 & 1.005 \\
    & SS2 & 0.902 & 0.826 & 0.937 & 0.825 & 0.750 & 0.959 & 0.955 & 1.012 & 1.005 \\
    & SS3 & 0.920 & 0.836 & 0.931 & 0.796 & 0.761 & 0.953 & 0.944 & 1.005 & 1.005 \\
    & SS4 & 0.910 & 0.833 & 0.933 & 0.824 & 0.744 & 0.965 & 0.940 & 0.997 & 1.005 \\
    & SS5 & 0.919 & 0.825 & 0.939 & 0.798 & 0.756 & 0.947 & 0.935 & 1.004 & 1.005 \\
    & SS6 & 0.919 & 0.819 & 0.935 & 0.828 & 0.743 & 0.954 & 0.930 & 1.009 & 1.005 \\
    & SS7 & 0.921 & 0.831 & 0.936 & 0.824 & 0.737 & 0.966 & 0.930 & 1.009 & 1.005 \\
    \addlinespace[1.5pt]
    5 & SC & 1.000 & 1.000 & 1.000 & 1.000 & 1.000 & 1.000 & 1.000 & 1.000 & 1.000 \\
    & SS1 & 0.819 & 0.763 & 0.837 & 1.017 & 0.960 & 1.010 & 1.014 & 1.001 & 1.005 \\
    & SS2 & 0.802 & 0.704 & 0.855 & 0.706 & 0.614 & 0.870 & 0.875 & 1.020 & 1.007 \\
    & SS3 & 0.802 & 0.713 & 0.863 & 0.683 & 0.658 & 0.883 & 0.880 & 1.004 & 1.006 \\
    & SS4 & 0.818 & 0.711 & 0.846 & 0.697 & 0.617 & 0.877 & 0.880 & 1.018 & 1.006 \\
    & SS5 & 0.801 & 0.720 & 0.833 & 0.672 & 0.666 & 0.891 & 0.891 & 1.004 & 1.006 \\
    & SS6 & 0.799 & 0.726 & 0.841 & 0.672 & 0.624 & 0.875 & 0.873 & 1.014 & 1.006 \\
    & SS7 & 0.812 & 0.733 & 0.832 & 0.670 & 0.635 & 0.885 & 0.872 & 1.012 & 1.006 \\
    \addlinespace[1.5pt]
    10 & SC & 1.000 & 1.000 & 1.000 & 1.000 & 1.000 & 1.000 & 1.000 & 1.000 & 1.000 \\
    & SS1 & 0.792 & 0.723 & 0.808 & 0.988 & 0.970 & 1.003 & 1.010 & 0.998 & 1.005 \\
    & SS2 & 0.787 & 0.679 & 0.817 & 0.672 & 0.606 & 0.869 & 0.836 & 1.021 & 1.006 \\
    & SS3 & 0.781 & 0.679 & 0.836 & 0.640 & 0.630 & 0.883 & 0.835 & 0.998 & 1.006 \\
    & SS4 & 0.781 & 0.676 & 0.804 & 0.662 & 0.610 & 0.894 & 0.839 & 1.012 & 1.006 \\
    & SS5 & 0.783 & 0.706 & 0.816 & 0.653 & 0.645 & 0.897 & 0.844 & 1.007 & 1.006 \\
    & SS6 & 0.789 & 0.685 & 0.809 & 0.669 & 0.620 & 0.892 & 0.831 & 1.011 & 1.006 \\
    & SS7 & 0.791 & 0.682 & 0.801 & 0.647 & 0.619 & 0.892 & 0.838 & 1.009 & 1.006 \\
    \bottomrule
  \end{tabular*}
\end{table}

Averaged over the eight DGPs and twelve sample configurations, SS1 has coverage variance ratio $0.921$, whereas SS2 and SS3 have ratios $0.831$ and $0.838$. The combined specifications SS4--SS7 have averages between $0.829$ and $0.834$, a range of $0.005$. The additional coordinates increase the asymptotic variance reduction, but no combined specification has the smallest finite-sample coverage variance ratio in every DGP--configuration combination.

For SS2--SS7, the method-specific average ratios range from $0.881$ to $0.889$ at $m/n=1$ and from $0.796$ to $0.808$ at $m/n=10$. Each of these methods benefits on average from the larger unlabeled sample. Changing $p$ from 5 to 10 changes each method-specific average by at most $0.015$. Across individual DGP--configuration combinations, mean coverage is between $0.900$ and $0.904$ for SS1--SS7, and their relative mean lengths range from $1.000$ to $1.021$. The method-specific average mean interval length variance ratios range from $0.837$ to $0.941$ and follow the same DGP-level patterns as the coverage variance ratios.

\FloatBarrier

\subsubsection{Sensitivity to Noise in the Covariate Representation}
\label{sec:supp-semi-quality}

We next vary the informativeness of the covariate representation while holding the sample configuration fixed. The experiment uses DGP1 with $n=250$, $p=10$, and $m=2500$. Let $\widetilde A(X)$ be the coordinatewise standardization of the covariate representation $A(X)$ in Section~5.2, using fixed means and standard deviations from the fixed independent training data. For $\eta\in\{0,0.2,0.4,0.6,0.8,1\}$, define
\[
  A_\eta(X)=\eta\widetilde A(X)+(1-\eta^2)^{1/2}Z, \qquad Z\sim N(0,I_5).
\]
Independent noise vectors are generated for the calibration and unlabeled data points, and the same draws are reused across the six values of $\eta$ within a replication. Thus $\eta=0$ gives an uninformative covariate representation, whereas $\eta=1$ recovers the standardized representation. We apply SS1, SS2, SS3, and SS7 to $A_\eta$ in place of $A$ and centre each resulting coordinate at its mean over the pooled calibration and unlabeled data.

\begin{figure}[!htbp]
  \centering
  \linespread{1}\selectfont
  \revisedfigure{\textwidth}{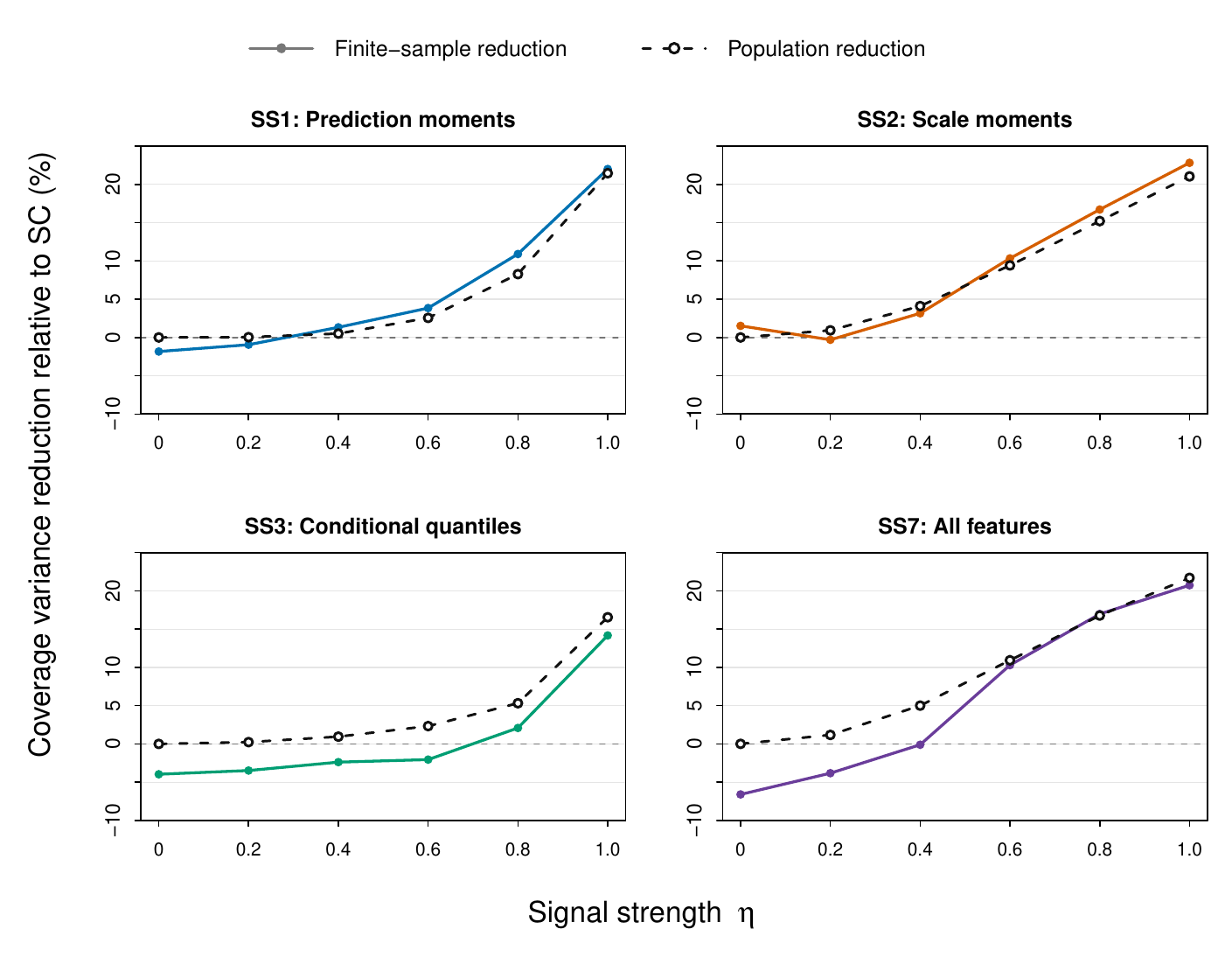}
  \caption{Relative reductions in the limiting variance of training-conditional coverage and in empirical coverage variance with noise in the covariate representation under DGP1 at $n=250$, $p=10$, and $m=2500$. Each panel corresponds to one specification.}
  \label{fig:supp-semi-quality-eta}
\end{figure}

Table~\ref{tab:supp-semi-information-quality} reports the asymptotic variance ratios for training-conditional coverage and the empirical coverage variance ratios and mean interval length variance ratios. Figure~\ref{fig:supp-semi-quality-eta} plots the relative reductions in the limiting variance of training-conditional coverage and in empirical coverage variance as functions of the signal strength $\eta$.

\begin{table}[!htbp]
  \centering
  \linespread{1}\selectfont
  \setlength{\belowcaptionskip}{6pt}
  \caption{Sensitivity to noise in the covariate representation under DGP1 at $n=250$, $p=10$, and $m=2500$. Entries are variance ratios relative to SC.}
  \label{tab:supp-semi-information-quality}
  \linespread{1}\fontsize{8.5}{9.5}\selectfont
  \setlength{\tabcolsep}{3.2pt}
  \renewcommand{\arraystretch}{1}
  \begin{tabular}{llcccccc}
    \toprule
    & & $\eta=0$ & $0.2$ & $0.4$ & $0.6$ & $0.8$ & $1.0$ \\
    \midrule
    \multicolumn{8}{l}{Asymptotic variance ratio for training-conditional coverage relative to SC} \\
    SS1 & Prediction moments & 1.000 & 1.000 & 0.995 & 0.974 & 0.917 & 0.785 \\
    SS2 & Residual scale moments & 1.000 & 0.991 & 0.959 & 0.906 & 0.848 & 0.790 \\
    SS3 & Conditional quantiles & 1.000 & 0.998 & 0.990 & 0.977 & 0.947 & 0.834 \\
    SS7 & All moment functions & 1.000 & 0.988 & 0.950 & 0.891 & 0.832 & 0.783 \\
    \midrule
    \multicolumn{8}{l}{Coverage variance ratio relative to SC} \\
    SS1 & Prediction moments & 1.018 & 1.009 & 0.987 & 0.962 & 0.891 & 0.780 \\
    SS2 & Residual scale moments & 0.985 & 1.003 & 0.968 & 0.897 & 0.833 & 0.772 \\
    SS3 & Conditional quantiles & 1.040 & 1.035 & 1.024 & 1.020 & 0.979 & 0.858 \\
    SS7 & All moment functions & 1.066 & 1.038 & 1.001 & 0.897 & 0.830 & 0.793 \\
    \midrule
    \multicolumn{8}{l}{Mean interval length variance ratio relative to SC} \\
    SS1 & Prediction moments & 1.058 & 1.038 & 1.022 & 0.986 & 0.923 & 0.828 \\
    SS2 & Residual scale moments & 1.012 & 1.048 & 1.030 & 0.952 & 0.887 & 0.837 \\
    SS3 & Conditional quantiles & 1.068 & 1.088 & 1.086 & 1.082 & 1.028 & 0.907 \\
    SS7 & All moment functions & 1.122 & 1.101 & 1.037 & 0.925 & 0.896 & 0.881 \\
    \bottomrule
  \end{tabular}
\end{table}

\FloatBarrier

At $\eta=0$, the asymptotic variance ratios equal one and the empirical coverage variance ratios range from $0.985$ to $1.066$. The finite-sample variance reductions generally increase with signal strength. At $\eta=1$, the asymptotic variance ratios range from $0.783$ to $0.834$, the empirical coverage variance ratios from $0.772$ to $0.858$, and the mean interval length variance ratios from $0.828$ to $0.907$. Mean coverage remains between $0.899$ and $0.902$ throughout. At $\eta=1$, SS7 has the smallest asymptotic variance ratio and SS2 has the smallest empirical coverage variance ratio.

\FloatBarrier

\subsection{Semi-Supervised Calibration with Learned Moment Functions}
\label{sec:supp-estimated-features}

This section provides the implementation details, asymptotic variance benchmarks, and finite-sample results for the additional settings considered in Section~5.3 of the main paper.

\paragraph{Cross-fitted moment functions.}
We implement the construction in Section~3.2 with five folds. The calibration and unlabeled indices are partitioned into corresponding folds. For each held-out calibration fold, the pilot conformal quantile, probabilistic classifier, and centring mean over the held-out calibration and unlabeled folds are computed as specified there. The same fold assignments are used for all three learners within a replication. Thus each calibration data point is excluded from the classifier used to compute its moment function value.

CF-Logit uses the five coordinates of $A$, their squares, and all pairwise interactions. The predictors are standardized using the calibration data outside the held-out fold, and the ridge penalty is selected by three-fold cross-validation over 18 logarithmically spaced values from $10^{1.5}$ to $10^{-4}$. CF-RF is a probability forest with the five coordinates of $A$ as predictors. Writing $n_{\rm tr}$ for the number of calibration data points outside the held-out fold, its two candidate specifications use $(m_{\rm try},n_{\rm leaf},B)=(2, \max\{8,\lfloor.06n_{\rm tr}\rfloor\},24)$ and $(3, \max\{12,\lfloor.10n_{\rm tr}\rfloor\},36)$, respectively. Trees have maximum depth three, and the specification is selected by out-of-bag logistic loss. CF-GB uses gradient-boosted classification trees with logistic loss. The candidate $(\text{depth},\text{learning rate})$ pairs are $(1,.12)$ and $(2,.07)$, and the number of iterations is selected from $\{10,20\}$ by two-fold cross-validated logistic loss. All tuning uses only calibration data points outside the held-out fold.

\paragraph{Asymptotic variance benchmarks.}
For a fitted scalar moment function $g(A)$, the semi-supervised projection reduction is
\[
  R_{g,\mathrm{ss}}^2=\frac{m}{n+m}\frac{\Cov^2\{\psi_\alpha(Z),g(A)\}}{\alpha(1-\alpha)\Var\{g(A)\}}, \qquad
  \psi_\alpha(Z)=\ind(S\le q_\alpha)-(1-\alpha).
\]
The reported asymptotic variance ratio for training-conditional coverage relative to SC is $1-R_{g,\mathrm{ss}}^2$. For SS4, the reduction is $R_{h,\mathrm{ss}}^2$ from Section~\ref{sec:supp-ss}. For each learned method, the criterion is evaluated on an independent sample of size 50,000 and averaged over 5000 independent calibration feature-learning fits under the fixed training fit. For the oracle transformation $h_\alpha^*(A)=\P\{S\le q_\alpha\mid A\}-(1-\alpha)$, the projection term reduces to $\Var\{h_\alpha^*(A)\}$. We approximate this transformation by regressing the DGP-implied conditional probability of $S\le q_\alpha$ on $A$ in an independent sample of size 50,000.
Evaluation uses a second independent sample of size 50,000. The resulting numerical oracle approximation is obtained from 5000 Monte Carlo repetitions per DGP and estimates the smallest asymptotic variance ratio attainable by a square-integrable scalar transformation of the chosen covariate representation.

Table~\ref{tab:supp-estimated-theory} reports these ratios for the main design. The asymptotic variance benchmarks give smaller ratios for SS4 and CF-RF than for CF-Logit and CF-GB. Their average ratios are $0.754$ and $0.769$, respectively, compared with $0.737$ for the numerical approximation based on $h_\alpha^*(A)$.

\begin{table}[!htbp]
  \centering
  \linespread{1}\selectfont
  \setlength{\belowcaptionskip}{6pt}
  \caption{Asymptotic variance ratios for training-conditional coverage relative to SC for SS4, the three learned moment functions, and a numerical approximation based on $h_\alpha^*(A)$ at $n=250$, $p=5$, and $m=2500$. The reported ratio is $1-R_{h,\mathrm{ss}}^2$ for SS4 and $1-R_{g,\mathrm{ss}}^2$ for the scalar moment functions.}
  \label{tab:supp-estimated-theory}
  \linespread{1}\fontsize{8.5}{9.5}\selectfont
  \setlength{\tabcolsep}{3.2pt}
  \renewcommand{\arraystretch}{1}
  \begin{tabular}{lccccccccc}
    \toprule
    Method
    & DGP1 & DGP2 & DGP3 & DGP4
    & DGP5 & DGP6 & DGP7 & DGP8 & Avg.\\
    \midrule
    SS4 & 0.739 & 0.681 & 0.772 & 0.617 & 0.631 & 0.829 & 0.762 & 1.000 & 0.754 \\
    CF-Logit & 0.787 & 0.698 & 0.808 & 0.691 & 0.686 & 0.874 & 0.804 & 1.000 & 0.793 \\
    CF-RF & 0.777 & 0.658 & 0.797 & 0.639 & 0.652 & 0.850 & 0.781 & 1.000 & 0.769 \\
    CF-GB & 0.821 & 0.702 & 0.834 & 0.677 & 0.684 & 0.862 & 0.797 & 1.000 & 0.797 \\
    Oracle & 0.728 & 0.635 & 0.766 & 0.603 & 0.597 & 0.815 & 0.749 & 1.000 & 0.737 \\
    \bottomrule
  \end{tabular}
\end{table}

\paragraph{Finite-sample results.}
Tables~\ref{tab:supp-estimated-n250-p5}--\ref{tab:supp-estimated-n500-p10} give the complete finite-sample results. Each table fixes $(n,p)$ and reports $m/n=1,5,$ and $10$. Entries for coverage variance and mean interval length variance are ratios relative to SC using the same data in each replication. The final column reports mean coverage in the upper block and mean interval length relative to SC in the lower block, both averaged over the eight DGPs.

\begin{table}[!htbp]
  \centering
  \linespread{1}\selectfont
  \setlength{\belowcaptionskip}{6pt}
  \caption{Results for SS4 and learned moment functions at $n=250$ and $p=5$. The three blocks within each criterion correspond to $m/n=1,5,$ and $10$. Variance entries are ratios relative to SC.}
  \label{tab:supp-estimated-n250-p5}
  \linespread{1}\fontsize{8.5}{9.5}\selectfont
  \setlength{\tabcolsep}{3.2pt}
  \renewcommand{\arraystretch}{1.2}
  \begin{tabular}{clccccccccc}
    \toprule
    $m/n$ & Method
    & DGP1 & DGP2 & DGP3 & DGP4
    & DGP5 & DGP6 & DGP7 & DGP8 & Mean\\
    \midrule
    \multicolumn{11}{l}{Coverage variance ratio relative to SC (final column: mean coverage)}\\
    \midrule
    1 & SC & 1.000 & 1.000 & 1.000 & 1.000 & 1.000 & 1.000 & 1.000 & 1.000 & 0.901 \\
    1 & SS4 & 0.882 & 0.877 & 0.919 & 0.809 & 0.841 & 0.973 & 0.886 & 0.992 & 0.902 \\
    1 & CF-Logit & 0.866 & 0.825 & 0.907 & 0.817 & 0.885 & 0.967 & 0.885 & 1.008 & 0.902 \\
    1 & CF-RF & 0.886 & 0.818 & 0.917 & 0.828 & 0.832 & 0.988 & 0.891 & 0.984 & 0.902 \\
    1 & CF-GB & 0.915 & 0.845 & 0.923 & 0.826 & 0.855 & 0.971 & 0.912 & 0.991 & 0.902 \\
    \addlinespace
    5 & SC & 1.000 & 1.000 & 1.000 & 1.000 & 1.000 & 1.000 & 1.000 & 1.000 & 0.901 \\
    5 & SS4 & 0.780 & 0.760 & 0.846 & 0.690 & 0.764 & 0.939 & 0.817 & 1.003 & 0.903 \\
    5 & CF-Logit & 0.811 & 0.763 & 0.855 & 0.706 & 0.827 & 0.977 & 0.811 & 1.001 & 0.902 \\
    5 & CF-RF & 0.748 & 0.691 & 0.839 & 0.699 & 0.739 & 0.950 & 0.781 & 1.011 & 0.902 \\
    5 & CF-GB & 0.836 & 0.715 & 0.858 & 0.777 & 0.825 & 0.937 & 0.871 & 1.017 & 0.902 \\
    \addlinespace
    10 & SC & 1.000 & 1.000 & 1.000 & 1.000 & 1.000 & 1.000 & 1.000 & 1.000 & 0.901 \\
    10 & SS4 & 0.768 & 0.750 & 0.834 & 0.687 & 0.723 & 0.913 & 0.772 & 1.005 & 0.903 \\
    10 & CF-Logit & 0.800 & 0.740 & 0.853 & 0.701 & 0.796 & 0.943 & 0.793 & 1.001 & 0.902 \\
    10 & CF-RF & 0.749 & 0.693 & 0.828 & 0.710 & 0.730 & 0.929 & 0.767 & 1.007 & 0.903 \\
    10 & CF-GB & 0.799 & 0.729 & 0.844 & 0.727 & 0.820 & 0.920 & 0.865 & 1.012 & 0.902 \\
    \midrule
    \multicolumn{11}{l}{Mean interval length variance ratio relative to SC (final column: relative mean length)}\\
    \midrule
    1 & SC & 1.000 & 1.000 & 1.000 & 1.000 & 1.000 & 1.000 & 1.000 & 1.000 & 1.000 \\
    1 & SS4 & 0.909 & 0.944 & 0.967 & 0.899 & 0.856 & 0.992 & 0.930 & 1.021 & 1.011 \\
    1 & CF-Logit & 0.878 & 0.888 & 0.932 & 0.888 & 0.855 & 0.997 & 0.923 & 1.025 & 1.008 \\
    1 & CF-RF & 0.899 & 0.879 & 0.974 & 0.895 & 0.841 & 1.022 & 0.915 & 1.010 & 1.009 \\
    1 & CF-GB & 0.930 & 0.911 & 0.979 & 0.894 & 0.838 & 1.010 & 0.949 & 1.010 & 1.008 \\
    \addlinespace
    5 & SC & 1.000 & 1.000 & 1.000 & 1.000 & 1.000 & 1.000 & 1.000 & 1.000 & 1.000 \\
    5 & SS4 & 0.805 & 0.800 & 0.951 & 0.809 & 0.814 & 0.986 & 0.858 & 1.035 & 1.012 \\
    5 & CF-Logit & 0.806 & 0.808 & 0.921 & 0.809 & 0.842 & 1.008 & 0.832 & 1.020 & 1.008 \\
    5 & CF-RF & 0.772 & 0.732 & 0.885 & 0.817 & 0.794 & 1.000 & 0.802 & 1.041 & 1.009 \\
    5 & CF-GB & 0.860 & 0.759 & 0.889 & 0.865 & 0.815 & 0.991 & 0.901 & 1.036 & 1.007 \\
    \addlinespace
    10 & SC & 1.000 & 1.000 & 1.000 & 1.000 & 1.000 & 1.000 & 1.000 & 1.000 & 1.000 \\
    10 & SS4 & 0.796 & 0.806 & 0.935 & 0.793 & 0.748 & 0.943 & 0.814 & 1.034 & 1.013 \\
    10 & CF-Logit & 0.785 & 0.789 & 0.906 & 0.791 & 0.792 & 0.963 & 0.821 & 1.018 & 1.007 \\
    10 & CF-RF & 0.765 & 0.740 & 0.881 & 0.816 & 0.777 & 0.975 & 0.791 & 1.038 & 1.010 \\
    10 & CF-GB & 0.808 & 0.765 & 0.882 & 0.808 & 0.819 & 0.966 & 0.895 & 1.027 & 1.008 \\
    \bottomrule
  \end{tabular}
\end{table}

The performance ordering varies across the twelve $(n,p,m/n)$ configurations. After averaging over DGP1--DGP8 within each configuration, SS4 has the smallest coverage variance ratio in seven configurations and CF-RF in five. No learned procedure has uniformly smaller coverage variance ratios than SS4 across DGPs or sample configurations.

\begin{table}[!htbp]
  \centering
  \linespread{1}\selectfont
  \setlength{\belowcaptionskip}{6pt}
  \caption{Results for SS4 and learned moment functions at $n=250$ and $p=10$. The three blocks within each criterion correspond to $m/n=1,5,$ and $10$. Variance entries are ratios relative to SC.}
  \label{tab:supp-estimated-n250-p10}
  \linespread{1}\fontsize{8.5}{9.5}\selectfont
  \setlength{\tabcolsep}{3.2pt}
  \renewcommand{\arraystretch}{1.2}
  \begin{tabular}{clccccccccc}
    \toprule
    $m/n$ & Method
    & DGP1 & DGP2 & DGP3 & DGP4
    & DGP5 & DGP6 & DGP7 & DGP8 & Mean\\
    \midrule
    \multicolumn{11}{l}{Coverage variance ratio relative to SC (final column: mean coverage)}\\
    \midrule
    1 & SC & 1.000 & 1.000 & 1.000 & 1.000 & 1.000 & 1.000 & 1.000 & 1.000 & 0.900 \\
    1 & SS4 & 0.870 & 0.835 & 0.859 & 0.870 & 0.788 & 0.925 & 0.845 & 1.008 & 0.902 \\
    1 & CF-Logit & 0.902 & 0.833 & 0.873 & 0.864 & 0.793 & 0.948 & 0.921 & 1.000 & 0.902 \\
    1 & CF-RF & 0.906 & 0.792 & 0.853 & 0.862 & 0.812 & 0.945 & 0.866 & 0.984 & 0.902 \\
    1 & CF-GB & 0.910 & 0.833 & 0.890 & 0.916 & 0.815 & 0.918 & 0.881 & 0.998 & 0.902 \\
    \addlinespace
    5 & SC & 1.000 & 1.000 & 1.000 & 1.000 & 1.000 & 1.000 & 1.000 & 1.000 & 0.900 \\
    5 & SS4 & 0.792 & 0.734 & 0.827 & 0.724 & 0.684 & 0.846 & 0.799 & 1.004 & 0.902 \\
    5 & CF-Logit & 0.831 & 0.714 & 0.848 & 0.741 & 0.707 & 0.878 & 0.850 & 1.018 & 0.902 \\
    5 & CF-RF & 0.843 & 0.690 & 0.825 & 0.719 & 0.686 & 0.894 & 0.798 & 0.987 & 0.902 \\
    5 & CF-GB & 0.860 & 0.717 & 0.866 & 0.769 & 0.743 & 0.885 & 0.800 & 1.005 & 0.902 \\
    \addlinespace
    10 & SC & 1.000 & 1.000 & 1.000 & 1.000 & 1.000 & 1.000 & 1.000 & 1.000 & 0.900 \\
    10 & SS4 & 0.795 & 0.734 & 0.821 & 0.666 & 0.661 & 0.852 & 0.781 & 1.000 & 0.902 \\
    10 & CF-Logit & 0.821 & 0.724 & 0.848 & 0.710 & 0.670 & 0.891 & 0.812 & 1.000 & 0.902 \\
    10 & CF-RF & 0.800 & 0.684 & 0.845 & 0.663 & 0.649 & 0.891 & 0.770 & 0.987 & 0.902 \\
    10 & CF-GB & 0.844 & 0.714 & 0.855 & 0.742 & 0.718 & 0.892 & 0.782 & 1.004 & 0.902 \\
    \midrule
    \multicolumn{11}{l}{Mean interval length variance ratio relative to SC (final column: relative mean length)}\\
    \midrule
    1 & SC & 1.000 & 1.000 & 1.000 & 1.000 & 1.000 & 1.000 & 1.000 & 1.000 & 1.000 \\
    1 & SS4 & 0.909 & 0.826 & 0.872 & 0.960 & 0.866 & 0.960 & 0.914 & 1.042 & 1.010 \\
    1 & CF-Logit & 0.922 & 0.818 & 0.882 & 0.932 & 0.868 & 0.972 & 0.979 & 1.019 & 1.008 \\
    1 & CF-RF & 0.921 & 0.783 & 0.878 & 0.945 & 0.853 & 0.976 & 0.924 & 0.998 & 1.008 \\
    1 & CF-GB & 0.948 & 0.847 & 0.899 & 0.990 & 0.887 & 0.942 & 0.930 & 1.012 & 1.008 \\
    \addlinespace
    5 & SC & 1.000 & 1.000 & 1.000 & 1.000 & 1.000 & 1.000 & 1.000 & 1.000 & 1.000 \\
    5 & SS4 & 0.833 & 0.728 & 0.824 & 0.820 & 0.766 & 0.826 & 0.867 & 1.042 & 1.009 \\
    5 & CF-Logit & 0.871 & 0.697 & 0.826 & 0.769 & 0.771 & 0.867 & 0.909 & 1.037 & 1.005 \\
    5 & CF-RF & 0.859 & 0.661 & 0.834 & 0.762 & 0.728 & 0.890 & 0.856 & 1.019 & 1.007 \\
    5 & CF-GB & 0.916 & 0.713 & 0.861 & 0.806 & 0.802 & 0.867 & 0.848 & 1.017 & 1.006 \\
    \addlinespace
    10 & SC & 1.000 & 1.000 & 1.000 & 1.000 & 1.000 & 1.000 & 1.000 & 1.000 & 1.000 \\
    10 & SS4 & 0.850 & 0.717 & 0.805 & 0.753 & 0.767 & 0.837 & 0.853 & 1.035 & 1.009 \\
    10 & CF-Logit & 0.870 & 0.706 & 0.822 & 0.763 & 0.738 & 0.879 & 0.855 & 1.022 & 1.006 \\
    10 & CF-RF & 0.833 & 0.636 & 0.835 & 0.740 & 0.702 & 0.898 & 0.824 & 1.018 & 1.007 \\
    10 & CF-GB & 0.896 & 0.682 & 0.846 & 0.787 & 0.802 & 0.882 & 0.846 & 1.018 & 1.006 \\
    \bottomrule
  \end{tabular}
\end{table}

The reductions generally become larger as $m/n$ increases, consistent with more accurate estimation of the mean of the moment function from the pooled calibration and unlabeled data: averaged over methods, DGPs, $n$, and $p$, the coverage variance ratio decreases from $0.884$ at $m/n=1$ to $0.813$ at $m/n=5$ and $0.799$ at $m/n=10$. Average coverage variance ratios are smaller at $n=500$ than at $n=250$, with ratios $0.822$ and $0.842$, respectively. Changing $p$ from 5 to 10 has a modest effect: the corresponding averages are $0.834$ and $0.830$, while all four procedures continue to operate on the same five-dimensional covariate representation.

\begin{table}[!htbp]
  \centering
  \linespread{1}\selectfont
  \setlength{\belowcaptionskip}{6pt}
  \caption{Results for SS4 and learned moment functions at $n=500$ and $p=5$. The three blocks within each criterion correspond to $m/n=1,5,$ and $10$. Variance entries are ratios relative to SC.}
  \label{tab:supp-estimated-n500-p5}
  \linespread{1}\fontsize{8.5}{9.5}\selectfont
  \setlength{\tabcolsep}{3.2pt}
  \renewcommand{\arraystretch}{1.2}
  \begin{tabular}{clccccccccc}
    \toprule
    $m/n$ & Method
    & DGP1 & DGP2 & DGP3 & DGP4
    & DGP5 & DGP6 & DGP7 & DGP8 & Mean\\
    \midrule
    \multicolumn{11}{l}{Coverage variance ratio relative to SC (final column: mean coverage)}\\
    \midrule
    1 & SC & 1.000 & 1.000 & 1.000 & 1.000 & 1.000 & 1.000 & 1.000 & 1.000 & 0.900 \\
    1 & SS4 & 0.838 & 0.859 & 0.867 & 0.778 & 0.843 & 0.879 & 0.862 & 1.002 & 0.901 \\
    1 & CF-Logit & 0.841 & 0.860 & 0.875 & 0.787 & 0.838 & 0.869 & 0.861 & 1.010 & 0.901 \\
    1 & CF-RF & 0.879 & 0.847 & 0.894 & 0.793 & 0.833 & 0.892 & 0.881 & 1.008 & 0.901 \\
    1 & CF-GB & 0.892 & 0.872 & 0.880 & 0.816 & 0.836 & 0.883 & 0.883 & 1.008 & 0.901 \\
    \addlinespace
    5 & SC & 1.000 & 1.000 & 1.000 & 1.000 & 1.000 & 1.000 & 1.000 & 1.000 & 0.900 \\
    5 & SS4 & 0.765 & 0.745 & 0.791 & 0.674 & 0.696 & 0.804 & 0.805 & 1.010 & 0.901 \\
    5 & CF-Logit & 0.766 & 0.736 & 0.812 & 0.679 & 0.691 & 0.831 & 0.803 & 1.005 & 0.901 \\
    5 & CF-RF & 0.808 & 0.745 & 0.810 & 0.693 & 0.704 & 0.807 & 0.836 & 1.012 & 0.901 \\
    5 & CF-GB & 0.819 & 0.777 & 0.819 & 0.699 & 0.687 & 0.830 & 0.839 & 1.004 & 0.901 \\
    \addlinespace
    10 & SC & 1.000 & 1.000 & 1.000 & 1.000 & 1.000 & 1.000 & 1.000 & 1.000 & 0.900 \\
    10 & SS4 & 0.736 & 0.737 & 0.782 & 0.679 & 0.676 & 0.801 & 0.786 & 1.016 & 0.901 \\
    10 & CF-Logit & 0.767 & 0.734 & 0.769 & 0.675 & 0.677 & 0.822 & 0.806 & 1.017 & 0.901 \\
    10 & CF-RF & 0.784 & 0.703 & 0.787 & 0.688 & 0.680 & 0.794 & 0.810 & 1.007 & 0.901 \\
    10 & CF-GB & 0.819 & 0.762 & 0.815 & 0.698 & 0.669 & 0.826 & 0.837 & 1.006 & 0.901 \\
    \midrule
    \multicolumn{11}{l}{Mean interval length variance ratio relative to SC (final column: relative mean length)}\\
    \midrule
    1 & SC & 1.000 & 1.000 & 1.000 & 1.000 & 1.000 & 1.000 & 1.000 & 1.000 & 1.000 \\
    1 & SS4 & 0.826 & 0.853 & 0.853 & 0.766 & 0.825 & 0.872 & 0.878 & 1.017 & 1.005 \\
    1 & CF-Logit & 0.831 & 0.841 & 0.854 & 0.782 & 0.819 & 0.863 & 0.868 & 1.026 & 1.004 \\
    1 & CF-RF & 0.877 & 0.831 & 0.882 & 0.767 & 0.813 & 0.883 & 0.891 & 1.018 & 1.004 \\
    1 & CF-GB & 0.885 & 0.866 & 0.867 & 0.789 & 0.815 & 0.859 & 0.899 & 1.016 & 1.003 \\
    \addlinespace
    5 & SC & 1.000 & 1.000 & 1.000 & 1.000 & 1.000 & 1.000 & 1.000 & 1.000 & 1.000 \\
    5 & SS4 & 0.771 & 0.773 & 0.761 & 0.653 & 0.703 & 0.811 & 0.838 & 1.022 & 1.005 \\
    5 & CF-Logit & 0.761 & 0.738 & 0.778 & 0.648 & 0.684 & 0.835 & 0.805 & 1.017 & 1.003 \\
    5 & CF-RF & 0.803 & 0.768 & 0.775 & 0.661 & 0.717 & 0.817 & 0.858 & 1.023 & 1.004 \\
    5 & CF-GB & 0.815 & 0.780 & 0.797 & 0.645 & 0.681 & 0.817 & 0.865 & 1.021 & 1.002 \\
    \addlinespace
    10 & SC & 1.000 & 1.000 & 1.000 & 1.000 & 1.000 & 1.000 & 1.000 & 1.000 & 1.000 \\
    10 & SS4 & 0.758 & 0.751 & 0.748 & 0.661 & 0.672 & 0.807 & 0.816 & 1.031 & 1.005 \\
    10 & CF-Logit & 0.775 & 0.728 & 0.732 & 0.648 & 0.669 & 0.822 & 0.804 & 1.027 & 1.003 \\
    10 & CF-RF & 0.792 & 0.699 & 0.751 & 0.660 & 0.678 & 0.801 & 0.834 & 1.014 & 1.004 \\
    10 & CF-GB & 0.819 & 0.764 & 0.784 & 0.644 & 0.660 & 0.813 & 0.857 & 1.022 & 1.003 \\
    \bottomrule
  \end{tabular}
\end{table}

Mean coverage remains between $0.899$ and $0.904$ across all scenarios. Mean interval length variance follows the same broad pattern as coverage variance: its average ratio decreases from $0.903$ at $m/n=1$ to $0.826$ at $m/n=5$ and $0.810$ at $m/n=10$. Relative mean lengths remain between $0.996$ and $1.021$. The variance reductions are accompanied by small changes in mean interval length.

\begin{table}[!htbp]
  \centering
  \linespread{1}\selectfont
  \setlength{\belowcaptionskip}{6pt}
  \caption{Results for SS4 and learned moment functions at $n=500$ and $p=10$. The three blocks within each criterion correspond to $m/n=1,5,$ and $10$. Variance entries are ratios relative to SC.}
  \label{tab:supp-estimated-n500-p10}
  \linespread{1}\fontsize{8.5}{9.5}\selectfont
  \setlength{\tabcolsep}{3.2pt}
  \renewcommand{\arraystretch}{1.2}
  \begin{tabular}{clccccccccc}
    \toprule
    $m/n$ & Method
    & DGP1 & DGP2 & DGP3 & DGP4
    & DGP5 & DGP6 & DGP7 & DGP8 & Mean\\
    \midrule
    \multicolumn{11}{l}{Coverage variance ratio relative to SC (final column: mean coverage)}\\
    \midrule
    1 & SC & 1.000 & 1.000 & 1.000 & 1.000 & 1.000 & 1.000 & 1.000 & 1.000 & 0.900 \\
    1 & SS4 & 0.865 & 0.862 & 0.886 & 0.777 & 0.786 & 0.923 & 0.868 & 1.002 & 0.901 \\
    1 & CF-Logit & 0.883 & 0.878 & 0.874 & 0.789 & 0.811 & 0.924 & 0.883 & 0.988 & 0.901 \\
    1 & CF-RF & 0.880 & 0.860 & 0.894 & 0.807 & 0.792 & 0.943 & 0.882 & 1.002 & 0.901 \\
    1 & CF-GB & 0.872 & 0.893 & 0.913 & 0.788 & 0.781 & 0.949 & 0.904 & 1.009 & 0.901 \\
    \addlinespace
    5 & SC & 1.000 & 1.000 & 1.000 & 1.000 & 1.000 & 1.000 & 1.000 & 1.000 & 0.900 \\
    5 & SS4 & 0.829 & 0.765 & 0.831 & 0.697 & 0.652 & 0.830 & 0.758 & 1.002 & 0.901 \\
    5 & CF-Logit & 0.835 & 0.752 & 0.845 & 0.714 & 0.682 & 0.858 & 0.811 & 1.002 & 0.901 \\
    5 & CF-RF & 0.823 & 0.740 & 0.818 & 0.674 & 0.674 & 0.856 & 0.770 & 1.004 & 0.901 \\
    5 & CF-GB & 0.819 & 0.804 & 0.842 & 0.746 & 0.677 & 0.873 & 0.828 & 1.006 & 0.901 \\
    \addlinespace
    10 & SC & 1.000 & 1.000 & 1.000 & 1.000 & 1.000 & 1.000 & 1.000 & 1.000 & 0.900 \\
    10 & SS4 & 0.812 & 0.736 & 0.802 & 0.664 & 0.611 & 0.828 & 0.742 & 1.008 & 0.901 \\
    10 & CF-Logit & 0.831 & 0.750 & 0.811 & 0.693 & 0.641 & 0.864 & 0.770 & 0.997 & 0.901 \\
    10 & CF-RF & 0.801 & 0.725 & 0.788 & 0.660 & 0.640 & 0.863 & 0.749 & 0.998 & 0.901 \\
    10 & CF-GB & 0.814 & 0.776 & 0.823 & 0.711 & 0.647 & 0.887 & 0.805 & 1.005 & 0.901 \\
    \midrule
    \multicolumn{11}{l}{Mean interval length variance ratio relative to SC (final column: relative mean length)}\\
    \midrule
    1 & SC & 1.000 & 1.000 & 1.000 & 1.000 & 1.000 & 1.000 & 1.000 & 1.000 & 1.000 \\
    1 & SS4 & 0.879 & 0.830 & 0.932 & 0.794 & 0.789 & 0.933 & 0.889 & 1.018 & 1.005 \\
    1 & CF-Logit & 0.888 & 0.833 & 0.919 & 0.820 & 0.822 & 0.923 & 0.896 & 0.999 & 1.003 \\
    1 & CF-RF & 0.887 & 0.832 & 0.932 & 0.801 & 0.794 & 0.958 & 0.894 & 1.004 & 1.003 \\
    1 & CF-GB & 0.873 & 0.863 & 0.941 & 0.821 & 0.789 & 0.965 & 0.927 & 1.011 & 1.003 \\
    \addlinespace
    5 & SC & 1.000 & 1.000 & 1.000 & 1.000 & 1.000 & 1.000 & 1.000 & 1.000 & 1.000 \\
    5 & SS4 & 0.826 & 0.706 & 0.848 & 0.708 & 0.610 & 0.834 & 0.742 & 1.016 & 1.004 \\
    5 & CF-Logit & 0.822 & 0.672 & 0.861 & 0.718 & 0.642 & 0.843 & 0.786 & 1.018 & 1.001 \\
    5 & CF-RF & 0.815 & 0.680 & 0.830 & 0.680 & 0.616 & 0.851 & 0.743 & 1.017 & 1.002 \\
    5 & CF-GB & 0.805 & 0.741 & 0.855 & 0.747 & 0.638 & 0.873 & 0.802 & 1.014 & 1.002 \\
    \addlinespace
    10 & SC & 1.000 & 1.000 & 1.000 & 1.000 & 1.000 & 1.000 & 1.000 & 1.000 & 1.000 \\
    10 & SS4 & 0.799 & 0.664 & 0.809 & 0.662 & 0.586 & 0.827 & 0.714 & 1.022 & 1.004 \\
    10 & CF-Logit & 0.808 & 0.671 & 0.818 & 0.701 & 0.609 & 0.857 & 0.737 & 1.013 & 1.002 \\
    10 & CF-RF & 0.780 & 0.659 & 0.798 & 0.650 & 0.592 & 0.856 & 0.719 & 1.004 & 1.002 \\
    10 & CF-GB & 0.790 & 0.709 & 0.822 & 0.700 & 0.611 & 0.875 & 0.779 & 1.016 & 1.001 \\
    \bottomrule
  \end{tabular}
\end{table}

\FloatBarrier

\subsection{Conditional Calibration}
\label{sec:supp-conditional}

This section reports additional finite-sample results for the conditional-calibration experiment in Section~5.4 of the main paper. The conformal score, group indicator $B$, feature map $\Phi$, and six calibration procedures are as defined there.

In addition to the main setting $n=250$, $p=5$, and $m=2500$, we consider $n\in\{250,500\}$, $p\in\{5,10\}$, and $m/n\in\{1,5,10\}$. The external-reference and test data sets each contain 5000 observations. The external-reference data are used only by Marg90 and Group90, and SS2-Marg and SS2-Group use the unlabeled data only through their covariates.

\paragraph{Detailed moment definitions.}
Let $\mathcal C_{\mathrm{ext}}(V)$ denote centring at the empirical mean over the external-reference data, as in Section~5.1 of the main paper. For a covariate-only quantity evaluated on the calibration and unlabeled data, write $\overline V_N=\bigl(\sum_{i=1}^n V_i+\sum_{j=1}^m V_j^{\rm u}\bigr)/(n+m)$ and $\mathcal C_N(V)=V-\overline V_N$. Let $Q_S^{\mathrm{ext}}(0.9)$ be the empirical marginal score quantile computed from the external-reference data, and let $Q_{S\mid B}^{\mathrm{ext}}(0.9\mid g)$ be the corresponding quantile within group $g$. With $$E_{90}^{\mathrm{ext}}(Z)=\ind\{S\le Q_{S\mid B}^{\mathrm{ext}}(0.9\mid B(X))\}-0.9,$$ the four moment functions are
\[
  \begin{aligned}
    h_{\rm Marg90}(Z)
    &=\mathcal C_{\mathrm{ext}}
    \left(\ind\{S\le Q_S^{\mathrm{ext}}(0.9)\}\right),\\
    h_{\rm Group90}(Z)
    &=\mathcal C_{\mathrm{ext}}
    \left(
    \left\{
    \{1-B(X)\}E_{90}^{\mathrm{ext}}(Z),
    B(X)E_{90}^{\mathrm{ext}}(Z)
    \right\}^{\top}
    \right),\\
    h_{\rm SS2\text{-}Marg}(X)
    &=\mathcal C_N
    \left(
    \left\{\widehat\rho(X),\widehat\rho^2(X)\right\}^{\top}
    \right),\\
    h_{\rm SS2\text{-}Group}(X)
    &=\mathcal C_N
    \left(
    \left\{
    \{1-B(X)\}\widehat\rho(X),
    \{1-B(X)\}\widehat\rho^2(X),
    B(X)\widehat\rho(X),
    B(X)\widehat\rho^2(X)
    \right\}^{\top}
    \right).
  \end{aligned}
\]

For Marg90, Group90, SS2-Marg, and SS2-Group, weights are computed once for all calibration data points using the empirical likelihood criterion. The estimated threshold for the conformal score in group $g$ is
\[
  \widehat q_g=\inf\left\{q:\frac{\sum_{\{i:B(X_i)=g\}}\widehat w_i\ind(S_i\le q)}{\sum_{\{i:B(X_i)=g\}}\widehat w_i}\ge0.9\right\}, \qquad g=0,1.
\]
This is a solution of the weighted quantile-regression objective in Section~4 of the main paper.

\paragraph{Additional empirical summaries.}
$\|\widehat M_b\|_1$, the coverage variance ratio, and the mean interval length variance ratio are defined in Section~5.4 of the main paper. For the two coverage moments, the coverage variance ratio is computed using the normalized trace that compares their covariance matrix with that of OQR. The mean interval length variance ratio also uses OQR as its reference. In addition, Table~\ref{tab:supp-conditional-main} reports marginal coverage, the two group coverages, $\mathrm{Worst}_b=\max_{g\in\{0,1\}}|\widehat C_b(g)-0.9|$, and the variance ratios for the group-specific coverage moments
\[
  R_g^{\rm emp}(H)
  =
  \frac{e_g^\top\widehat\Omega_H e_g}
  {e_g^\top\widehat\Omega_{\rm OQR}e_g},
  \qquad g=0,1.
\]
Here $\widehat C_b(g)$ is the empirical coverage among test data points in group $g$, $e_0=(1,0)^\top$, $e_1=(0,1)^\top$, and $\widehat\Omega_H$ is the across-replication covariance matrix of the two coverage moments under method $H$. Coverage, $\|\widehat M_b\|_1$, and Worst are averaged over replications.

\begin{table}[!htbp]
  \centering
  \linespread{1}\selectfont
  \setlength{\belowcaptionskip}{6pt}
  \caption{Complete conditional-calibration results at $n=250$, $p=5$, and $m=2500$, based on 500 Monte Carlo replications. The coverage variance ratio is computed using the normalized trace, and the mean length variance ratio denotes the mean interval length variance ratio. Both ratios are relative to OQR.}
  \label{tab:supp-conditional-main}
  \linespread{1}\fontsize{8.5}{9.5}\selectfont
  \setlength{\tabcolsep}{2.2pt}
  \renewcommand{\arraystretch}{1}
  \begin{tabular}{llccccccccc}
    \toprule
    DGP & Method &
    \shortstack{Marg.\\cov.} & \shortstack{Cov.\\$B=0$} & \shortstack{Cov.\\$B=1$} &
    $\|\widehat M_b\|_1$ & Worst &
    $R_0^{\rm emp}$ & $R_1^{\rm emp}$ &
    \shortstack{Coverage\\variance\\ratio} & \shortstack{Mean length\\variance\\ratio}\\
    \midrule
    DGP1 & SC & 0.900 & 0.901 & 0.900 & 0.016 & 0.019 & -- & -- & -- & 0.955 \\
    & OQR & 0.896 & 0.896 & 0.896 & 0.023 & 0.033 & 1.000 & 1.000 & 1.000 & 1.000 \\
    & Marg90 & 0.897 & 0.897 & 0.898 & 0.019 & 0.027 & 0.767 & 0.713 & 0.713 & 0.197 \\
    & Group90 & 0.905 & 0.905 & 0.904 & 0.011 & 0.016 & 0.239 & 0.186 & 0.214 & 0.267 \\
    & SS2-Marg & 0.897 & 0.898 & 0.897 & 0.021 & 0.031 & 0.857 & 0.902 & 0.874 & 0.750 \\
    & SS2-Group & 0.898 & 0.898 & 0.897 & 0.020 & 0.029 & 0.836 & 0.745 & 0.790 & 0.777 \\
    \addlinespace
    DGP2 & SC & 0.900 & 0.970 & 0.829 & 0.071 & 0.087 & -- & -- & -- & 0.972 \\
    & OQR & 0.896 & 0.897 & 0.895 & 0.022 & 0.032 & 1.000 & 1.000 & 1.000 & 1.000 \\
    & Marg90 & 0.899 & 0.898 & 0.900 & 0.019 & 0.028 & 0.989 & 0.553 & 0.782 & 0.598 \\
    & Group90 & 0.905 & 0.905 & 0.905 & 0.011 & 0.015 & 0.192 & 0.221 & 0.206 & 0.247 \\
    & SS2-Marg & 0.897 & 0.898 & 0.897 & 0.022 & 0.031 & 0.988 & 0.907 & 0.948 & 0.945 \\
    & SS2-Group & 0.898 & 0.900 & 0.896 & 0.021 & 0.030 & 0.993 & 0.820 & 0.906 & 0.863 \\
    \addlinespace
    DGP3 & SC & 0.900 & 0.964 & 0.837 & 0.064 & 0.079 & -- & -- & -- & 1.188 \\
    & OQR & 0.896 & 0.897 & 0.896 & 0.021 & 0.031 & 1.000 & 1.000 & 1.000 & 1.000 \\
    & Marg90 & 0.899 & 0.898 & 0.899 & 0.019 & 0.027 & 0.938 & 0.587 & 0.755 & 0.588 \\
    & Group90 & 0.903 & 0.903 & 0.904 & 0.010 & 0.015 & 0.230 & 0.268 & 0.250 & 0.321 \\
    & SS2-Marg & 0.897 & 0.897 & 0.897 & 0.021 & 0.030 & 0.977 & 0.948 & 0.962 & 0.920 \\
    & SS2-Group & 0.898 & 0.899 & 0.896 & 0.021 & 0.030 & 0.920 & 0.979 & 0.950 & 0.948 \\
    \addlinespace
    DGP4 & SC & 0.900 & 0.978 & 0.823 & 0.077 & 0.093 & -- & -- & -- & 1.305 \\
    & OQR & 0.896 & 0.897 & 0.896 & 0.023 & 0.033 & 1.000 & 1.000 & 1.000 & 1.000 \\
    & Marg90 & 0.897 & 0.896 & 0.899 & 0.020 & 0.029 & 0.985 & 0.574 & 0.776 & 0.627 \\
    & Group90 & 0.905 & 0.905 & 0.905 & 0.011 & 0.015 & 0.214 & 0.209 & 0.213 & 0.271 \\
    & SS2-Marg & 0.897 & 0.897 & 0.898 & 0.022 & 0.031 & 0.982 & 0.855 & 0.917 & 0.862 \\
    & SS2-Group & 0.898 & 0.899 & 0.897 & 0.021 & 0.030 & 0.974 & 0.829 & 0.902 & 0.954 \\
    \addlinespace
    DGP5 & SC & 0.899 & 0.898 & 0.900 & 0.016 & 0.019 & -- & -- & -- & 1.042 \\
    & OQR & 0.895 & 0.895 & 0.895 & 0.023 & 0.032 & 1.000 & 1.000 & 1.000 & 1.000 \\
    & Marg90 & 0.898 & 0.898 & 0.898 & 0.018 & 0.025 & 0.620 & 0.735 & 0.676 & 0.276 \\
    & Group90 & 0.904 & 0.904 & 0.904 & 0.011 & 0.015 & 0.177 & 0.252 & 0.214 & 0.267 \\
    & SS2-Marg & 0.896 & 0.896 & 0.896 & 0.022 & 0.030 & 0.891 & 0.892 & 0.891 & 0.788 \\
    & SS2-Group & 0.898 & 0.897 & 0.898 & 0.019 & 0.027 & 0.675 & 0.772 & 0.723 & 0.808 \\
    \addlinespace
    DGP6 & SC & 0.899 & 0.959 & 0.839 & 0.060 & 0.075 & -- & -- & -- & 1.293 \\
    & OQR & 0.896 & 0.896 & 0.896 & 0.022 & 0.032 & 1.000 & 1.000 & 1.000 & 1.000 \\
    & Marg90 & 0.898 & 0.896 & 0.900 & 0.020 & 0.027 & 0.932 & 0.618 & 0.756 & 0.556 \\
    & Group90 & 0.904 & 0.904 & 0.905 & 0.011 & 0.016 & 0.233 & 0.230 & 0.238 & 0.267 \\
    & SS2-Marg & 0.897 & 0.896 & 0.897 & 0.022 & 0.031 & 1.000 & 0.980 & 0.992 & 0.987 \\
    & SS2-Group & 0.896 & 0.896 & 0.896 & 0.022 & 0.031 & 0.952 & 0.930 & 0.944 & 0.952 \\
    \addlinespace
    DGP7 & SC & 0.901 & 0.969 & 0.832 & 0.068 & 0.084 & -- & -- & -- & 1.222 \\
    & OQR & 0.896 & 0.898 & 0.894 & 0.022 & 0.032 & 1.000 & 1.000 & 1.000 & 1.000 \\
    & Marg90 & 0.898 & 0.898 & 0.898 & 0.019 & 0.027 & 0.975 & 0.572 & 0.771 & 0.576 \\
    & Group90 & 0.904 & 0.904 & 0.905 & 0.011 & 0.015 & 0.254 & 0.205 & 0.230 & 0.237 \\
    & SS2-Marg & 0.897 & 0.898 & 0.896 & 0.021 & 0.030 & 0.982 & 0.891 & 0.936 & 0.941 \\
    & SS2-Group & 0.897 & 0.899 & 0.896 & 0.021 & 0.030 & 0.959 & 0.856 & 0.908 & 0.929 \\
    \addlinespace
    DGP8 & SC & 0.900 & 0.900 & 0.900 & 0.016 & 0.019 & -- & -- & -- & 0.999 \\
    & OQR & 0.896 & 0.896 & 0.896 & 0.023 & 0.032 & 1.000 & 1.000 & 1.000 & 1.000 \\
    & Marg90 & 0.898 & 0.898 & 0.897 & 0.019 & 0.026 & 0.674 & 0.800 & 0.715 & 0.210 \\
    & Group90 & 0.905 & 0.905 & 0.905 & 0.010 & 0.015 & 0.189 & 0.216 & 0.204 & 0.236 \\
    & SS2-Marg & 0.896 & 0.896 & 0.897 & 0.023 & 0.032 & 0.996 & 1.009 & 1.003 & 1.018 \\
    & SS2-Group & 0.896 & 0.896 & 0.896 & 0.023 & 0.032 & 0.989 & 1.011 & 1.000 & 1.019 \\
    \bottomrule
  \end{tabular}
\end{table}

\paragraph{Results for all scenarios.}
Tables~\ref{tab:supp-conditional-n250-p5}--\ref{tab:supp-conditional-n500-p10} report all four $(n,p)$ combinations. Each table reports SC, OQR, Marg90, and Group90 once, followed by SS2-Marg and SS2-Group for $m/n=1,5,$ and $10$. Marg90 and Group90 do not use the unlabeled sample and hence do not vary with $m/n$. The across-replication variance of mean interval length is reported relative to OQR.

\begin{table}[!htbp]
  \centering
  \linespread{1}\selectfont
  \setlength{\belowcaptionskip}{6pt}
  \caption{Conditional-calibration results at $n=250$ and $p=5$. Rows with a specified $m/n$ use unlabeled data. A dash denotes methods that do not depend on $m$.}
  \label{tab:supp-conditional-n250-p5}
  \linespread{1}\fontsize{8.5}{9.5}\selectfont
  \setlength{\tabcolsep}{3.2pt}
  \renewcommand{\arraystretch}{1.2}
  \begin{tabular}{clccccccccc}
    \toprule
    $m/n$ & Method &
    DGP1 & DGP2 & DGP3 & DGP4 & DGP5 & DGP6 & DGP7 & DGP8 & Avg.\\
    \midrule
    \multicolumn{11}{l}{Mean of $\|\widehat M_b\|_1$ across replications}\\
    -- & SC & 0.016 & 0.071 & 0.064 & 0.077 & 0.016 & 0.060 & 0.068 & 0.016 & 0.049 \\
    -- & OQR & 0.023 & 0.022 & 0.021 & 0.023 & 0.023 & 0.022 & 0.022 & 0.023 & 0.022 \\
    -- & Marg90 & 0.019 & 0.019 & 0.019 & 0.020 & 0.018 & 0.020 & 0.019 & 0.019 & 0.019 \\
    -- & Group90 & 0.011 & 0.011 & 0.010 & 0.011 & 0.011 & 0.011 & 0.011 & 0.010 & 0.011 \\
    1 & SS2-Marg & 0.022 & 0.022 & 0.021 & 0.022 & 0.022 & 0.022 & 0.022 & 0.023 & 0.022 \\
    1 & SS2-Group & 0.021 & 0.022 & 0.021 & 0.022 & 0.021 & 0.022 & 0.022 & 0.023 & 0.022 \\
    5 & SS2-Marg & 0.022 & 0.022 & 0.021 & 0.022 & 0.022 & 0.022 & 0.021 & 0.023 & 0.022 \\
    5 & SS2-Group & 0.021 & 0.021 & 0.021 & 0.022 & 0.019 & 0.022 & 0.021 & 0.023 & 0.021 \\
    10 & SS2-Marg & 0.021 & 0.022 & 0.021 & 0.022 & 0.022 & 0.022 & 0.021 & 0.023 & 0.022 \\
    10 & SS2-Group & 0.020 & 0.021 & 0.021 & 0.021 & 0.019 & 0.022 & 0.021 & 0.023 & 0.021 \\
    \midrule
    \multicolumn{11}{l}{Coverage variance ratio (normalized trace) relative to OQR}\\
    -- & OQR & 1.000 & 1.000 & 1.000 & 1.000 & 1.000 & 1.000 & 1.000 & 1.000 & 1.000 \\
    -- & Marg90 & 0.713 & 0.782 & 0.755 & 0.776 & 0.676 & 0.756 & 0.771 & 0.715 & 0.743 \\
    -- & Group90 & 0.214 & 0.206 & 0.250 & 0.213 & 0.214 & 0.238 & 0.230 & 0.204 & 0.221 \\
    1 & SS2-Marg & 0.925 & 0.972 & 0.979 & 0.937 & 0.904 & 0.996 & 0.961 & 1.000 & 0.959 \\
    1 & SS2-Group & 0.867 & 0.946 & 0.974 & 0.938 & 0.828 & 0.995 & 0.962 & 1.009 & 0.940 \\
    5 & SS2-Marg & 0.884 & 0.945 & 0.971 & 0.910 & 0.893 & 0.974 & 0.938 & 1.002 & 0.940 \\
    5 & SS2-Group & 0.811 & 0.915 & 0.954 & 0.931 & 0.736 & 0.952 & 0.902 & 1.007 & 0.901 \\
    10 & SS2-Marg & 0.874 & 0.948 & 0.962 & 0.917 & 0.891 & 0.992 & 0.936 & 1.003 & 0.940 \\
    10 & SS2-Group & 0.790 & 0.906 & 0.950 & 0.902 & 0.723 & 0.944 & 0.908 & 1.000 & 0.891 \\
    \midrule
    \multicolumn{11}{l}{Mean interval length variance ratio relative to OQR}\\
    -- & SC & 0.955 & 0.972 & 1.188 & 1.305 & 1.042 & 1.293 & 1.222 & 0.999 & 1.122 \\
    -- & OQR & 1.000 & 1.000 & 1.000 & 1.000 & 1.000 & 1.000 & 1.000 & 1.000 & 1.000 \\
    -- & Marg90 & 0.197 & 0.598 & 0.588 & 0.627 & 0.276 & 0.556 & 0.576 & 0.210 & 0.453 \\
    -- & Group90 & 0.267 & 0.247 & 0.321 & 0.271 & 0.267 & 0.267 & 0.237 & 0.236 & 0.264 \\
    1 & SS2-Marg & 0.896 & 0.966 & 0.983 & 0.895 & 0.884 & 1.023 & 0.949 & 1.006 & 0.950 \\
    1 & SS2-Group & 0.897 & 0.940 & 0.979 & 0.926 & 0.914 & 1.018 & 0.945 & 1.003 & 0.953 \\
    5 & SS2-Marg & 0.764 & 0.943 & 0.944 & 0.843 & 0.801 & 0.976 & 0.929 & 1.011 & 0.901 \\
    5 & SS2-Group & 0.787 & 0.876 & 0.963 & 0.976 & 0.812 & 0.960 & 0.911 & 1.015 & 0.913 \\
    10 & SS2-Marg & 0.750 & 0.945 & 0.920 & 0.862 & 0.788 & 0.987 & 0.941 & 1.018 & 0.901 \\
    10 & SS2-Group & 0.777 & 0.863 & 0.948 & 0.954 & 0.808 & 0.952 & 0.929 & 1.019 & 0.906 \\
    \bottomrule
  \end{tabular}
\end{table}

Across all scenarios, SC has mean values of $\|\widehat M_b\|_1$ between $0.011$ and $0.080$, whereas OQR gives $0.015$--$0.023$. Marg90 has coverage variance ratios between $0.656$ and $0.823$, while Group90 gives $0.192$--$0.272$. Their mean interval length variance ratio ranges are $0.185$--$0.673$ and $0.162$--$0.379$, respectively.

\begin{table}[!htbp]
  \centering
  \linespread{1}\selectfont
  \setlength{\belowcaptionskip}{6pt}
  \caption{Conditional-calibration results at $n=250$ and $p=10$. Rows with a specified $m/n$ use unlabeled data. A dash denotes methods that do not depend on $m$.}
  \label{tab:supp-conditional-n250-p10}
  \linespread{1}\fontsize{8.5}{9.5}\selectfont
  \setlength{\tabcolsep}{3.2pt}
  \renewcommand{\arraystretch}{1.2}
  \begin{tabular}{clccccccccc}
    \toprule
    $m/n$ & Method &
    DGP1 & DGP2 & DGP3 & DGP4 & DGP5 & DGP6 & DGP7 & DGP8 & Avg.\\
    \midrule
    \multicolumn{11}{l}{Mean of $\|\widehat M_b\|_1$ across replications}\\
    -- & SC & 0.015 & 0.070 & 0.061 & 0.074 & 0.016 & 0.056 & 0.071 & 0.016 & 0.047 \\
    -- & OQR & 0.022 & 0.023 & 0.022 & 0.023 & 0.023 & 0.021 & 0.022 & 0.023 & 0.022 \\
    -- & Marg90 & 0.019 & 0.020 & 0.020 & 0.019 & 0.018 & 0.019 & 0.019 & 0.020 & 0.019 \\
    -- & Group90 & 0.011 & 0.010 & 0.011 & 0.011 & 0.010 & 0.011 & 0.011 & 0.011 & 0.011 \\
    1 & SS2-Marg & 0.021 & 0.022 & 0.022 & 0.022 & 0.021 & 0.021 & 0.022 & 0.023 & 0.022 \\
    1 & SS2-Group & 0.021 & 0.022 & 0.022 & 0.022 & 0.020 & 0.021 & 0.022 & 0.023 & 0.022 \\
    5 & SS2-Marg & 0.021 & 0.022 & 0.022 & 0.021 & 0.020 & 0.021 & 0.022 & 0.023 & 0.022 \\
    5 & SS2-Group & 0.020 & 0.021 & 0.022 & 0.022 & 0.018 & 0.021 & 0.022 & 0.023 & 0.021 \\
    10 & SS2-Marg & 0.021 & 0.021 & 0.022 & 0.021 & 0.020 & 0.021 & 0.022 & 0.023 & 0.021 \\
    10 & SS2-Group & 0.020 & 0.021 & 0.022 & 0.022 & 0.018 & 0.021 & 0.022 & 0.023 & 0.021 \\
    \midrule
    \multicolumn{11}{l}{Coverage variance ratio (normalized trace) relative to OQR}\\
    -- & OQR & 1.000 & 1.000 & 1.000 & 1.000 & 1.000 & 1.000 & 1.000 & 1.000 & 1.000 \\
    -- & Marg90 & 0.764 & 0.746 & 0.771 & 0.773 & 0.678 & 0.799 & 0.762 & 0.708 & 0.750 \\
    -- & Group90 & 0.262 & 0.196 & 0.227 & 0.220 & 0.192 & 0.267 & 0.259 & 0.207 & 0.229 \\
    1 & SS2-Marg & 0.969 & 0.883 & 0.985 & 0.963 & 0.874 & 0.981 & 0.962 & 0.992 & 0.951 \\
    1 & SS2-Group & 0.945 & 0.892 & 0.970 & 0.982 & 0.777 & 0.982 & 0.981 & 0.995 & 0.941 \\
    5 & SS2-Marg & 0.971 & 0.877 & 0.991 & 0.913 & 0.822 & 0.987 & 0.945 & 0.990 & 0.937 \\
    5 & SS2-Group & 0.890 & 0.867 & 0.948 & 0.966 & 0.666 & 0.963 & 1.020 & 0.986 & 0.913 \\
    10 & SS2-Marg & 0.939 & 0.854 & 0.957 & 0.917 & 0.820 & 0.980 & 0.940 & 0.995 & 0.925 \\
    10 & SS2-Group & 0.885 & 0.854 & 0.943 & 0.994 & 0.643 & 0.952 & 1.040 & 0.999 & 0.914 \\
    \midrule
    \multicolumn{11}{l}{Mean interval length variance ratio relative to OQR}\\
    -- & SC & 1.024 & 0.968 & 1.063 & 0.891 & 0.994 & 1.038 & 1.171 & 1.072 & 1.028 \\
    -- & OQR & 1.000 & 1.000 & 1.000 & 1.000 & 1.000 & 1.000 & 1.000 & 1.000 & 1.000 \\
    -- & Marg90 & 0.280 & 0.548 & 0.561 & 0.601 & 0.185 & 0.564 & 0.537 & 0.200 & 0.434 \\
    -- & Group90 & 0.379 & 0.199 & 0.251 & 0.290 & 0.210 & 0.283 & 0.284 & 0.253 & 0.269 \\
    1 & SS2-Marg & 0.969 & 0.840 & 1.019 & 0.950 & 0.702 & 0.956 & 0.981 & 0.992 & 0.926 \\
    1 & SS2-Group & 1.032 & 0.828 & 1.010 & 0.950 & 0.725 & 0.977 & 0.961 & 0.998 & 0.935 \\
    5 & SS2-Marg & 0.955 & 0.812 & 1.027 & 0.856 & 0.605 & 0.936 & 0.901 & 0.996 & 0.886 \\
    5 & SS2-Group & 0.983 & 0.839 & 0.966 & 0.883 & 0.624 & 0.944 & 0.920 & 1.011 & 0.896 \\
    10 & SS2-Marg & 0.915 & 0.768 & 0.987 & 0.881 & 0.616 & 0.942 & 0.911 & 1.007 & 0.878 \\
    10 & SS2-Group & 1.011 & 0.799 & 0.962 & 0.873 & 0.615 & 0.922 & 0.927 & 1.010 & 0.890 \\
    \bottomrule
  \end{tabular}
\end{table}

Averaged over DGPs and the four $(n,p)$ combinations, the SS2-Marg coverage variance ratio changes from $0.953$ at $m/n=1$ to $0.921$ at $m/n=10$, and the corresponding SS2-Group averages change from $0.929$ to $0.878$. SS2-Group has the smaller coverage variance ratio in 76 of the 96 configurations. Increasing $n$ from 250 to 500 changes the average coverage variance ratio from $0.942$ to $0.925$ for SS2-Marg and from $0.917$ to $0.879$ for SS2-Group. Changing $p$ from 5 to 10 changes the corresponding averages from $0.932$ to $0.935$ and from $0.892$ to $0.904$.

\begin{table}[!htbp]
  \centering
  \linespread{1}\selectfont
  \setlength{\belowcaptionskip}{6pt}
  \caption{Conditional-calibration results at $n=500$ and $p=5$. Rows with a specified $m/n$ use unlabeled data. A dash denotes methods that do not depend on $m$.}
  \label{tab:supp-conditional-n500-p5}
  \linespread{1}\fontsize{8.5}{9.5}\selectfont
  \setlength{\tabcolsep}{3.2pt}
  \renewcommand{\arraystretch}{1.2}
  \begin{tabular}{clccccccccc}
    \toprule
    $m/n$ & Method &
    DGP1 & DGP2 & DGP3 & DGP4 & DGP5 & DGP6 & DGP7 & DGP8 & Avg.\\
    \midrule
    \multicolumn{11}{l}{Mean of $\|\widehat M_b\|_1$ across replications}\\
    -- & SC & 0.012 & 0.069 & 0.066 & 0.080 & 0.011 & 0.061 & 0.070 & 0.012 & 0.048 \\
    -- & OQR & 0.016 & 0.016 & 0.016 & 0.016 & 0.016 & 0.017 & 0.016 & 0.016 & 0.016 \\
    -- & Marg90 & 0.013 & 0.014 & 0.014 & 0.014 & 0.014 & 0.015 & 0.014 & 0.013 & 0.014 \\
    -- & Group90 & 0.008 & 0.008 & 0.008 & 0.008 & 0.008 & 0.008 & 0.008 & 0.008 & 0.008 \\
    1 & SS2-Marg & 0.016 & 0.015 & 0.015 & 0.016 & 0.015 & 0.017 & 0.016 & 0.016 & 0.016 \\
    1 & SS2-Group & 0.015 & 0.015 & 0.015 & 0.016 & 0.014 & 0.016 & 0.015 & 0.016 & 0.015 \\
    5 & SS2-Marg & 0.015 & 0.015 & 0.015 & 0.016 & 0.014 & 0.016 & 0.015 & 0.016 & 0.015 \\
    5 & SS2-Group & 0.014 & 0.015 & 0.015 & 0.015 & 0.013 & 0.016 & 0.015 & 0.016 & 0.015 \\
    10 & SS2-Marg & 0.015 & 0.015 & 0.015 & 0.016 & 0.014 & 0.016 & 0.015 & 0.016 & 0.015 \\
    10 & SS2-Group & 0.014 & 0.015 & 0.015 & 0.015 & 0.013 & 0.016 & 0.015 & 0.016 & 0.015 \\
    \midrule
    \multicolumn{11}{l}{Coverage variance ratio (normalized trace) relative to OQR}\\
    -- & OQR & 1.000 & 1.000 & 1.000 & 1.000 & 1.000 & 1.000 & 1.000 & 1.000 & 1.000 \\
    -- & Marg90 & 0.691 & 0.794 & 0.816 & 0.789 & 0.680 & 0.820 & 0.796 & 0.683 & 0.759 \\
    -- & Group90 & 0.234 & 0.224 & 0.272 & 0.224 & 0.235 & 0.223 & 0.256 & 0.244 & 0.239 \\
    1 & SS2-Marg & 0.911 & 0.915 & 0.952 & 0.950 & 0.862 & 0.955 & 0.971 & 0.994 & 0.939 \\
    1 & SS2-Group & 0.869 & 0.888 & 0.918 & 0.934 & 0.787 & 0.931 & 0.949 & 0.997 & 0.909 \\
    5 & SS2-Marg & 0.872 & 0.888 & 0.939 & 0.909 & 0.787 & 0.925 & 0.949 & 1.001 & 0.909 \\
    5 & SS2-Group & 0.778 & 0.842 & 0.885 & 0.876 & 0.701 & 0.882 & 0.934 & 0.994 & 0.862 \\
    10 & SS2-Marg & 0.868 & 0.884 & 0.935 & 0.887 & 0.790 & 0.926 & 0.952 & 1.003 & 0.906 \\
    10 & SS2-Group & 0.767 & 0.829 & 0.862 & 0.850 & 0.671 & 0.875 & 0.925 & 1.007 & 0.848 \\
    \midrule
    \multicolumn{11}{l}{Mean interval length variance ratio relative to OQR}\\
    -- & SC & 0.978 & 0.818 & 1.143 & 1.232 & 1.015 & 1.115 & 1.341 & 0.986 & 1.079 \\
    -- & OQR & 1.000 & 1.000 & 1.000 & 1.000 & 1.000 & 1.000 & 1.000 & 1.000 & 1.000 \\
    -- & Marg90 & 0.187 & 0.635 & 0.673 & 0.593 & 0.257 & 0.634 & 0.585 & 0.201 & 0.471 \\
    -- & Group90 & 0.180 & 0.176 & 0.234 & 0.162 & 0.230 & 0.176 & 0.211 & 0.203 & 0.197 \\
    1 & SS2-Marg & 0.839 & 0.847 & 0.882 & 0.925 & 0.785 & 0.935 & 0.954 & 1.007 & 0.897 \\
    1 & SS2-Group & 0.831 & 0.832 & 0.904 & 0.923 & 0.791 & 0.909 & 0.962 & 1.009 & 0.895 \\
    5 & SS2-Marg & 0.729 & 0.810 & 0.908 & 0.833 & 0.641 & 0.882 & 0.932 & 1.003 & 0.842 \\
    5 & SS2-Group & 0.698 & 0.757 & 0.876 & 0.831 & 0.635 & 0.845 & 0.903 & 0.993 & 0.817 \\
    10 & SS2-Marg & 0.737 & 0.775 & 0.903 & 0.812 & 0.646 & 0.860 & 0.928 & 0.998 & 0.833 \\
    10 & SS2-Group & 0.699 & 0.729 & 0.853 & 0.812 & 0.619 & 0.822 & 0.907 & 1.004 & 0.806 \\
    \bottomrule
  \end{tabular}
\end{table}

For the mean interval length variance ratio relative to OQR, the SS2-Marg averages change from $0.924$ at $m/n=1$ to $0.863$ at $m/n=10$, and the SS2-Group averages change from $0.925$ to $0.857$. Under DGP8, the SS2 ratios remain close to one because the true conditional scale is constant. Marg90 and Group90 continue to reduce variance under DGP8 through moment functions involving the score $S$. Across all reported procedures, mean group coverages range from $0.819$ to $0.981$.

\begin{table}[!htbp]
  \centering
  \linespread{1}\selectfont
  \setlength{\belowcaptionskip}{6pt}
  \caption{Conditional-calibration results at $n=500$ and $p=10$. Rows with a specified $m/n$ use unlabeled data. A dash denotes methods that do not depend on $m$.}
  \label{tab:supp-conditional-n500-p10}
  \linespread{1}\fontsize{8.5}{9.5}\selectfont
  \setlength{\tabcolsep}{3.2pt}
  \renewcommand{\arraystretch}{1.2}
  \begin{tabular}{clccccccccc}
    \toprule
    $m/n$ & Method &
    DGP1 & DGP2 & DGP3 & DGP4 & DGP5 & DGP6 & DGP7 & DGP8 & Avg.\\
    \midrule
    \multicolumn{11}{l}{Mean of $\|\widehat M_b\|_1$ across replications}\\
    -- & SC & 0.011 & 0.070 & 0.066 & 0.080 & 0.012 & 0.061 & 0.072 & 0.012 & 0.048 \\
    -- & OQR & 0.016 & 0.016 & 0.015 & 0.017 & 0.016 & 0.016 & 0.016 & 0.017 & 0.016 \\
    -- & Marg90 & 0.013 & 0.015 & 0.013 & 0.015 & 0.013 & 0.014 & 0.014 & 0.014 & 0.014 \\
    -- & Group90 & 0.008 & 0.008 & 0.008 & 0.008 & 0.008 & 0.008 & 0.008 & 0.008 & 0.008 \\
    1 & SS2-Marg & 0.015 & 0.016 & 0.015 & 0.016 & 0.016 & 0.015 & 0.016 & 0.017 & 0.016 \\
    1 & SS2-Group & 0.015 & 0.016 & 0.015 & 0.016 & 0.015 & 0.015 & 0.015 & 0.017 & 0.015 \\
    5 & SS2-Marg & 0.015 & 0.016 & 0.015 & 0.016 & 0.015 & 0.015 & 0.015 & 0.017 & 0.015 \\
    5 & SS2-Group & 0.015 & 0.015 & 0.014 & 0.016 & 0.014 & 0.015 & 0.015 & 0.017 & 0.015 \\
    10 & SS2-Marg & 0.015 & 0.016 & 0.015 & 0.016 & 0.015 & 0.015 & 0.015 & 0.017 & 0.015 \\
    10 & SS2-Group & 0.014 & 0.015 & 0.014 & 0.015 & 0.013 & 0.015 & 0.015 & 0.017 & 0.015 \\
    \midrule
    \multicolumn{11}{l}{Coverage variance ratio (normalized trace) relative to OQR}\\
    -- & OQR & 1.000 & 1.000 & 1.000 & 1.000 & 1.000 & 1.000 & 1.000 & 1.000 & 1.000 \\
    -- & Marg90 & 0.697 & 0.817 & 0.784 & 0.823 & 0.724 & 0.779 & 0.797 & 0.656 & 0.760 \\
    -- & Group90 & 0.261 & 0.234 & 0.241 & 0.222 & 0.229 & 0.249 & 0.263 & 0.227 & 0.241 \\
    1 & SS2-Marg & 0.969 & 0.946 & 0.958 & 0.950 & 0.928 & 0.981 & 0.976 & 0.993 & 0.963 \\
    1 & SS2-Group & 0.923 & 0.922 & 0.896 & 0.926 & 0.829 & 0.942 & 0.973 & 0.995 & 0.926 \\
    5 & SS2-Marg & 0.904 & 0.918 & 0.906 & 0.908 & 0.858 & 0.939 & 0.946 & 0.998 & 0.922 \\
    5 & SS2-Group & 0.843 & 0.866 & 0.849 & 0.875 & 0.698 & 0.903 & 0.906 & 1.005 & 0.868 \\
    10 & SS2-Marg & 0.892 & 0.928 & 0.904 & 0.898 & 0.839 & 0.925 & 0.921 & 1.006 & 0.914 \\
    10 & SS2-Group & 0.832 & 0.874 & 0.836 & 0.860 & 0.672 & 0.902 & 0.905 & 1.003 & 0.860 \\
    \midrule
    \multicolumn{11}{l}{Mean interval length variance ratio relative to OQR}\\
    -- & SC & 1.052 & 0.945 & 1.034 & 1.126 & 1.061 & 1.139 & 1.344 & 0.999 & 1.088 \\
    -- & OQR & 1.000 & 1.000 & 1.000 & 1.000 & 1.000 & 1.000 & 1.000 & 1.000 & 1.000 \\
    -- & Marg90 & 0.241 & 0.604 & 0.587 & 0.643 & 0.225 & 0.562 & 0.606 & 0.186 & 0.457 \\
    -- & Group90 & 0.243 & 0.193 & 0.175 & 0.189 & 0.187 & 0.184 & 0.199 & 0.178 & 0.194 \\
    1 & SS2-Marg & 0.920 & 0.888 & 0.912 & 0.910 & 0.798 & 0.991 & 0.974 & 0.990 & 0.923 \\
    1 & SS2-Group & 0.941 & 0.864 & 0.895 & 0.886 & 0.800 & 0.994 & 0.976 & 0.987 & 0.918 \\
    5 & SS2-Marg & 0.758 & 0.820 & 0.855 & 0.852 & 0.661 & 0.928 & 0.935 & 1.000 & 0.851 \\
    5 & SS2-Group & 0.775 & 0.810 & 0.819 & 0.825 & 0.672 & 0.920 & 0.910 & 1.011 & 0.843 \\
    10 & SS2-Marg & 0.728 & 0.833 & 0.846 & 0.842 & 0.627 & 0.920 & 0.912 & 1.005 & 0.839 \\
    10 & SS2-Group & 0.766 & 0.798 & 0.791 & 0.799 & 0.649 & 0.932 & 0.875 & 1.007 & 0.827 \\
    \bottomrule
  \end{tabular}
\end{table}

\FloatBarrier

\section[Implementation and Additional Results for the Real-data Analysis]{Implementation and Additional Results for the\\ Real-data Analysis}
\label{sec:supp-applications}

This section provides the data-processing and implementation details for Section~6 of the main paper, together with additional results that complement the variance ratios reported there.

\subsection{Data Preparation and Repeated Splits}
\label{sec:supp-applications-data}

We use the repeated-split design with fixed training and test subsets, as described in Section~6.1 of the main paper. Table~\ref{tab:supp-applications-allocation} gives the sample allocation for each data set. The calibration, external-reference, and unlabeled subsets have the same sizes across applications.

\begin{table}[!htbp]
  \centering
  \linespread{1}\selectfont
  \setlength{\belowcaptionskip}{6pt}
  \caption{Sample allocation for the real-data analysis. Ext. ref. and Unlab. denote the external-reference and unlabeled subsets, respectively.}
  \label{tab:supp-applications-allocation}
  \linespread{1}\fontsize{8.5}{9.5}\selectfont
  \setlength{\tabcolsep}{3.2pt}
  \renewcommand{\arraystretch}{1}
  \begin{tabular}{lrrrrrl}
    \toprule
    Data set & Train & Calibration & Ext. ref. & Unlab. & Test & Transform\\
    \midrule
    Steel industry energy
    & 250 & 250 & 2500 & 2500 & 5000 & identity\\
    Seoul bike sharing
    & 250 & 250 & 2500 & 2500 & 3000 & identity\\
    SGEMM runtime
    & 250 & 250 & 2500 & 2500 & 5000 & identity\\
    \bottomrule
  \end{tabular}
\end{table}

For Steel Industry Energy, the predictors include the recorded power measurements and time-of-day variables, with weekdays defining the conditional group. For Seoul Bike Sharing, eight weather measurements are combined with cyclic encodings of hour and day of year and indicators for holidays and system operation. Winter defines the conditional group. For SGEMM, the response is the average of the four recorded kernel runtimes and the 14 kernel-configuration variables are used as predictors. The conditional group has both workgroup dimensions at least 64. All responses are analysed on their original scales. No additional centring, scaling, or truncation is applied to the predictor matrices.

\FloatBarrier

\subsection{Prediction and Side-Information Construction}
\label{sec:supp-applications-construction}

\paragraph{Prediction and covariate representation.}
The outcome predictor is a random forest fitted once on the fixed training subset with 32 resampling trees and maximum depth six. At each node, eight candidate cutpoints are generated for each eligible predictor, and the minimum leaf size is the larger of 20 and $0.8\%$ of the number of training data points. The number of candidate predictors at each split is $\min\{p,\max(4,\lceil\sqrt p\rceil)\}$. The conformal score is $S=|Y-\widehat\mu(X)|$.

The residual-scale estimate $\widehat\rho(X)$ is fitted once from five-fold out-of-fold absolute residuals on the fixed training subset. Conditional $0.1$, $0.5$, and $0.9$ quantiles of $Y$ given $X$ are estimated by an ensemble of regression trees fitted once on the same fixed subset, averaging the corresponding within-leaf empirical quantiles across trees. Together with $\widehat\mu$ and $\widehat\rho$, these estimates form the covariate representation $A$ used in Sections~5.2 and 5.3 of the main paper. Interval endpoints are truncated below at zero.

\paragraph{Semi-supervised marginal calibration.}
SS4 is implemented as in Section~5.2 of the main paper, with the required means estimated from the pooled calibration and unlabeled covariates. CF-RF uses the cross-fitting construction in Section~S2.3 with the covariate representation $A$ defined above. The unlabeled subset is used only through its covariates for both procedures.

\paragraph{Conditional calibration.}
The six procedures follow Section~5.4 of the main paper, using the application-specific group indicator $B(X)$ defined in Section~6.3. Marg90 and Group90 use score summaries from the external-reference subset, while the required means for SS2-Marg and SS2-Group are estimated from the pooled calibration and unlabeled covariates. These four side-informed procedures compute weights for all calibration data points once using the empirical likelihood criterion and then minimize the weighted check loss at level $0.9$, as in Section~S2.4.

\subsection{Evaluation Criteria}
\label{sec:supp-applications-criteria}

For marginal calibration, coverage and mean interval length on the fixed test sample are computed in each repetition as in Section~5.1 of the main paper. Tables 7 and 8 of the main paper report the corresponding coverage variance ratios and mean interval length variance ratios. Here we additionally report the mean and standard deviation (SD) of coverage across repetitions, and mean interval length relative to standard split conformal prediction (Len/Base). In the tables, SC denotes standard split conformal prediction. In Tables~\ref{tab:supp-applications-external} and \ref{tab:supp-applications-semi}, the coverage variance ratio and mean interval length variance ratio are relative to standard split conformal prediction.

For conditional calibration, $\|\widehat M_b\|_1$, the coverage variance ratio, and the mean interval length variance ratio are defined as in Section~5.4 of the main paper, with $n_{\mathrm{test}}$ given in Table~\ref{tab:supp-applications-allocation}. Both ratios use OQR as the reference, and the coverage variance ratio is computed using the normalized trace. In addition to the quantities reported in Table~9 of the main paper, we give global and group-specific mean coverage, the two variance ratios for the group-specific coverage moments from Section~S2.4, and Len/Base relative to OQR. The variance ratios for the group-specific coverage moments compare procedures using the same class of group-specific thresholds as OQR.

\subsection{Detailed Results}
\label{sec:supp-applications-results}

\paragraph{Externally specified marginal moments.}
Table~\ref{tab:supp-applications-external} supplements the externally specified-moment results in Section~6.2 of the main paper with mean coverage, coverage standard deviation, and relative mean length. Mean coverage for the four specifications differs from the corresponding value for standard split conformal prediction by at most $0.006$. Their relative mean interval lengths range from $0.992$ to $1.012$.

\begin{table}[!htbp]
  \centering
  \linespread{1}\selectfont
  \setlength{\belowcaptionskip}{6pt}
  \caption{Results for externally specified marginal moments over 200 repeated splits. Each ratio is relative to standard split conformal prediction within the same data set. The mean length variance ratio denotes the mean interval length variance ratio.}
  \label{tab:supp-applications-external}
  \linespread{1}\fontsize{8.5}{9.5}\selectfont
  \setlength{\tabcolsep}{3.2pt}
  \renewcommand{\arraystretch}{1}
  \begin{tabular}{llrrrrr}
    \toprule
    Data set & Method & Coverage & SD & \shortstack{Coverage\\variance\\ratio} & \shortstack{Mean length\\variance\\ratio} & Len/Base\\
    \midrule
    Steel industry energy & SC & 0.898 & 0.020 & 1.000 & 1.000 & 1.000\\
    & Ext-Scale & 0.899 & 0.017 & 0.726 & 0.788 & 1.005\\
    & Ext-ScoreMean & 0.900 & 0.012 & 0.344 & 0.331 & 1.001\\
    & Ext-Score95 & 0.899 & 0.015 & 0.603 & 0.562 & 0.999\\
    & Ext-Score90 & 0.902 & 0.008 & 0.144 & 0.140 & 1.008\\
    \addlinespace
    Seoul bike sharing & SC & 0.893 & 0.019 & 1.000 & 1.000 & 1.000\\
    & Ext-Scale & 0.895 & 0.017 & 0.767 & 0.802 & 1.005\\
    & Ext-ScoreMean & 0.896 & 0.012 & 0.357 & 0.376 & 1.007\\
    & Ext-Score95 & 0.895 & 0.014 & 0.493 & 0.488 & 1.004\\
    & Ext-Score90 & 0.898 & 0.006 & 0.111 & 0.107 & 1.010\\
    \addlinespace
    SGEMM runtime & SC & 0.904 & 0.019 & 1.000 & 1.000 & 1.000\\
    & Ext-Scale & 0.906 & 0.017 & 0.790 & 0.853 & 1.012\\
    & Ext-ScoreMean & 0.906 & 0.014 & 0.486 & 0.422 & 0.994\\
    & Ext-Score95 & 0.905 & 0.014 & 0.497 & 0.459 & 0.992\\
    & Ext-Score90 & 0.909 & 0.008 & 0.165 & 0.163 & 1.000\\
    \bottomrule
  \end{tabular}
\end{table}

\paragraph{Semi-supervised marginal calibration.}
Table~\ref{tab:supp-applications-semi} supplements the semi-supervised results in Section~6.2 of the main paper with mean coverage, coverage standard deviation, and relative mean length. Across standard split conformal prediction and the two semi-supervised procedures, mean coverage ranges from $0.893$ to $0.906$. Relative mean interval lengths for SS4 and CF-RF are between $1.000$ and $1.007$. The coverage standard deviations also reflect the variance-ratio comparison: for example, SD decreases from $0.019$ under standard split conformal prediction to $0.016$ under SS4 and $0.017$ under CF-RF for Seoul Bike Sharing.

\begin{table}[!htbp]
  \centering
  \linespread{1}\selectfont
  \setlength{\belowcaptionskip}{6pt}
  \caption{Detailed semi-supervised marginal-calibration results over 200 repeated splits. All ratios are relative to standard split conformal prediction within the same data set. The mean length variance ratio denotes the mean interval length variance ratio.   }
  \label{tab:supp-applications-semi}
  \linespread{1}\fontsize{8.5}{9.5}\selectfont
  \setlength{\tabcolsep}{3.2pt}
  \renewcommand{\arraystretch}{1}
  \begin{tabular}{llrrrrr}
    \toprule
    Data set & Method & Coverage & SD & \shortstack{Coverage\\variance\\ratio} & \shortstack{Mean length\\variance\\ratio} & Len/Base\\
    \midrule
    Steel industry energy & SC & 0.898 & 0.020 & 1.000 & 1.000 & 1.000\\
    & SS4 & 0.899 & 0.017 & 0.705 & 0.742 & 1.004\\
    & CF-RF & 0.900 & 0.017 & 0.749 & 0.795 & 1.007\\
    \addlinespace
    Seoul bike sharing & SC & 0.893 & 0.019 & 1.000 & 1.000 & 1.000\\
    & SS4 & 0.895 & 0.016 & 0.706 & 0.740 & 1.006\\
    & CF-RF & 0.894 & 0.017 & 0.796 & 0.835 & 1.004\\
    \addlinespace
    SGEMM runtime & SC & 0.904 & 0.019 & 1.000 & 1.000 & 1.000\\
    & SS4 & 0.906 & 0.016 & 0.702 & 0.743 & 1.002\\
    & CF-RF & 0.905 & 0.017 & 0.720 & 0.692 & 1.000\\
    \bottomrule
  \end{tabular}
\end{table}

\paragraph{Conditional calibration.}
Table~\ref{tab:supp-applications-conditional} reports the group-specific coverages and variance ratios for the group-specific coverage moments that complement Table~9 of the main paper.

\begin{table}[!htbp]
  \centering
  \linespread{1}\selectfont
  \setlength{\belowcaptionskip}{6pt}
  \caption{Conditional-calibration results over 200 repeated splits. All ratios are relative to OQR within the same data set. The coverage variance ratio is computed using the normalized trace, and the mean length variance ratio denotes the mean interval length variance ratio.}
  \label{tab:supp-applications-conditional}
  \linespread{1}\fontsize{8.5}{9.5}\selectfont
  \setlength{\tabcolsep}{2.2pt}
  \renewcommand{\arraystretch}{1}
  \begin{tabular}{llrrrrrrrrr}
    \toprule
    Data set & Method & \shortstack{Global\\coverage} & \shortstack{Cov.\\$B=0$} & \shortstack{Cov.\\$B=1$} & $\|\widehat M_b\|_1$ & $R_0^{\rm emp}$ & $R_1^{\rm emp}$ & \shortstack{Coverage\\variance\\ratio} & \shortstack{Mean length\\variance\\ratio} & Len/Base\\
    \midrule
    \shortstack[l]{Steel industry\\energy} & SC & 0.898 & 0.952 & 0.876 & 0.035 & -- & -- & -- & 0.956 & 0.978\\
    & OQR & 0.893 & 0.882 & 0.898 & 0.023 & 1.000 & 1.000 & 1.000 & 1.000 & 1.000\\
    & Marg90 & 0.896 & 0.882 & 0.901 & 0.018 & 0.980 & 0.374 & 0.675 & 0.413 & 1.003\\
    & Group90 & 0.900 & 0.893 & 0.903 & 0.012 & 0.287 & 0.176 & 0.232 & 0.234 & 1.016\\
    & SS2-Marg & 0.894 & 0.883 & 0.899 & 0.021 & 0.932 & 0.655 & 0.793 & 0.742 & 0.998\\
    & SS2-Group & 0.894 & 0.884 & 0.899 & 0.019 & 0.692 & 0.617 & 0.655 & 0.686 & 0.995\\
    \addlinespace
    \shortstack[l]{Seoul bike\\sharing} & SC & 0.893 & 0.858 & 1.000 & 0.056 & -- & -- & -- & 1.122 & 1.037\\
    & OQR & 0.884 & 0.888 & 0.874 & 0.025 & 1.000 & 1.000 & 1.000 & 1.000 & 1.000\\
    & Marg90 & 0.886 & 0.890 & 0.874 & 0.019 & 0.230 & 1.000 & 0.616 & 0.289 & 1.004\\
    & Group90 & 0.892 & 0.894 & 0.888 & 0.012 & 0.125 & 0.384 & 0.254 & 0.177 & 1.018\\
    & SS2-Marg & 0.885 & 0.888 & 0.875 & 0.023 & 0.760 & 0.987 & 0.874 & 0.810 & 1.000\\
    & SS2-Group & 0.885 & 0.888 & 0.876 & 0.023 & 0.798 & 0.922 & 0.860 & 0.852 & 1.001\\
    \addlinespace
    \shortstack[l]{SGEMM\\runtime} & SC & 0.904 & 0.976 & 0.839 & 0.068 & -- & -- & -- & 1.003 & 0.852\\
    & OQR & 0.896 & 0.893 & 0.900 & 0.021 & 1.000 & 1.000 & 1.000 & 1.000 & 1.000\\
    & Marg90 & 0.900 & 0.893 & 0.906 & 0.018 & 0.966 & 0.506 & 0.737 & 0.553 & 1.024\\
    & Group90 & 0.904 & 0.900 & 0.908 & 0.010 & 0.216 & 0.205 & 0.210 & 0.264 & 1.044\\
    & SS2-Marg & 0.897 & 0.893 & 0.900 & 0.019 & 0.883 & 0.687 & 0.787 & 0.653 & 0.998\\
    & SS2-Group & 0.898 & 0.894 & 0.901 & 0.017 & 0.762 & 0.557 & 0.661 & 0.565 & 1.005\\
    \bottomrule
  \end{tabular}
\end{table}

The variance ratios for the group-specific coverage moments clarify the covariance comparisons summarized by the coverage variance ratio. Group90 reduces both group-specific coverage-moment variances in every application, with $R_0^{\rm emp}$ and $R_1^{\rm emp}$ between $0.125$ and {$0.384$}. For Marg90, the reductions differ between groups. For example, its two ratios are {$0.980$ and $0.374$ for Steel Industry Energy.} The semi-supervised specifications reduce the coverage variance ratio in every application, with SS2-Group giving values between $0.655$ and $0.860$. Relative mean interval lengths remain between $0.995$ and $1.044$ across all side-informed conditional procedures.

\bibliographystyle{agsm-noin}
\bibliography{refs}